\PassOptionsToPackage{dvipsnames}{xcolor}
\documentclass[a4paper,11pt]{article}
\usepackage[utf8]{inputenc}
\usepackage{appendix}
\usepackage{jheparxiv}
\usepackage{amsfonts,amsmath,amssymb,amsthm,mathrsfs,mathtools,bm}
\usepackage[final]{showkeys}
\usepackage{comment}
\usepackage[normalem]{ulem}
\usepackage{faktor}
\usepackage{bigints}
\usepackage{braket}
\usepackage[inline]{enumitem}
\usepackage{tabularx}
\usepackage{tikz-cd}
\usepackage{mathabx}
\usepackage[pass]{geometry}

\makeatletter
\newsavebox{\@brx}
\newcommand{\llangle}[1][]{\savebox{\@brx}{\(\m@th{#1\langle}\)}%
  \mathopen{\copy\@brx\kern-0.5\wd\@brx\usebox{\@brx}}}
\newcommand{\rrangle}[1][]{\savebox{\@brx}{\(\m@th{#1\rangle}\)}%
  \mathclose{\copy\@brx\kern-0.5\wd\@brx\usebox{\@brx}}}
\makeatother

\RequirePackage[dvipsnames]{xcolor}
\definecolor{arXiv}{named}{OliveGreen}
\definecolor{ColorCite}{named}{BrickRed}
\definecolor{ColorLink}{named}{NavyBlue}
\definecolor{ColorURL}{named}{RoyalBlue}

\newtheorem{proposition}{Proposition}[section]
\newtheorem{definition}[proposition]{Definition}

\newtheorem{theorem}[proposition]{Theorem}
\newtheorem{corollary}[proposition]{Corollary}
\theoremstyle{definition}
\newtheorem{remark}[proposition]{Remark}

\newcommand{\E}{\mathcal{E}}
\newcommand{\T}{\mathcal{T}}
 \newcommand{\q}{q}
\renewcommand{\P}{\mathcal{P}}
   \newcommand{\N}{\mathcal{N}}
     \newcommand{\K}{\mathcal{K}}
 \newcommand{\C}{\mathcal{C}}
 
  \newcommand{\zetab}{\bar{\zeta}}
\newcommand{\betab}{\bar{\beta}}
 \newcommand{\bz}{\bar{z}}

 \newcommand{\p}{\partial}
 \newcommand{\ISO}[1][2]{\widetilde{ISO}(#1)}

\newlength{\blength}
\renewcommand{\proof}[1]{\vspace{-.05cm}
\begin{list}{\bf Proof:}
{\listparindent=\parindent\parsep=0pt \labelwidth=-0.5cm
\labelsep=\parindent \addtolength{\labelsep}{-\blength}
\addtolength{\labelsep}{1.5cm} \itemindent=-\blength
\addtolength{\itemindent}{\parindent} \leftmargin=1.0cm} \item
#1~$\qedsymbol$\end{list} \vspace{.0cm}}

\title{BMS representations for generic supermomentum}

\author{Xavier Bekaert and Yannick Herfray}

\affiliation{Institut Denis Poisson UMR-CNRS 7013, Université de Tours, 
Parc de Grandmont, 37200 Tours, France}

\emailAdd{xavier.bekaert@univ-tours.fr, yannick.herfray@univ-tours.fr}

 \abstract{We revisit the classification, and give explicit realisations, of unitary irreducible representations of the BMS group. As compared to McCarthy's seminal work, we make use of a unique, Lorentz-invariant, decomposition of supermomenta into a hard and a soft piece, that we introduce and properly define, to investigate the extent to which generic representations depart from usual Poincaré particles and highlight their relations to gravitational infrared physics. We insist on making wavefunctions as explicit as possible. Similarly, we explain how branching to a Poincaré subgroup works in practice: this is physically relevant because this amounts to reading off the field content of a given BMS state in terms of a choice of gravity vacuum. In particular, we emphasise how different gravity vacua differ in their interpretation of the same BMS state, here again providing concrete examples as well as the general procedure. Finally, we demonstrate on an example that generic BMS particles are flexible enough to encode memory, as opposed to usual Poincaré particles.}

\begin{document}

\maketitle
 
\section{Introduction}

The Bondi-van der Burg--Metzner-Sachs group, i.e. $BMS_4 \simeq SO_0(3,1) \ltimes \E[1]$,\footnote{The vector space $\E[w]$ is spanned by the conformal densities of weight $w$ on the celestial sphere $S^2$, see our conventions in Section \ref{Section: supertranslationsmomenta}.} was introduced in \cite{bondi_gravitational_1962,Sachs:1962wk} as the group of asymptotic symmetries associated to an asymptotically flat spacetime. It can also be realised as conformal isometries of null infinity, see e.g. \cite{Geroch1977}, alternatively known as conformal Carroll symmetries \cite{Duval:2014uva}. Importantly, this is the smallest possible gravitational symmetry group in the following sense:  fixing a Poincaré subgroup $ISO_0(3,1) \subset BMS_4$ completely freezes the gravitational degrees of freedom. Accordingly, the space $\mathbb{V} = BMS_4 \,/\, ISO_0(3,1)$ of gravity vacua \cite{Ashtekar:1987tt}, which gathers all the Poincaré subgroups that could be realised in this way, corresponds to all possible inequivalent choices of flat backgrounds, which gravity treats on equal footing. More recently, it has been realised \cite{Strominger:2013jfa,He:2014laa} that the BMS group is a symmetry of the gravitational scattering problem -- at least in the perturbative framework of Quantum Field Theory (QFT) -- and, more generally, of any scattering problem coupled to gravity. This crucially relies on Strominger's observation \cite{Strominger:2013jfa} that the BMS group can be realised as a global symmetry, acting at past and future null infinity simultaneously. This group being the symmetry group of any QFT coupled to gravity, one is lead to reconsider the admissible notions of particles. Here, BMS particles will be understood as unitary irreducible representations (UIRs) of the BMS group.

The idea that elementary particles could be defined as unitary irreducible representations of the BMS group goes back to McCarthy \cite{McCarthy_71,Mccarthy:1972ry}. In a series of pioneering works, he studied and classified the UIRs of $BMS_4$ \cite{McCarthy_71,Mccarthy:1972ry,McCarthy_72-I,McCarthy_73-II,McCarthy_73-III,McCarthy:1974aw,McCarthy_75,McCarthy_76-IV,McCarthy_78,McCarthy_78errata} (see also the important works \cite{Cantoni:1966,Cantoni_1967,Cantoni:1967,Girardello:1974sq}).\footnote{The exotic representation theory of the split signature group $SO(2,2) \ltimes \E[1]$ was studied by Melas (see e.g. the reviews \cite{Melas:2002xt,Melas_2011,Melas:2016fqi} and refs therein).} The UIRs of $BMS_3$, including super-rotations, were studied in \cite{Barnich:2014kra,Barnich:2015uva,Oblak:2016eij,Melas:2017jzb}. For the generalised BMS group in 4D, $Diff(S^2)\ltimes \E[1]$,  much fewer results are available, see \cite{Freidel:2024jyf}. 

Among BMS representations, Sachs' \cite{Sachs:1962wk} and Longhi--Materassi's \cite{Longhi:1997zt} representations stands out as being the lift to BMS of Poincaré UIRs. Despite the fact that McCarthy's classification demonstrated that these representations are rather special, and leave aside infinitely many others, very few progresses have been made until recently on the physical understanding of generic BMS representations. In a recent work with Laura Donnay \cite{Bekaert:2024jxs}, the authors showed that wavefunctions of any BMS particle can be explicitly realised in terms of wavefunctions $\Psi_{\C}(p)$ of usual Poincaré particles, each associated to a different gravity vacuum $\partial_z^2\C \in \mathbb{V}$. As such, BMS particles can always be thought of as quantum superpositions of usual Poincaré particles, each of them propagating on a different background. The present work intends to further decipher the physical meaning of generic BMS representations by reconsidering McCarthy's results in view of a decomposition of supermomenta into hard and soft pieces that we discuss below. We also provide provide some proofs for results that where hinted at in \cite{Bekaert:2024jxs}. Another goal of the article is to try to make all constructions (such as the wavefunctions, norms, actions of the group, branching rules, etc) as explicit as possible.\\

Supermomenta $\P(z,\bz) \in \E[-3]$ generalise the usual momenta of Poincaré particles. They are dual to supertranslations $\T(z,\bz) \in \E[1]$ and closely related to BMS charges and fluxes (see e.g. \cite{Barnich:2021dta, Donnay:2021wrk}). The present article relies on the following result, here stated in a loose form (see Section \ref{Section: Hard/Softdecompositionofsupermomenta} 
for details): supermomenta $\P(z,\bz)$ can always be decomposed in a unique, Lorentz-invariant, way as 
\begin{equation}\label{decomphard+soft}
    \P(z,\bz) = P(z,\bz) + \partial^2_z \partial^2_{\bz} \N(z,\bz)\,,
\end{equation}
where the soft charge $\partial^2_z \partial^2_{\bz} \N(z,\bz)$, with $\N \in \E[1]$, parametrises the departure of the representation from the hard ones, while the hard part $P(z,\bz)$ of the supermomentum is a non-linear function of the momentum $p_{\mu} \in (\mathbb{R}^{3,1})^*$. More explicitly,
\begin{equation}
   P(z,\bz) = \left \{ \hspace*{0.75cm} \begin{array}{crcrl}
       -\frac{m^4}{\pi}\big(p_{\mu}q^{\mu}\big(z,\bz)\big)^{-3} &\hspace*{0.75cm}&\text{``massive BMS reps''} &\hspace*{0.75cm}& p^2 =-m^2 \\[0.6em]
    \omega\,\delta^{(2)}\big(z-\zeta, \bz - \zetab\big) &&\text{``massless BMS reps''} && p_{\mu} = \omega\,q_{\mu}(\zeta,\zetab)
   \end{array} 
   \right. \,.
\end{equation}
Here $q^{\mu}(z,\bz) \in\E[1] \otimes \mathbb{R}^{3,1}$ is the weighted null vector realising the canonical inclusion of translations into supertranslations (see Section \ref{Section: supertranslationsmomenta} for our conventions). 

The physical relevance of the above decomposition \eqref{decomphard+soft} is the following. First of all, when $p_\mu\neq 0$ and $\N(z,\bz)=0$ one recovers Sachs' \cite{sachs_asymptotic_1962} (massless case) and Longhi--Materassi's representations \cite{Longhi:1997zt} (massive case) - see \cite{He:2014laa,Campiglia:2015kxa} for relations to the corresponding soft theorems. These hard representations are the simplest UIRs of BMS group and, as we already highlighted, they coincide with the usual UIRs of the Poincaré group. More generally, the decomposition allows to discriminate between trivial $(p_{\mu}=0$, $\N(z,\bz)=0$), hard  $(p_{\mu}\neq0$, $\N(z,\bz)=0$), soft $(p_{\mu}=0$, $\N(z,\bz)\neq 0$) and generic $(p_{\mu}\neq 0$, $\N(z,\bz)\neq 0$) representations of the BMS group. Hard representations are very special: \emph{for generic representations the soft charge will be non-zero and parametrises to which extent a BMS particle differs from a usual Poincar\'e particle}. The relevance of the more exotic (i.e. soft and generic) representations of $BMS_4$ comes from the fact that hard representations cannot, by themselves, preserve supermomentum: if $\{p_i^{\mu}\}$, $\{p_f^{\mu}\}$ are families of momenta satisfying momentum conservation, then the corresponding hard supermomenta will not fulfil the corresponding conservation law\footnote{This point has been especially emphasised, for the massive case, in \cite{Chatterjee:2017zeb}.}
\begin{equation}
\sum_i p_i^{\mu} = \sum_f p_f^{\mu}\quad\Longrightarrow  \quad  \sum_i P_i(z,\bz) \neq \sum_f P_f(z,\bz).
\end{equation}
Therefore, in order for a process to satisfy supermomentum conservation, representations with $\N (z,\bz) \neq 0$ are needed. This a representation-theoretic rephrasing of the result of \cite{Kapec:2017tkm,Choi:2017ylo} that hard particles cannot satisfy the conservation of BMS charges by themselves and that, in the language of these works, soft particles carrying non-vanishing soft charges need to be included in a scattering process -- infrared divergences being the price to pay for not doing so. However, while the corresponding dressed states \cite{Kulish:1970ut} are typically plagued with ambiguities, and the extent to which they belong to a properly defined Fock space is a subtle question, representation theory offers a very different perspective: BMS particles with non-zero soft charges are just as good as usual hard particles. For example, the corresponding Hilbert spaces are separable. What is more, generic representations typically contain states which, when interpreted in a given gravity vacuum -- i.e. when decomposed into UIRs of the corresponding Poincaré subgroup -- genuinely coincide with usual particles. The new feature, due to the presence of the soft charge, is that such states are tied up to a given gravity vacuum $\C_0$ : in any other vacuum $\C$, the BMS particle will appear as a quantum superposition of particles with all possible -- including continuous -- spins (see Section \ref{Ssection: Branching in different gravity vacua}). Here again, the aim is to investigate those generic representations in a way which is as concrete as possible and provide in particular explicit wavefunctions, norms, transformation laws, as well as branching rules under restriction to a Poincaré subgroup.

There is another sense in which the representations needed for conservation of momentum are ``generic''. It is clear that one should not expect the corresponding supermomentum $\P$ to have any particular symmetry. Typically, it will not be preserved by any Lorentz transformation. In other words, the (BMS) little group will be trivial. This is in contrast with McCarthy's emphasis on representations with non-trivial little groups, which is natural from the perspective of classifying all representations. Generic supermomenta will have trivial little group and the corresponding representations might be the most physically relevant (and the other representations, oddities). As such, they deserve a close inspection, which is another goal of this article. 

Finally, we wish to comment on an important point, emphasised very early on \cite{Ashtekar:1987tt} by Ashtekar: usual (hard) representations of the BMS group cannot accommodate the memory effect. This is because the $L^2$ norm on the states is too stringent to allow for the energy pole in the wavefunction that would encode memory (both of which relate \cite{Strominger:2014pwa} to Weinberg's soft theorems in QFT). As we shall exhibit on a particular example of  BMS representation, 
it is nevertheless possible to construct explicitly BMS particle states which bypass this problem (see Section \ref{Section: memoryeffect}).\\

The plan of the paper is as follows. In Section \ref{Section: supertranslationsmomenta}, we review basic definitions and properties of supertranslations and supermomenta. In Section \ref{Section: Hard/Softdecompositionofsupermomenta}, we introduce the decomposition of supermomenta into hard and soft pieces. In Section \ref{Section: McCarthy}, we review Wigner's and McCarthy's classification of the UIRs of the Poincar\'e and BMS groups, respectively, as well as the branching rules from one to the other. In Section \ref{Section: HardrepresentationsoftheBMSgroup}, we discuss hard representations of the BMS group in some details (little groups, branching rules, interpretation as scattering data). In Sections \ref{Section: MasslessUIRsBMS} and \ref{Section: MassiveUIRsBMS}, we work out explicitly some aspects (wavefunctions, norms, actions of the group and branching rules) of, respectively massless and massive, BMS particles. In Section \ref{Section: BMS wavefunctions}, we make contact with \cite{Bekaert:2024jxs} by discussing off-shell and on-shell wavefunctions of BMS particles as functions on suitable homogeneous spaces. In Section \ref{Section: memoryeffect}, we shed some new light on the memory effect in the context of UIRs of the BMS group by discussing explicitly a particular example.
Finally, in Section \ref{Section: concl} we briefly summarise our main results.
Some technical proofs (Appendices \ref{Appendix: proofsPaneitz}-\ref{Appendix: Distributional identity for hard supermomenta}) and computations (Appendices \ref{Appendix: Haar mesure}-\ref{Appendix: transformedsupermomentum}), as well as a review on continuous spin representations (Appendix \ref{Appendix: CSP}),
have been moved to the appendix.

\section{Supertranslations and supermomenta}\label{Section: supertranslationsmomenta}

We will here review some basic definitions and properties of conformal densities, supertranslations and supermomenta.

Everywhere, $(z,\bar z)$ stands for stereographic coordinates on the celestial sphere $S^2$, identified with the complex projective line $\mathbb{C}P^1$. Whenever an explicit expression appears for conformal densities\footnote{By default, all fields on the celestial sphere considered in this paper will always be assumed real, globally defined and smooth everywhere. This will most often be understood implicitly to lighten the text.} they are implicitly written with respect to the flat metric\footnote{This unusual normalisation has been chosen for later purpose.}: $h = 4d z d\bar z$. With this definition and the embedding $q^{\mu} : S^2 \hookrightarrow \mathbb{R}^{3,1}$ of the celestial sphere into the null cone, defined by \eqref{q definition}, this metric also coincides with the pull back $q^* \eta$  of the Minkowksi metric: $h = d q^{\mu} dq^{\nu} \eta_{\mu\nu} =  4d z d\bar z$.

\subsection{Conformal densities on the celestial sphere}

\subsubsection{Conformal densities}\label{confdens}

\begin{definition}\label{confdensity}
A conformal density of weight $w\in\mathbb R$ on the celestial sphere is defined as an equivalence class $[\,\tilde g,\tilde f\,]$ of pairs $(\tilde g,\tilde f)$ of (necessarily conformally flat) metrics $\tilde g$ on $S^2$ and of real functions $\tilde f$ on $S^2$ with respect to the equivalence relation
\begin{equation}\label{equivrelWeyl}
   \big( \tilde g(z,\bar z) ,\tilde f(z,\bar z) \big) =\left( \Omega(z,\bar z)^2 4dz\,d\bar z ,\, \Omega(z,\bar z)^w f(z,\bar z) \right) \sim \big( 4dz\,d\bar z , f(z,\bar z) \big)
\end{equation}
where $\Omega(z,\bar z)>0$ is a nowhere-vanishing conformal factor.
The vector space of smooth conformal densities of weight $w$ will be denoted $\mathcal{E}[w]$. 
\end{definition}

\noindent With a standard abuse of notation, the equivalence class $[\tilde g,\tilde f]$ will simply be denoted by its representative $f(z,\bz) \in \mathcal{E}[w]$ for the flat metric $4dz d\bz$, as in \eqref{equivrelWeyl}, with its transformation law under Weyl transformations implied. Then $f(z,\bar z)$ has a constrained behavior as $z\to \infty$. To see this, consider the change of chart $\hat{z}= z^{-1}$ and let us note $\left( 4d\hat z\, d\hat{\bar z} , \hat f(\hat{z},\bar {\hat z}) \right)$ the expression of the density in this chart. Since we suppose that the density is defined on the whole celestial sphere, this implies in particular that $\hat f(0,0)$ must be finite. Now by definition
\begin{align*}
    \left( 4d\hat z d\hat{\bar z} , \hat f(\hat{z},\bar {\hat z}) \right) = \left( \frac{1}{|z|^4} 4d z d\bar z , \hat f\left(z^{-1},\bar{z}^{-1}\right) \right) \sim \left( 4d z d\bar z , |z|^{2w}\hat f\left(z^{-1},\bar{z}^{-1}\right) \right)= \left( 4d z d\bar z , f(z,\bar z)\right)
\end{align*}
therefore 
\begin{equation}\label{Conformal density: asymptotic behavior}
    f(z,\bar z) = |z|^{2w}\hat f\left(z^{-1},\bar{z}^{-1}\right) \quad \stackrel{z \to \infty}{\sim} \quad  |z|^{2w}\hat f(0,0).
\end{equation}

\vspace{2mm}

The M\"obius group $PSL(2,\mathbb{C})\simeq \frac{SL(2,\mathbb{C})}{\mathbb{Z}_2}\simeq SO_0(1,3)$ acts on the celestial sphere $S^2\simeq\mathbb{C}P^1$ via fractional linear transformations
\begin{equation}\label{Mobiustransfo}
    \phi :  z  \to z'(z)= \frac{az+b}{cz+d}\,,\quad\text{with}\quad M =\begin{pmatrix}
        a & b \\ c & d
    \end{pmatrix}\in SL(2,\mathbb{C})\,.
\end{equation} 
Its action on a conformal density $f$
is given by the pullback by $\phi$. More explicitly,
\begin{align*}
    \Big( \phi^*(4dz\,d\bar z), (\phi^*f)(z,\bar z) \Big) &= \left( \left|\frac{\partial z'}{\partial z}\right|^2 4dz\,d\bar z , f(z',\bar z') \right)\\
    &\qquad \sim \left( 4dz\,d\bar z , \left|\frac{\partial z'}{\partial z}\right|^{-w} f(z',\bar z') \right) =:\Big( 4d z d\bar z , \big(f\cdot\phi\big)(z,\bar z)\Big)\,\nonumber
\end{align*}
Note that, because it is a pull-back, this is a \emph{right action}, as indicated by the notation $f\cdot \phi$ in the above. For the fractional linear transformation  \eqref{Mobiustransfo}, one has \begin{equation}\label{JacobianSL2C}
\frac{\partial z'}{\partial z} = \frac1{(cz+d)^2}
\end{equation} and its action on a conformal density of weight $w$ will be written
 \begin{equation}\label{SL2Caction}
   f(z,\bar z) \quad \mapsto \quad \big(f\cdot M\big) (z,\bar z) = |cz+ d|^{2w} f\left( \frac{az+b}{cz+d} \,, \frac{\bar a\bar z+\bar b}{\bar c \bar z+\bar d}\right).
\end{equation}   
   Note again that this is a right action, i.e.
 \begin{equation*}
    \Big(f\cdot( M_1M_2 )\Big)(z,\bar z) = \Big((f\cdot M_1) \cdot M_2 \Big)(z,\bar z),\quad\text{for}\quad  M_1,M_2\in SL(2,\mathbb{C}).
\end{equation*}
From time to time it will be useful to make use of the left action given by
\begin{equation}\label{SL2Caction on the left}
   \big(M\cdot f\big) (z,\bar z) := \big(f\cdot M^{-1}\big) (z,\bar z).
\end{equation}
Finally, the definition \ref{confdensity} easily generalises to conformal density tensors by merely replacing the functions $\tilde f$ by tensor fields.

\subsubsection{Spin-weighted densities}

In order to make manifest some $SL(2,\mathbb{C})$-invariance properties, it is useful to introduce spin-weighted functions and related ``edth'' operators (see \cite{Newman:1966ub,Eastwood_Tod_1982} for classical references). We will try to minimise their use in the body of this article but one should keep in mind that most expressions that will subsequently appear, despite being written in a fixed coordinate system $z,\bz$ on the celestial sphere, can be invariantly expressed in terms of suitable, conformally-invariant, combination of edth operators. Accordingly, we review here some basic definitions and properties of these operators. We closely follow \cite{Newman:1966ub,Eastwood_Tod_1982}, up to a sign convention.
\begin{definition}
A spin-weighted density of spin-weight $s\in\mathbb R$ and conformal-weight $w\in\mathbb R$ is defined by an equivalence class $[\tilde m, \tilde \phi]$ of pairs $(\tilde m,\tilde \phi)$ made of a null co-frame $\{\tilde{m}= m_z dz ,\bar{\tilde{m}}=  \bar{m}_{\bz} d \bz\}$ for the metric  $\tilde g = |m_z|^2 dz \,d\bz$ on $S^2$ and of a complex function $\tilde \phi$ on $S^2$, with respect to the equivalence relation
\begin{equation}
    \big(\tilde m(z,\bar z),\tilde \phi(z,\bar z)\big)=\left( e^{i\theta(z,\bar z)}\Omega(z,\bar z) 2dz \,,\, e^{i s \theta(z,\bar z)} \Omega(z,\bar z)^{w} \phi(z,\bar z) \right)
    \sim \big( 2d z , \phi(z,\bar z) \big) 
\end{equation}
where $\Omega(z,\bar z)>0$ is a nowhere-vanishing conformal factor and $\theta(z,\bar z)$ a real phase.
The vector space of densities of smooth spin-weight $s$ and conformal-weight $w$ is denoted $\mathcal{O}(w+s , w-s)$.  Taking $s=0$, one recovers the bundle $\E_{\mathbb{C}}[w] = \mathcal{O}(w,w)$ of complexified weight-$w$ conformal densities .
\end{definition}

By construction, a spin-weighted\footnote{As compare to \cite{Newman:1966ub,Eastwood_Tod_1982} this convention means that $s_{here} = - s_{there}$. The convention chosen here (taken from \cite{Adamo:2021lrv,Adamo:2022mev}), however, has the advantage that it fits nicely with twistor theory and scattering data at null infinity : the asymptotic shear $\bar \sigma \in \mathcal{O}(1,-3)$, with spin weight $+2$, is then the scattering data for a \emph{self-dual} spacetime, the deformation parameter of the corresponding asymptotic twistor space and, when quantised, creates a \emph{positive} helicity graviton.} density $\phi(z,\bar z) \in \mathcal{O}\left(w+ s,w - s\right)$ defines a conformal density \emph{tensor} on the sphere $\bm{\phi}$ of weight $w+s$,
\begin{equation*}
    \bm{\phi} = \phi(z,\bar z) (2d\bz)^{s},
\end{equation*}
i.e. an equivalence class $[\,\tilde g,\bm{\tilde\phi}\,]$ of pairs $(\tilde g,\bm{\tilde\phi})=( \tilde{m}\,\bar{\tilde{m}},\tilde\phi\,\bar{\tilde{m}}^s)$ with respect to the equivalence relation
$(\tilde g,\bm{\tilde\phi})=(\Omega^2 4dz\,d\bar z,\Omega^{w+s}\bm{\phi})\sim(4dz\,d\bar z,\bm{\phi})$.

Introducing the homogeneous coordinates $[\lambda_\alpha]=\begin{bmatrix} \lambda_1 \\ \lambda_2 \end{bmatrix}$ on $\mathbb{C}P^1$, spin-weighted densities $\phi\in\mathcal{O}\left(w+ s , w - s \right)$ are in one-to-one correspondence with functions $\Phi(\lambda_{\alpha}, \bar \lambda_{\dot\alpha})$ on $\mathbb{C}^2\setminus\{\,0\,\}$ (where $\alpha,\dot\alpha=1,2$) satisfying the following homogeneity condition\footnote{\label{Gelfandfootnote}The space of all such homogeneous functions which are smooth everywhere, except perhaps at the origin, is denoted $D_{(w+s+1,w-s+1)}$ by Gelfand et al \cite[Section III.2.2]{Gelfand2} in their textbook on generalised functions and representations of $SL(2,\mathbb{C})$. This is the framework used by McCarthy (see e.g. \cite{McCarthy_75} for a concise review).}
\begin{equation*}
    \Phi(t\,\lambda_{\alpha}, \bar{t}\,\bar \lambda_{\dot\alpha})= t^{w +s} \,\bar{t}^{\,w -s} \Phi(\lambda_{\alpha}, \bar \lambda_{\dot\alpha}).
\end{equation*}
The relationship with the previous definition is obtained by taking $[\lambda_\alpha]= \left[e^{i\frac{\theta}{2}} \sqrt{\Omega} \begin{pmatrix} 1 \\ z \end{pmatrix} \right]$ and

\begin{equation*}
    \big(\tilde m(z,\bar z),\tilde \phi(z,\bar z)\big)= \big( 2\lambda_{\alpha} d\lambda^{\alpha} , \Phi(\lambda_{\alpha}, \bar \lambda_{\dot\alpha})\big) = \left( e^{i\theta(z,\bar z)}\Omega(z,\bar z) 2dz \,,\, e^{i s \theta(z,\bar z)} \Omega(z,\bar z)^{w} \phi(z,\bar z) \right)
\end{equation*}
where indices have been raised using the Levi-Civita antisymmetric tensor $\epsilon^{\alpha\beta}$ such that $\epsilon^{12}=1$ and
\begin{equation}\label{homogeneouscorrespondence}
\phi(z,\bar z) = \Phi(1,z,1,\bar z)
\end{equation}
in a patch where $\lambda_1\neq 0$.

Let $\gamma^{-2} 4dz d\bz$ be the round sphere metric with $\gamma = (1+ |z|^2)$. The edth operators $\eth$ and $\bar \eth$ are defined as\footnote{We here take the definition of \cite[p.323]{Eastwood_Tod_1982}.  It differs from the one in \cite{Newman:1966ub} by a factor $\gamma = P_{there}$ and is such that $\eth^k f = \gamma^{-k}(\eth_{there})^k f$. See also \cite[Sections III.3.3 and III.5.2]{Gelfand2} for the same construction with different notations.}
\begin{align}
   \eth &\left| \begin{array}{ccc}
         \mathcal{O}\left( h , \bar h \right) & \to & \mathcal{O}\left( h-2 , \bar h \right)  \\[0.3em]
         \phi(z,\bar z) & \mapsto 
         & \gamma^{h} \partial_{z} \left( \gamma^{-h}\phi(z,\bar z) \right)
    \end{array}\right.,
\end{align} 
and the corresponding complex conjugated expression, where $(h,\bar h)\in\mathbb R^2$ is any couple of real numbers. The edth operator  $\eth$ is $SU(2)$-covariant but, due to the explicit appearance of $\gamma$, it is not $SL(2,\mathbb{C})$-covariant unless $h=0$. Nevertheless, for any non-negative integer $k\in\mathbb N$, the following powers
\begin{align}
   \eth^{k} &\left| \begin{array}{ccc}
         \mathcal{O}\left( k-1, \bar h \right) & \to & \mathcal{O}\left( -k-1, \bar h \right)  \\[0.3em]
         \phi(z,\bar z) & \mapsto 
         & \p_{z}^{k} \phi(z,\bar z)
    \end{array}\right.,
\end{align} 
are $SL(2,\mathbb{C})$-covariant, as manifested here by the fact that $\gamma$ drops out from the expressions (which can be checked explicitly using that $\partial_z^2\gamma=0$). To make the $SL(2,\mathbb{C})$-covariance manifest, it is useful to keep in mind the following identity for spin-weighted densities $\phi\in\mathcal{O}\left( k-1 , \bar h \right)$
\begin{equation}
    \frac{\partial }{\partial \lambda_{\beta_1}}\,\cdots\,\frac{\partial }{\partial \lambda_{\beta_k}}\, \Phi(\lambda_{\alpha}, \bar \lambda_{\dot\alpha}) = (-1)^k
    \lambda^{\beta_1}\cdots\lambda^{\beta_k} \Big(\eth^k \Phi \Big) (\lambda_{\alpha}, \bar \lambda_{\dot\alpha}),
\end{equation}
where undotted indices have been raised using the Levi-Civita antisymmetric tensor $\epsilon^{\alpha\beta}$.

\subsection{Basic properties of supertranslations and supermomenta}

In this subsection, we briefly review the definitions of supertranslations and supermomenta, as well as their properties that will be useful in the sequel.

\subsubsection{Basic properties of supertranslations}\label{basicpropertiesofsupertranslations}

\begin{definition}
A supertranslation $\T(z,\bar z) \in \E[1]$ is a conformal density of weight $1$ on the celestial sphere. A translation $T^{\mu}$ is an element of Minkowski spacetime $\mathbb{R}^{3,1}$ with Cartesian coordinates $x^\mu$ ($\mu=0,1,2,3$). The BMS group is the semi-direct product $BMS_4 \,\simeq\, SO_0(3,1)\ltimes\E[1]$ of the Lorentz group acting on the additive group of supertranslations.\footnote{We consider the connected component of the Poincar\'e and BMS groups, $ISO_0(3,1)$ and $BMS_4$ respectively. All projective UIRs of these groups lift to linear UIRs of their universal (i.e. double) covering \cite{McCarthy_78}. Therefore, as is usual in physics literature, the adjective ``projective''  will be understood implicitly below. Hence, the representations are either ``single-valued'' or ``double-valued'' whether they involve integer or half-integer helicities.}
\end{definition}

\begin{proposition}\label{Proposition: Injection translation -> super}
There is a canonical, $SL(2,\mathbb{C})$-equivariant, linear injection
\begin{equation}\label{Injection translation -> super}
    q_{\mu}\left|\; \begin{matrix}
        &\mathbb{R}^{3,1} &  \hookrightarrow &  \mathcal{E}[1] \\[1em]
        & T^{\mu} & \mapsto \quad& \T(z,\bar z)
    \end{matrix} \right.
\end{equation}
of translations into supertranslations given by
\begin{equation}\label{q definition}
    q_{\mu}(z,\bar z ) = \begin{pmatrix}
        -1- |z|^2 ,& z+\bar z ,& -i(z-\bar z) ,&   1 -|z|^2
    \end{pmatrix}
\end{equation}
i.e.
\begin{equation}\label{Injection translation: image}
\T(z,\bar z)=T^\mu q_{\mu}(z,\bar z )= (-T^0+T^3) + (T^1-i\,T^2) \; z + (T^1+i\,T^2) \;\bar z + (-T^0-T^3)\;|z|^2.
\end{equation}
The metric on $\mathbb{R}^{3,1}$ is obtained by computing the scalar curvature $R$ of the metric $\frac{4dz d\bar z}{\T^{2}(z,\bar z)}$, i.e.
\begin{align*}
   -\frac{R}{2}&= -(T^0)^2 + (T^1)^2 +(T^2)^2 + (T^3)^2\,.
\end{align*}

\end{proposition}

\proof{
Let us introduce the two-component spinor $\lambda_{\alpha} = (1,z)$ of $SL(2,\mathbb{C})$ as well as
\begin{equation*}
    T^{\alpha\dot \alpha} = \begin{pmatrix}  -T^0+T^3  & T^1+iT^2 \\ T^1-iT^2  & -T^0-T^3 \end{pmatrix}\,.
\end{equation*}
Then, one can rewrite \eqref{Injection translation: image} as $\T(z,\bar z) = T^{\alpha\dot \alpha}\lambda_{\alpha}\bar\lambda_{\dot\alpha}$. Since this function is homogeneous of degree one in both in $\lambda$ and $\bar \lambda$, it defines an elment of $\mathcal{O}(1,1) = \E[1]$, i.e. a conformal density of weight one. Alternatively, upon acting on $T^{\alpha\dot \alpha}$ with the group $SL(2,\mathbb{C})$, then one can show that $\T(z,\bar z)$ transforms as a weight-one density under M\"obius transformations. 

 In dimension $n=2$, the scalar curvature $\hat R$ of a metric $\hat g= \T^{-2} g$ is related to the scalar curvature $R$ of $g$ and the Laplacian $\Delta= g^{\mu\nu}\nabla_\mu\nabla_\nu$
  as 
\begin{equation*}
   - \hat R = -\T\left(2\Delta +  R  \right)\T + 2 g^{\mu\nu}\partial_{\mu}\T \partial_{\nu}\T
\end{equation*}
taking $g = 4dz\, d\bar z$ gives $\Delta= \partial_z\partial_{\bar z}$ hence
\begin{align*}
    -\frac{\hat R}{2} &= - \T \partial_z\partial_{\bar z}\T + \partial_{z}\T \partial_{\bar z}\T
    = -(T^0)^2 + (T^1)^2 +(T^2)^2 + (T^3)^2.
\end{align*}
}

\begin{remark}
\label{Remark 2.1:}
Note that the four-vector $q^{\mu}(z,\bar z)$ is null everywhere, $q^2(z,\bar z) =0$, and therefore the zeroes of $\T(z,\bar z) =T^{\mu}q_{\mu}(z,\bar z)$ are related to the type of the four-vector $T^\mu$. If $T^{\mu}$ is timelike, then the density $\T(z,\bar z)$ is nowhere vanishing : it is negative if $T^{\mu}$ is future-oriented and positive if it is past-oriented. If $T^{\mu}$ is null, then $\T(z,\bar z)$ has an isolated zero. For instance, $z=0$ is the only zero of $\T(z,\bar z)$ when $T^\mu=(T^0,0,0,T^0)$, as follows from \eqref{Injection translation: image}. If $T^{\mu}$ is spacelike then $\T(z,\bar z)$ vanishes on a circle.
For instance, $|z|=1$ are the zeroes of $\T(z,\bar z)$ when $T^\mu=(0,0,0,T^3)$.
\end{remark}

With a standard abuse of notation and terminology, the space $\mathbb{R}^{3,1}$ of translations will often be identified with its image inside $\mathcal{E}[1]$ and translations are seen as specific types of supertranslations. The canonical injection \eqref{Injection translation -> super} suggests to consider its cokernel $\mathcal{E}[1]/\mathbb{R}^{3,1}$, i.e. the quotient space of the codomain of this map by its image. This quotient space will frequently appear in what follows. Its elements are supertranslations modulo translations. Physically, they can be interpreted as shifts of gravitational vacua (cf. Definition \ref{gravacua}), which motivates the following

\begin{definition}\label{quotientgravityshifts}
The elements of the quotient vector space $\mathcal{E}[1]\big/\mathbb{R}^{3,1}$ will be called shifts of vacua. In other words, a shift of vacua is an equivalence class of supertranslations $\T \in \E[1]$ modulo elements of the form $T^\mu q_{\mu}$ for some translation $T^\mu\in\mathbb{R}^{3,1}$.
\end{definition}
\noindent Such equivalence classes $[\T] \in \E[1]/\mathbb{R}^{3,1}$
are in one-to-one correspondence with the spin-weighted densities of the form $\eth^2\T \in \mathcal{O}\left(-3,1\right)$ to which they will sometimes be loosely identified later on.

\subsubsection{Basic properties of supermomenta}\label{basispropertiesofsupermomenta}

The proper framework for supermomenta is the language of distributions. Just like usual momenta $p_{\mu} \in (\mathbb{R}^{3,1})^*$ are in the space dual to translations, supermomenta $\P\in \E[1]^\prime$ are in the space dual to supertranslations. Since supertranslations are smooth conformal densities on the sphere, supermomenta are distributions. This technical subtlety cannot be avoided since the space of supertranslations is infinite-dimensional and some care is in order. 

\begin{definition}
A supermomentum $\P(z,\bar z) \in \mathcal{E}[1]^\prime$ is a distribution (i.e. a continuous linear functional) on the space of supertranslations. In particular, any conformal density of weight $-3$ is a supermomentum, $\mathcal{E}[-3]\subset\E[1]^\prime$. By definition, supermomenta are dual to supertranslations and we will write
\begin{equation}\label{pairing}
    \langle \P, \T \rangle = 
    \int_{S^2} \frac{i}{2} dz\wedge d\bar z \;\P(z,\bar z) \T(z,\bar z) \qquad  \in \mathbb{R}\,,
\end{equation}
where we used\footnote{This is the volume forme of the flat metric $dzd\bz$. This normalisation of the pairing turns out to be more convenient than using the volume form of $4dzd\bz$. We can get away with it, and other potential convention-related confusions, if one remembers that the meaningful quantity for distributions really is the pairing $\langle \P, \T \rangle$. For example the delta function $\delta^{(2)}(z,\bz)$ will always mean, without possible confusion, $\langle \delta^{(2)}, \T \rangle = \T(0,0)$.} the volume form $dx\wedge dy=\frac{i}{2} dz\wedge d\bar z$.
A momentum $p_{\mu}$ is an element of the dual $\left(\mathbb{R}^{3,1}\right)^*$ of the vector space $\mathbb{R}^{3,1}$ of translations $T^\mu$.
\end{definition}

\begin{remark}\label{distrib} We will not delve too much into subtleties of functional analysis, see e.g. the papers \cite{McCarthy_71,Mccarthy:1972ry,McCarthy_72-I,McCarthy_73-II,McCarthy_73-III,McCarthy:1974aw,McCarthy_75,McCarthy_76-IV,McCarthy_78,McCarthy_78errata,Girardello:1974sq} where such issues are addressed carefully (see also Footnote \ref{Gelfandfootnote}). From now on, we will most often loosely denote the space of supermomenta as $\mathcal{E}[-3]$ with a slight abuse of notation. Despite this, most results discussed in this paper straightforwardly extend from $\mathcal{E}[-3]$ to $\E[1]^\prime$. We will warn the reader, and comment on that point in due place, when the distinction between regular distributions (i.e. smooth densities) and genuine distributions (such as singular functions admitting a regularisation in the sense of distributions) is important.    
\end{remark}

\begin{remark}
It will be useful to keep in mind that, with our definitions \eqref{SL2Caction}-\eqref{SL2Caction on the left} for the action of an element $M$ of $SL(2,\mathbb{C})$, one has the identity $\langle \P\cdot M, \T \rangle= \langle \P, M\cdot\T \rangle$\,.
\end{remark}

\begin{proposition}\label{projectiontomomenta}
There is a canonical, $SL(2,\mathbb{C})$-equivariant, linear surjection
\begin{equation}\label{Projection super -> momenta}
    \pi_{\mu}\left|\; \begin{matrix}
        &\mathcal{E}[-3] &  \twoheadrightarrow & \left(\mathbb{R}^{3,1}\right)^* \\[1em]
        &\P(z,\bar z) & \mapsto& p_{\mu}
    \end{matrix} \right.
\end{equation}
from supermomenta onto
momenta. It is given by
\begin{align*}
    p_{\mu} &= 
    \int_{S^2} \frac{i}{2}dz\wedge d\bar z \;\P(z,\bar z)\,q_{\mu}(z,\bar z)
\end{align*}
i.e.
\begin{align*}
    p_{0} &= -\int_{S^2}\frac{i}{2} dz\wedge d\bar z \;(1+|z|^2)\P(z,\bar z), & p_{1} &= 
    \int_{S^2} \frac{i}{2} dz\wedge d\bar z  \;(z+ \bar z)\P(z,\bar z), \\
    p_{2} &= 
    i\int_{S^2} \frac{i}{2} dz\wedge d\bar z \;(z- \bar z)\P(z,\bar z), & p_{3} &= 
    \int_{S^2} \frac{i}{2} dz\wedge d\bar z  \;(1-|z|^2)\P(z,\bar z).
\end{align*}
\end{proposition}

\proof{
Let $T^{\mu} \in \mathbb{R}^{3,1}$ be a translation and let $\T(z,\bar z)$ be the density \eqref{Injection translation: image} obtained by the injection  \eqref{Injection translation -> super}. Contracting this with the supermomenta $\P$ gives
\begin{align*}
    \langle \P, \T \rangle &= \int_{S^2} \frac{i}{2}
    dz\wedge d\bar z \;\P(z,\bar z) \T(z,\bar z)\\
    &=-\, T^0 \; \int_{S^2} \frac{i}{2}
    dz\wedge d\bar z \;(1+|z|^2)\P(z,\bar z) + T^1 \; \int_{S^2} \frac{i}{2}
    dz\wedge d\bar z\;(z+ \bar z)\P(z,\bar z) \\
    &\quad + T^2 \;i\int_{S^2} \frac{i}{2}
    dz\wedge d\bar z \;(z- \bar z)\P(z,\bar z)+ T^3 \;\int_{S^2} \frac{i}{2}
    dz\wedge d\bar z \;(1-|z|^2)\P(z,\bar z)\,,
\end{align*}
which defines an element $p_{\mu} \in (\mathbb{R}^{3,1})^*$ of the dual. The equivariance of the projection  \eqref{Projection super -> momenta} directly comes from the fact that the injection \eqref{Injection translation -> super} is equivariant.
}

\begin{definition}
We will call $p_{\mu} = \pi_{\mu}\left(\P\right)$ the momentum associated to the supermomentum $\P(z,\bar z)$. In particular, supermomenta with zero momentum will be called soft supermomenta.
\end{definition}
\noindent The kernel of the projection operator \eqref{Projection super -> momenta} is the space $\text{Ker}\,\pi$ of soft supermomenta. Equivalently, a supermomentum is soft if and only if it annihilates all translations $\T=T^\mu q_\mu$.

The Paneitz operator \cite{Paneitz_2008,Fradkin:1982xc} is the conformal completion of the square of the Laplacian. There exists an analogue of the Paneitz operator in two dimensions: the only difference with its higher dimension counterpart is that it is only covariant under the action of M\"obius transformations and not under Weyl transformations.
\begin{definition}
    The Paneitz operator in two dimensions is the natural, $SL(2,\mathbb{C})$-covariant, differential operator $\eth^2 \bar{\eth}^2$ of order 4 mapping $\mathcal{E}[1]$ to $\mathcal{E}[-3]$ whose expression, in the flat representative, reads
\begin{equation*}
    \eth^2 \bar{\eth}^2\left|\; \begin{matrix}
        &\mathcal{E}[1] &  \to & \mathcal{E}[-3] \\[1em]
        & f & \mapsto&  \partial_z^2\partial_{\bar z}^2 f
    \end{matrix} \right. .
\end{equation*}
\end{definition}

The following proposition is a particular case of classical results\footnote{See e.g. \cite[Sections III.6.4 and III.6.6]{Gelfand2} applied to the case $D_{(2,2)}$ (cf. Footnote \ref{Gelfandfootnote}).} in the representation theory of $SL(2,\mathbb{C})$: 

\begin{proposition}\label{innerproductprop}
The vector space $\mathcal{E}[1]\big/\mathbb{R}^{3,1}$ of shifts of vacua carries a unitary irreducible representation of $SL(2,\mathbb{C})$ belonging to the principal series. In fact, there exists an $SL(2,\mathbb{C})$-invariant inner product on the space $\mathcal{E}[1]$ of supertranslations:
\begin{equation}\label{innerprod}
    \llangle\T_1  , \T_2 \rrangle: = \langle   \eth^2 \bar{\eth}^2 \T_1 , \T_2 \rangle=\int_{S^2} \frac{i}{2}
    dz\wedge d\bar z \;  \T_2(z,\bar z) \;  \partial_{z}^2\partial_{\bar z}^2 \T_1(z,\bar z)
\qquad  \in \mathbb{R}\,.
\end{equation}
The inner product \eqref{innerprod} is positive semi-definite on $\mathcal{E}[1]$ but it
induces a scalar product (i.e. a positive-definite inner product) on the quotient $\mathcal{E}[1]\big/\mathbb{R}^{3,1}$.
\end{proposition}

\proof{Basically, the only thing to show is that the radical of the bilinear map \eqref{innerprod} is the image of the map $q$ in Proposition \ref{Proposition: Injection translation -> super}. If $\T$ sits in the kernel of $\eth^2 \bar{\eth}^2$, then $\llangle \T,\T \rrangle =0$. Therefore,
\begin{equation}
\int_{S^2} \frac{i}{2} 
dz\wedge d\bar z \; \T\, \partial_{z}^2\partial_{\bar z}^2
    \T   
    =  \int_{S^2} \frac{i}{2}
    dz\wedge d\bar z \; |\partial_z^2 \T |^2=0\,,
\end{equation}
hence
$\partial_z^2 \T=0$ and $\partial_{\bar z}^2 \T=0$. Thus $\T$ is in the image of \eqref{Injection translation -> super}, i.e. it takes the form
\eqref{Injection translation: image}. Moreover, it is clear that the converse is true: elements $\T \in \mathcal{E}[1]$ of the form \eqref{Injection translation: image} are in the kernel of $\eth^2 \bar{\eth}^2$. As a consequence, the scalar product \eqref{Proof, main lemma: hermitian product} is non-degenerate on the quotient $\mathcal{E}[1]/\mathbb{R}^{3,1}$.}

As we already pointed out, shift of vacua $[\T] \in \mathcal{E}[1]\big/\mathbb{R}^{3,1}$ are in one-to-one correspondence with spin-weighted densities
of the form $\eth^2 \T\in \mathcal{O}(1,-3)$. It will thus be convenient to write the above inner product as
\begin{equation}
    \llangle\T_1  , \T_2 \rrangle = \langle    \bar{\eth}^2 \T_1 , \eth^2\T_2 \rangle=\int_{S^2} \frac{i}{2}
    dz\wedge d\bar z \;  \partial_{z}^2\T_2(z,\bar z) \;  \partial_{\bar z}^2\T_1(z,\bar z)\, ,
\end{equation}
where here the bracket $\langle\,.\,,.\rangle$ should be understood as the natural duality paring between $\mathcal{O}(1,-3)$ and $\mathcal{O}(-3,1)$. This notation is consistent
with the possibility of integrating by part the edth operators on the sphere.

\begin{proposition}\label{Paneitz operator Lemma}
The Paneitz operator fits in the exact sequence
\begin{equation}\label{longexact}
0 \;\to\; \mathbb{R}^{3,1} \;\stackrel{\q_{\mu}}{\hookrightarrow}\; \mathcal{E}[1] \;\xrightarrow{\eth^2 \bar{\eth}^2} \mathcal{E}[-3] \; \stackrel{\pi_{\mu}}\twoheadrightarrow \;\left(\mathbb{R}^{3,1}\right)^* \;\to \; 0.
\end{equation}
In other words, the linear map
\begin{equation}
    \eth^2 \bar{\eth}^2 : \faktor{\E[1]}{\mathbb{R}^{3,1}} \stackrel{\sim}{\longrightarrow}\text{Ker}(\pi),
\end{equation}
is an isomorphism, i.e.
\begin{align*}
\pi_{\mu}(\P)=0,&\quad \P(z,\bar z)\in \mathcal{E}[3] &&\Longleftrightarrow & \exists \N(z,\bar z) \in \E[1], &\quad \text{such that}\quad  
\P(z,\bar z) = \partial^2_z\partial^2_{\bar z}\N(z,\bar z)\,,
\end{align*}
and
\begin{align*}
\partial^2_z\partial^2_{\bar z}\N(z,\bar z)=0,&\quad \N(z,\bar z)\in \mathcal{E}[-1] &&\Longleftrightarrow & \exists\, T^\mu\in\mathbb{R}^{3,1}, &\quad \text{such that}\quad  \N(z,\bar z) = T^\mu q_\mu(z,\bar z).
\end{align*}
\end{proposition}

\proof{See Appendix \ref{Appendix: proofsPaneitz}.}

For smooth densities this is a classical result which can be extracted, e.g., from \cite[Sections III.3.3 and III.5.2]{Gelfand2}. We discuss this point in detail in Appendix \ref{Appendix: proofsPaneitz}, as well as its generalisation to distributions, since this proposition is instrumental for later considerations.

\section{Hard/Soft decomposition of supermomenta}\label{Section: Hard/Softdecompositionofsupermomenta}

In this section, we introduce a decomposition of supermomenta into hard and soft pieces. This will be our main guide for reconsidering McCarthy's classification of BMS UIRs in Section \ref{Section: McCarthy}. 

\subsection{Decomposition of supermomenta}

In Subsection \ref{basispropertiesofsupermomenta}, we saw that soft supermomenta are always of the form $\partial_z^2\partial_{\bar z}^2 \N(z,\bar z)\in \text{Ker}\,\pi$ for some $\N(z,\bar z) \in \mathcal{E}[1]\big/\mathbb{R}^{3,1}$. This leads to the following proposition, which is central in the present paper.

\begin{proposition}\label{Prop: classification of supermomenta}
A supermomentum
\begin{equation*}
    \P(z,\bar z) \in \mathcal{E}[-3]
\end{equation*} is canonically equivalent to a pair 
\begin{equation*}
   \Big(\;p_{\mu}\,,\, \N(z, \bar z)\,\Big) \in \big(\mathbb{R}^{3,1}\big)^* \times \big(\mathcal{E}[1]\big/\mathbb{R}^{3,1}\big)
\end{equation*}
 where $p_{\mu}=\pi_\mu(\P) \in \big(\mathbb{R}^{3,1}\big)^*$ is the momentum associated to the supermomentum $\P(z,\bar z)$. The equivalence class $\N(z, \bar z) \in \mathcal{E}[1]\big/\mathbb{R}^{3,1}$ of weight-one conformal densities will be called the soft charge\footnote{Most of the time, we will try not to name $\N$ at all, in order to avoid possible confusions with $\partial_z^2 \partial_{\bz}^2 \N$, which should genuinely be called the soft charge.} of the supermomentum $\P(z,\bar z)$. Supermomenta with vanishing soft charge will be called hard supermomenta.

The supermomentum decomposes as
\begin{equation}\label{genericdecomppsition}
        \P(z,\bar z) = P(z,\bar z)+\partial_z^2 \partial_{\bar z}^2\N(z,\bar z) 
    \end{equation}
where the hard part $P(z,\bar z)$ of the supermomentum $\P(z,\bar z)$ satifies  $p_\mu=\pi_\mu(P)$ and is defined as follows:
 \begin{enumerate}
    \item If $p^2 <0$ (massive case) then, in the referential where
    \begin{equation*}
        p_{\mu}(\bm{k})= \pm\, m \begin{pmatrix}
        -\sqrt{1 + \bm{k}^2},& \bm{k}
    \end{pmatrix}\quad\text{with}\quad \bm{k}= \begin{pmatrix}
        k^1,&k^2,&k^3
    \end{pmatrix}\in\mathbb{R}^3,
    \end{equation*}
     one has
    \begin{equation}\label{massivedecomppsition}
        \P(z,\bar z) = P(z,\bar z)+\partial_z^2 \partial_{\bar z}^2\N(z,\bar z) \quad\text{where}\quad P(z,\bar z)= -\frac{m^4}{\pi}\Big(p^{\mu}(\bm{k}) q_{\mu}(z,\bar z)\Big)^{-3},
    \end{equation}
    for any positive mass $m>0$. 
    \item If $p^2 =0$ (massless case) then, in the referential where,
    \begin{equation*}
        p_{\mu}=\omega \,q_\mu(\zeta,\bar\zeta) =\omega \begin{pmatrix}
        -1-|\zeta|^2 ,& \zeta + \bar \zeta, &i(\zeta - \bar \zeta),&1-|\zeta|^2
    \end{pmatrix},
    \end{equation*} one has
    \begin{equation}\label{masslessdecomppsition}
        \P(z,\bar z) = P(z,\bar z)+\partial_z^2 \partial_{\bar z}^2 \N(z,\bar z)\quad\text{where}\quad P(z,\bar z)= \omega \,\delta^{(2)}(z-\zeta,\bar z-\bar\zeta)\,,
    \end{equation}
    for any non-vanishing energy $\omega\neq 0$.
\item If $p^2 >0$ (tachyonic case) then, in the referential where
    \begin{equation*}
        p_{\mu}(\bm{k})= \pm m \begin{pmatrix}
        -\sqrt{\bm{k}^2-1},& \bm{k}
    \end{pmatrix}\quad\text{where}\quad \lVert\bm{k}\rVert \geqslant 1,
    \end{equation*}
     one has
    \begin{equation}\label{tachyondecomppsition}
        \P(z,\bar z) = P(z,\bar z)+\partial_z^2 \partial_{\bar z}^2\N(z,\bar z)\quad\text{where}\quad P(z,\bar z)= -\frac{m^4}{\pi}\Big(p^{\mu}(\bm{k}) q_{\mu}(z,\bar z)\Big)^{-3},
    \end{equation}
    for any positive $m> 0$. 

 \item If $p_{\mu}=0$ (zero-momentum case) then
    \begin{equation}
        \P(z,\bar z) = \partial_z^2 \partial_{\bar z}^2 \N(z,\bar z).
    \end{equation}
    \end{enumerate}
\end{proposition}
\noindent Note that for any massive future (past) oriented momentum $p^{\mu}$ the corresponding hard momentum is positive (negative) and that the support of a massless hard supermomentum $P(z,\bar z)= \omega \,\delta^{(2)}(z-\zeta,\bar z-\bar\zeta)$ is a single point $(\zeta,\bar\zeta)$ on the celestial sphere (corresponding to the direction of its associated null momentum) since it is written in terms of the Dirac distribution (defined as $\delta^{(2)}(z,\bar z)=\delta(x)\delta(y)$ for $z=x+i y$). 

\proof{
Let us first consider the massive case $p^2 = -m^2<0$ (the tachyonic case $p^2=+m^2>0$ is treated similarly). If $\P$ is any supermomentum with associated momentum $p_{\mu}=\pi_\mu(\P)$ which is timelike, then one can always, through an action of $SL(2,\mathbb{C})$, transform it in such a way that $p_{\mu}= m \begin{pmatrix}
        \mp1 ,&0,&0,&0
    \end{pmatrix}$ with $m>0$.  Now, by performing explicitly the integrals in Proposition \ref{projectiontomomenta} in polar coordinates,
one can check that the corresponding hard supermomenta $P(z,\bz) =\pm\frac{m}{\pi(1+ |z|^2)^3}$ is projected on that same momentum $m \begin{pmatrix}
        \mp1 ,&0,&0,&0
    \end{pmatrix}$. It thus follows, by linearity of the projection, that
\begin{align*}
    \pi_{\mu}\left(\P(z,\bar z)\mp\frac{m}{\pi(1+ |z|^2)^3}\right)&=0.
\end{align*}
   By Proposition \ref{Paneitz operator Lemma}, this means that there exists an element $\N(z, \bar z) \in \mathcal{E}[1]\big/\mathbb{R}^{3,1}$ such that \begin{equation*}
        \P(z,\bar z)\mp\frac{m}{\pi(1+ |z|^2)^3} = \partial_z^2\partial_{\bar z}^2 \N(z, \bar z).
    \end{equation*} 

For the massless case, the reasoning goes exactly the same by noting that the projection of the hard supermomentum $\omega \delta^{(2)}(z-\zeta)$ is $\omega 
    \begin{pmatrix}
        -1-|\zeta|^2 ,& \zeta + \bar \zeta, &i(\zeta - \bar \zeta),&1-|\zeta|^2
    \end{pmatrix}$. 
Finally, for the tachyonic case,  one can check that  the projection of the hard supermomentum $P(z,\bz) =\mp\frac{m}{\pi(1- |z|^2)^3}$ is the momentum $P_{\mu} = m \begin{pmatrix}
        0,&0,&0,&\pm 1
    \end{pmatrix}$.
}

\begin{remark}
\label{Remark 2.2:}
As mentioned in Remark \ref{Remark 2.1:}, in the tachyonic case the density $P(z,\bar z)=p^{\mu}(\bm{k}) q_{\mu}(z,\bar z)$ has a whole circle of zeroes and therefore the hard momentum \eqref{tachyondecomppsition} is not globally well-defined on the celestial sphere but only well-defined defined on an open disk, as a smooth function. Nevertheless, the regularised version appears to be globally well-defined in the sense of distributions \cite{Girardello:1974sq,McCarthy_75}. Since tachyons are unlikely to be physically relevant, we do not address such technical subtleties (cf. Remark \ref{distrib}).
\end{remark}

\begin{remark}\label{masslessremark}
    Let us stress that the massless case is  distinguished because we cannot use the naive analogue of the massive and tachyonic cases, i.e. the tentative supermomenta $P(z,\bar z)=\big(p^{\mu}(\bm{k}) q_{\mu}(z,\bar z)\big)^{-3}$ where $p_{\mu}(\bm{k})=  
    \begin{pmatrix}
        \pm|\bm{k}|,& \bm{k}
    \end{pmatrix}$ is a null momentum. For instance, the momentum $p_\mu=K(-1,0,0,1)$ would lead to    
the candidate hard supermomentum $P(z,\bar z)=K$, for $K\in\mathbb{R}\setminus\{0\}$. However, the corresponding conformal density is not admissible because it does not decrease at infinity (see Subsection \ref{confdens}). Furthermore, the other hard momenta in the same orbit (for example, the null momentum $p_\mu=K(1,0,0,1)$ would lead to $\tilde P(z,\bar z)=K|z|^{-6}$) have an admissible behavior at infinity (see Subsection \ref{confdens}) but have a power-law singularity at an isolated point (e.g. at the origin in the example). Such singular densities are not admissible because they do not correspond to well-defined distributions.\footnote{Note, however, that MacCarthy mentioned that such singular densities appear to be well-defined ``hyperfunctions'' \cite{McCarthy_78errata}. But this corresponds to a rather exotic choice of topology on the vector space of supertranslations  which will not be explored here.} In fact, their action is not well-defined on all smooth supertranslations (in contrast to the Dirac delta distributions in the proposition).
Via the correspondence \eqref{homogeneouscorrespondence}, this last example defines the homogeneous function $\tilde\Phi(\lambda_\alpha,\bar\lambda_{\dot \alpha})=|\lambda_2|^{-6}$. But it is known (cf. \cite[Appendix B.2.2]{Gelfand}) that one regularisation of such homogeneous distributions of negative integer degrees ($n\in\mathbb N$) is precisely the partial derivatives of the Dirac distribution, in the sense that $|\lambda_2|^{-2(n+1)}\,\propto\, (\partial{}^{}_{\lambda_2}\partial_{\bar\lambda_{\dot 2}})^n\delta^{(2)}(\lambda_2,\bar\lambda_2)$ in the limit when $n$ approaches a non-negative integer value. In our case, $n=2$ hence $\alpha=2$. This corresponds to the distributional momentum $\partial_z^2 \partial_{\bz}^2 \delta^{(2)}(z,\bar z)$ which is soft.
\end{remark} 

\subsection{Some remarks about the decomposition}\label{remarksaboutthedecomposition}

Some further discussion, about the previous decomposition of supermomenta into hard and soft parts, is in order.

\subsubsection{Non-linearity of the decomposition}\label{Remark: non linearity}

We wish to emphasise that the decompositions \eqref{massivedecomppsition}, \eqref{masslessdecomppsition} and \eqref{tachyondecomppsition} are not linear in the supermomentum. Therefore, \textit{the above decompositions are not stable under the addition of supermomenta}, as explained in the introduction (see also Subsection \ref{nonlin_summary}). Nevertheless, the projection on the momentum is both linear and $SL(2,\mathbb{C})$-invariant. Furthermore, these decompositions are well-defined and, in particular, they are $SL(2,\mathbb{C})$-invariant. In fact, the hard momenta $P(z,\bar z)$ in \eqref{massivedecomppsition} and \eqref{tachyondecomppsition} only depend on the quantity $p^{\mu}(\bm{k}) q_{\mu}(z,\bar z)$ which is manifestly $SL(2,\mathbb{C})$-covariant. In the massless case, the hard momenta $P(z,\bar z)= \omega \,\delta^{(2)}(z-\zeta,\bar z-\bar\zeta)$ is also $SL(2,\mathbb{C})$-covariant and the transformation laws of $\omega$ and $\zeta$ under a homothety $z'= \lambda z +z_0$ with $\lambda>0$ and $z_0\in\mathbb C$ reads $\omega'=\lambda \omega$ and $\zeta'=\lambda^{-1}(\zeta-z_0)$.

\paragraph{Relation to spherical harmonics decomposition.}
In particular, note that, while soft supermomenta can be characterised as having only spherical harmonics higher or equal to $2$, the decompositions \eqref{massivedecomppsition}-\eqref{tachyondecomppsition} have essentially nothing to do with a spherical harmonics decomposition
\begin{equation}\label{shperharmdecomp}
    \P(z,\bz) = \sum_{\ell=0,1,m}\,\P_{\ell,m} Y_{\ell,m}(z,\bz)  +  \sum_{\ell\geq 2,m} \P_{\ell,m}\,Y_{\ell,m}(z,\bz)\,,
\end{equation}
into lower and higher spherical harmonics.
In fact the spherical harmonic decomposition above has just opposite properties as compare to the decompositions in Proposition \ref{Prop: classification of supermomenta}: the decomposition \eqref{shperharmdecomp} is linear but \emph{not} Lorentz-invariant, while the decompositions \eqref{massivedecomppsition}-\eqref{tachyondecomppsition} are Lorentz-invariant but \emph{not} linear.

\subsubsection{Uniqueness of the decomposition: the massless case}\label{Remark: uniqueness for massless case}

To avoid some confusion about the decomposition\eqref{masslessdecomppsition}, we now wish to comment on the identity 
\begin{equation}\label{usual identity}
    \delta^{(2)}(z-\zeta,\bar z - \bar \zeta)\stackrel{?}{=} \frac{1}{\pi}\,\partial_{\bar z}^2 \left(\frac{\bar z - \bar \zeta}{z-\zeta}\right)
\end{equation}
which commonly appears in the literature. Strictly speaking, as we shall see, it is not correct as a distributional identity, at least not for distributions acting on the space that we are considering, of weight-one conformal densities $\T(z,\bar z )$ on the celestial sphere. This is a subtle point because \eqref{usual identity} is for instance a valid identity on the space of test functions on the complex plane $\mathbb{C}$ (which, in particular, vanish at infinity). Furthermore, the equality \eqref{usual identity} is valid in the sense that
\begin{align}\label{delta identity}
    \int \frac{i}{2} dz\wedge d\bar z\; \T(z,\bar z) \,\delta^{(2)}(z-\zeta,\bar z - \bar \zeta)  =& \;\;\frac{1}{\pi}\int \frac{i}{2} dz\wedge d\bar z  \,\partial_{\bar z}^2 \T(z,\bar z ) \frac{\bar z - \bar \zeta}{z-\zeta}  \\&- \; \frac{1}{2\pi i}\oint \left(\frac{\T(z,\bar z ) }{z-\zeta} - \partial_{\bar z}\T(z,\bar z ) \frac{\bar z - \bar \zeta}{z-\zeta}\right)dz. \nonumber
\end{align}
where the above right-hand side is formally equal to  $\frac{1}{\pi} \int \frac{i}{2} dz\wedge d\bar z \, \T(z,\bar z) \,\partial_{\bar z}^2 \left(\frac{\bar z - \bar \zeta}{z-\zeta}\right)$ via integration by part. The  equality \eqref{delta identity} follows from the Cauchy--Pompeiu formula (see Appendix \ref{proof: identity for massless hard supermomenta} for a proof), with the implicit contour here taken to be around the point at infinity, $z=\infty$. Now, if $\T(z,\bar z )$ was a weight-$w$ conformal density on the celestial sphere with  $w<0$, the above boundary term would vanish and it would be correct to think of \eqref{usual identity} as an equality between distributions. However, here $\T(z,\bar z )$ is a weight-one conformal density on the celestial sphere, which implies that it behaves asymptotically as
\begin{align*}
    &\T(z,\bar z ) = |\hat z|^{-2}\hat{\T}(\hat z, \bar{ \hat{z}})\\
    &= z\bar{z}\,\hat{\T}(0, 0) +  \bar{ z}\,\partial_{\hat{z}}\hat{\T}(0, 0) +   z\,\partial_{\bar{\hat{z}}}\hat{\T}(0, 0) +\partial_{\hat{z}}\partial_{\bar{\hat{z}}}\hat{\T}(0, 0)  +\frac{z}{2\bar{z}}\partial_{\bar{\hat{z}}}^2\hat{\T}(0, 0)+\frac{\bar z}{2z}\partial_{\hat{z}}^2\hat{\T}(0, 0)+ \mathcal{O}\big(|z|^{-1}\big),
\end{align*} 
which will lead to a non-zero boundary term (again, see Appendix \ref{proof: identity for massless hard supermomenta}). The presence of this boundary term means that the right-hand side of \eqref{usual identity} cannot be interpreted as derivatives of a distribution and thus there is no contradiction with the uniqueness of the decomposition discussed in Proposition \ref{Prop: classification of supermomenta}. In fact, another way to phrase the problem is that \eqref{usual identity} seems to suggests that $\delta^{(2)}(z-\zeta,\bar z-\bar\zeta) = \partial_{\bar z}^2 \N_{zz}(z,\bar z)$ for some spin-weighted density $\N_{zz}\in\mathcal{O}(-3,1)$ on the celestial sphere. However this is not correct because $\N_{zz}(z,\bar z)=\frac1{\pi}\frac{\bar z - \bar \zeta}{z-\zeta}$ has an asymptotic behavior incompatible with the one of a spin-weighted density of conformal weight $-1$  on the celestial sphere:
\begin{equation}
    \frac{\bar z - \bar \zeta}{z-\zeta} = \frac{\bar z}{z} + \mathcal{O}\big(|z|^{-1}\big)
\end{equation}
(instead of the expected behavior in $|z|^{-2}$). If one wishes to maintain the interpretation of the expression $\frac{\bar z - \bar \zeta}{z-\zeta}$ in terms of a conformal density on the celestial sphere, this means that it has a singularity at $z=\infty$ (which we do not allow).

For all theses reasons, \eqref{usual identity} is not valid as an identity between distributions, i.e. 
\begin{align}\label{delta identity'}
    \int \frac{i}{2} dz\wedge d\bar z\; \T(z,\bar z) \,\delta^{(2)}(z-\zeta,\bar z - \bar \zeta)  \neq \;\;\frac{1}{\pi}\int \frac{i}{2} dz\wedge d\bar z  \,\partial_{\bar z}^2 \T(z,\bar z ) \frac{\bar z - \bar \zeta}{z-\zeta}  \,.
\end{align}
It can nonetheless be corrected by a boundary term to obtain the following proposition.\footnote{This is in line with \cite{IvancovichJ.1989Gfot} where it is highlighted that $\eth^2 F_{-2}= A_{0}$ has no solution if $A_0(z,\bz)$ contains $\ell=0,1$ spherical harmonics. Note that, as a consequence, since this last property is not $SL(2,\mathbb{C})$-invariant, the particular form for the Green functions proposed in \cite{IvancovichJ.1989Gfot} necessarily breaks $SL(2,\mathbb{C})$ invariance; the same is true for the kernel in Proposition \ref{Prop: massless hard supermomenta identity}, as manifested by the fact that $z=\infty$ plays a special role.}
\begin{proposition}\label{Prop: massless hard supermomenta identity}
Let $p_{\mu}= \omega\, q_{\mu}(\zeta,\bar{\zeta})$ be a massless momentum and let $P(z,\bar z) = \omega \,\delta^{(2)}(z-\zeta,\bar z - \bar \zeta)$ be the corresponding hard supermomentum. Then
\begin{equation}
         \omega \delta^{(2)}(z-\zeta,\bar z - \bar \zeta) =  \frac{1}{\pi}\partial_{\bar z}^2 \left( \omega\frac{\bar z - \bar \zeta}{z-\zeta}\right) +p^{\mu}\, \mathcal{D}_{\mu}\big(\delta^{(2)}(z-\infty,\bar z -\infty)\big)
\end{equation}
where $\mathcal{D}^{(2)}_{\mu}\big(\delta(z-\infty,\bar z -\infty)\big)$ is a distribution supported at $z=\infty$ (or equivalently at $\hat{z}=z^{-1}=0)$ and defined by
\begin{equation}\label{Thomas operator distribution}
    \int \frac{i}{2}dz\wedge d\bar z\; \T(z,\bar z) \, \mathcal{D}_{\mu}\big(\delta^{(2)}(z-\infty)\big) = - \Big(\mathcal{D}_{\mu}\T\Big)(\hat z=0)  = \frac{1}{2}\begin{pmatrix}
            \partial_{\hat z}\partial_{\bar{\hat{z}}}\hat\T + \hat\T\\ \partial_{\hat z}\hat\T + \partial_{\bar{\hat{z}}}\hat \T \\ i(\partial_{\bar{ \hat{z}}}\hat \T - \partial_{\hat z}\hat\T )\\  \partial_{\hat z}\partial_{\bar{\hat{z}}}\hat\T -\hat\T 
        \end{pmatrix}\big(\hat z=0\big).
\end{equation}
The identity here holds in the sense of distributions.
\end{proposition}

The operator $\mathcal{D}_{\mu}$ appearing in the above proposition can be phrased in more geometrical terms. As the notation suggests it results of evaluating a conformally-invariant differential operator (the ``Thomas operator'' of tractor calculus, see \cite{Curry_Gover_2018}) at $z=\infty$.

\proof{See Appendix \ref{proof: identity for massless hard supermomenta}.}

\subsubsection{Uniqueness of the decomposition: the massive case}\label{Remark: uniqueness for massive case}

We now comment on the identity
\begin{equation}\label{notquitecorrect2}
    \frac{m^4}{\pi(q \cdot p)^3} \stackrel{?}{=} \frac{1}{\pi}\partial_{\bar z}^2 \left(\frac{(\partial_z q \cdot p)^2 }{2\, q \cdot p}\right)
\end{equation}
appearing in the literature. Developing the right-hand side one finds
\begin{equation}
    \frac{1}{2\pi}\partial_{\bar z}^2 \left(\frac{(\partial_z q \cdot p)^2 }{q \cdot p}\right) = \frac{1}{\pi(\q\cdot p)^3} \Big( p^{\mu}\big( - q_{\mu}\partial_z \partial_{\bz}q_{\nu}  +  \partial_z q_{\mu} \partial_{\bz}q_{\nu} \big) p^{\nu} \Big)^2
 \end{equation}
and making use of $\eta_{\mu\nu} = -q_{(\mu}\partial_z \partial_{\bz}q_{\nu)}  + \partial_z q_{(\mu} \partial_{\bz}q_{\nu)}$ then yields the equality. Therefore this might suggests that $(q \cdot p)^{-3} = \partial_{\bar z}^2 \N_{zz}$ for some spin-weighted density $\N_{zz}\in\mathcal{O}(-3,1)$ on the celestial sphere. This last statement is however incorrect: 

To start, note that the combination under the derivative, $\frac{(\partial_z q \cdot p)^2 }{2 q \cdot p}$ cannot be properly interpreted as a spin-weighted density because, under change of the conformal representative $(4dzd\bz , q_{\mu}) \mapsto ( 4\Omega^2 dzd\bz , \Omega q_{\mu})$ one has
\begin{equation}
    \partial_z q_{\mu} \mapsto \Omega \big( \partial_z q_{\mu} + \Omega ^{-1}\partial_z \Omega \, q_{\mu} \big)
\end{equation}
where the inhomogeneous term prevents us to think of $\partial_z q(z)_{\mu}$ as a proper conformal density. This directly relates to the fact that polarisation tensors $\varepsilon_{\mu} := \frac{1}{\sqrt{2}} \partial_z q_{\mu}$ are \textit{not} gauge-invariant quantities. What is more, one has
\begin{equation}
    \frac{(\partial_z q \cdot p)^2 }{q \cdot p}\, =\, \frac{\bar{z}}{z}(...) + \mathcal{O}\big(|z|^{-1}\big).
\end{equation}
This is not the expected behavior $\N_{zz}(z,\bar z) = \mathcal{O}\big(|z|^{-2}\big)$ for a spin-weighted quantity of conformal weight $w=-1$ and it implies that the would-be spin-weighted density has a singularity at $z=\infty$. Finally, and more closely related to our subject, the equality $(q \cdot p)^{-3} = \partial_{\bar z}^2 \N_{zz}$ for some spin-weighted density $\N_{zz}\in\mathcal{O}(-3,1)$ on the celestial sphere would mean that the projection of this supermomentum on its associated momentum
\begin{align}
   p_{\mu} = -\frac{m^4}{\pi}\int_{S^2}\frac{i}{2} dz\wedge d\bar z \; \frac{q_{\mu}(z,\bar z)}{\big(q(z,\bar z) \cdot p\big)^3}
\end{align}
would vanish (due to integration by part), which would again be an incorrect conclusion. In practice, when using \eqref{notquitecorrect2}, one should be careful that integration by part produces an extra boundary term at $z=\infty$ due to the singular behaviour of  $\frac{(\partial_z q \cdot p)^2 }{2\, q \cdot p}$ at infinity (see Appendix \ref{proof: identity for massive hard supermomenta}) and this means that, here again, \eqref{notquitecorrect2} is not correct as an identity between distributions. Rather one has
\begin{proposition}\label{Prop: massive hard supermomenta identity}
Let $p_{\mu}$ be a massive momentum and let $P(z,\bar z) = -\frac{m^4}{\pi}\big(q(z,\bar z)\cdot p\big)^{-3}$ be the corresponding hard supermomentum.  Then
\begin{equation}
         \frac{-m^4}{\pi(q \cdot p)^3} =  -\frac{1}{\pi}\partial_{\bar z}^2 \left(\frac{(\partial_z q \cdot p)^2 }{2\,q \cdot p}\right) + p^{\mu}\, \mathcal{D}_{\mu}\big(\delta^{(2)}(z-\infty,\bar z -\infty)\big)
\end{equation}
where $\mathcal{D}_{\mu}\big(\delta^{(2)}(z-\infty,\bar z -\infty)\big)$ is the distribution supported at $z=\infty$ defined by \eqref{Thomas operator distribution}. Here again, the identity holds in the sense of distributions.
\end{proposition}
\proof{See Appendix \ref{proof: identity for massive hard supermomenta}.}

\subsubsection{Non-linearity and uniqueness of the decomposition: summary}\label{nonlin_summary}

With a bit more work, the results of the previous subsections can be summarised and presented in a different form as follows.
\begin{proposition}\label{Prop: final hard supermomenta identity}
Let $p_{\mu}$ be a massive or massless momentum and let $P(z,\bar z)$ be the corresponding hard supermomentum.  Then
\begin{align}
\begin{alignedat}{2}
         P(z,\bar z) &=  -\frac{1}{\pi}\partial_{\bar z}^2 \left(\frac{(\varepsilon \cdot p)^2 }{q \cdot p}\right) + p^{\mu}\, \mathcal{D}_{\mu}\big(\delta^{(2)}(z-\infty,\bar z -\infty)\big)\\[0.5em]
         &=  -\frac{1}{2\pi}\partial_{ z}^2\partial_{\bar z}^2 \Big( (q \cdot p)\; \ln|q \cdot p| \Big) + p^{\mu}\, \mathcal{D}_{\mu}\big(\delta^{(2)}(z-\infty,\bar z -\infty)\big).
\end{alignedat}
\end{align}
where $\varepsilon_{\mu}=\frac{1}{\sqrt{2}}\partial_z q_{\mu}$. These identities hold in the sense of distributions.
\end{proposition}

\proof{The first identity follows from Proposition \ref{Prop: massless hard supermomenta identity}-\ref{Prop: massive hard supermomenta identity} and from the fact that, when $p_\mu = \omega\, q_\mu(\zeta-\zeta,\bz-\zetab)$ then $-\frac{(\varepsilon \cdot p)^2 }{q \cdot p}= \omega\frac{\bar z - \bar \zeta}{z-\zeta}$. For the second identity, see Appendix \ref{proof: final identity for hard supermomenta}.}

This allows to reconsider the non-linearity of the decomposition.

\begin{corollary}\label{Corollary: sum of supermomenta}
    Let $\P_1$ and $\P_2$ be two massive or massless supermomenta and let 
\begin{align}
    \P_1(z,\bz) &= P_1(z,\bz) + \partial_z^2\partial_{\bz}^2 \N_1(z,\bz),&      \P_2(z,\bz) &= P_2(z,\bz) + \partial_z^2\partial_{\bz}^2 \N_2(z,\bz).
\end{align}
be their decomposition as given by Proposition \ref{Prop: classification of supermomenta}. Let $p_1^{\mu} = \pi^{\mu}\left( \P_1\right)$ and $p_2^{\mu} =\pi^{\mu}\left( \P_2\right)$ be the associated momenta. Then
\begin{align}
    \P_1(z,\bz) +\P_2(z,\bz) = P_3(z,\bz) + \partial_{z}^2\partial_{\bz}^2\left( \N_1(z,\bz) + \N_2(z,\bz) -\frac{1}{2\pi}\mathcal{S}(z,\bz)\right)
\end{align}
where $P_3(z,\bz)$ is the hard supermomentum corresponding to $p_3^{\mu} := p_1^{\mu} +p_2^{\mu}$ and 
\begin{align}\label{soft part supermomenta: sum of 2}
 \mathcal{S} := (q \cdot p_1) \ln\big|q \cdot p_1 \big| + (q \cdot p_2) \ln\big|q \cdot p_2 \big| - (q \cdot p_3) \ln\big|q \cdot p_3\big|
\end{align}
defines a conformal density of weight one on the sphere, which is smooth everywhere except at the zeroes of the functions $p_i\cdot q(z,\bz)$. If all supermomenta are massive, then \eqref{soft part supermomenta: sum of 2} is a smooth density: $\mathcal{S}\in\E[1]$.
\end{corollary}

\proof{The equality \eqref{soft part supermomenta: sum of 2} is an immediate consequence of the previous proposition. The smoothness of \eqref{soft part supermomenta: sum of 2} at any point $z$ away from the zeroes and from $z=\infty$ is clear. If all momentum $p^{\mu}_i$ are massive $q\cdot p_i$ has no zero and \eqref{soft part supermomenta: sum of 2} is in fact manifestly smooth at every point away from $z=\infty$. Finally, since \begin{equation}
\mathcal{S} = -\frac{1}{2\pi}\,\Big(P_1(z,\bz) + P_2(z,\bz) - P_3(z,\bz)\Big),
    \end{equation}  is Lorentz-invariant, a singularity cannot arise at $z=\infty$. See the end of Appendix \ref{proofcorollary} for an explicit check that the sum of all possible singular terms at $z=\infty$ vanishes as a result of momentum conservation.
}

More generally, one sees from Proposition \ref{Prop: final hard supermomenta identity} that, for a family $\{p_i^{\mu}\}_{i\in 1 ... N}$ of momentum satisfying momentum conservation, Weinberg's soft factor, 
\begin{align*}
    \mathcal{S}_{zz} := \sum_{i=1}^{N} \frac{(\varepsilon \cdot p_i)^2 }{q \cdot p_i} = \frac{1}{2}\,\partial_z^2\left( \sum_{i=1}^{N} (q\cdot p_i) \ln|q \cdot p_i| \right) \,,
\end{align*}
is directly related to the impossibility of hard representations to conserve supermomentum by themselves
\begin{equation}
\sum\limits_{i=1}^{N} p_i^{\mu}=0  \quad\Longrightarrow\quad   \sum_{i=1}^{N} P_i(z ,\bz) = -\frac{1}{\pi} \partial_{\bz}^2\mathcal{S}_{zz}(z ,\bz)\,.
\end{equation}

From Corollary \ref{Corollary: sum of supermomenta} we see that supermomenta, $\P(z,\bz) = P(z,\bz) + \partial_z^2 \partial_{\bz}^2 \N$, 
whose momenta are future-oriented, massive or null, and such that $\N$ is square-integrable define a space of supermomenta which can scatter into themselves, since they are stable under addition, and are parametrized by countably many quantities. These features seems appealing for a scattering program.

\subsection{BMS, Poincaré and soft little groups}\label{ssection: BMS, Poincaré and soft little groups}

The BMS little group, associated to a supermomentum, is a straightforward generalisation of the usual little group for momentum.
\begin{definition}\label{BMSlittlegroup}
The stabiliser $\ell_{\mathcal P}\subseteq SL(2,\mathbb{C})$ of a supermomentum $\P \in \mathcal{E}[-3]$ under the group $SL(2,\mathbb{C})$ will be called the BMS little group of $\P$.
\end{definition}
This is to be distinguished from the stabiliser $\ell_p$ of the associated momentum $p_{\mu}=\pi_{\mu}(\P)$.
\begin{definition}
The stabiliser $\ell_{p}\subseteq SL(2,\mathbb{C})$ of the momentum $p_{\mu}=\pi(\P) \in (\mathbb{R}^{3,1})^*$ associated to a supermomentum $\P \in \mathcal{E}[-3]$ will be called the Poincar\'e little group of $\P$.
\end{definition}
Making use of the decomposition from Proposition \ref{Prop: classification of supermomenta}:
\begin{equation}\label{little groups: hard/soft decomposition}
        \P(z,\bar z) = P(z,\bar z) + \Sigma(z,\bar z) \,,\quad\text{where}\quad \Sigma(z,\bar z)=\partial_z^2 \partial_{\bar z}^2\N(z,\bar z)\,,
    \end{equation}
  one can also define the soft little group $\ell_{\Sigma}$.
\begin{definition}
The stabiliser $\ell_{\Sigma}\subseteq SL(2,\mathbb{C})$ of the soft part $\Sigma=\partial_z^2 \partial_{\bar z}^2\N \in \mathcal{E}[-3]$ of a supermomentum $\P\in \mathcal{E}[-3]$ will be called the soft little group of $\P$.
\end{definition}
By Proposition \ref{Prop: classification of supermomenta}, the decomposition \eqref{little groups: hard/soft decomposition} is unique. It follows that an element of $SL(2,\mathbb{C})$ is in the BMS little group $\ell_{\mathcal P}$ if and only if it stabilises separately the hard and soft parts. In other words, we have
\begin{proposition}\label{Proposition: BMS, Poincaré and Soft little groups}
The BMS little group is the intersection of the Poincar\'e little group and the soft little group,
    \begin{align}\label{intersect}
    \ell_{\mathcal P} = \ell_{p} \cap \ell_{\Sigma}\,.
\end{align}
\end{proposition}

It is important to realise that the normal subgroup $\mathbb{Z}_2\subset SL(2,\mathbb{C})$ formed of the matrices $\pm I$ acts trivially on $S^2\simeq \mathbb{C}P^1$ via fractional linear transformations. In other words, only $PSL(2,\mathbb{C})\simeq \frac{SL(2,\mathbb{C})}{\mathbb{Z}_2}\simeq SO_0(1,3)$ acts effectively on the sphere and, hence, on the space of supermomenta. This motivates the following terminology.

\begin{definition}\label{smoothorbits}
Let $\ell_\P$ be the BMS little group of a supermomentum $\P \in \mathcal{E}[-3]$. Its quotient $\frac{\ell_{\mathcal P}}{\mathbb{Z}_2}\subseteq SO_0(1,3)$ by the normal subgroup $\mathbb{Z}_2=\{I,-I\}$
will be called the effective BMS little group.
\end{definition}

For the sake of simplicity, the effective BMS little groups considered in the paper will always be assumed to be connected Lie subgroups of the Lorentz group (in other words, we only consider the connected component of effective little groups). In terms of the double cover, this means that the BMS little groups $\ell_\P\subseteq SL(2,\mathbb{C})$ are assumed to be either connected Lie groups, or at least connected Lie groups times a factor $\mathbb{Z}_2=\{I,-I\}$.

\section{McCarthy classification of the unitary irreducible representations of the BMS group}\label{Section: McCarthy}

In this section, we review the classification of BMS group UIRs by McCarthy \cite{McCarthy_71,Mccarthy:1972ry,McCarthy_72-I,McCarthy_73-II,McCarthy_73-III,McCarthy:1974aw,McCarthy_75,McCarthy_76-IV,McCarthy_78,McCarthy_78errata}. In order to be self-contained and fix the terminology, we first recall the classification of Poincar\'e group UIRs by Wigner \cite{Wigner:1939cj}.

\subsection{Wigner's classification of the unitary representations of the Poincar\'e group}

\begin{theorem}[Induced representations of Poincar\'e group; Wigner 1939]\label{Wignertheo}\mbox{}\\
All UIRs of the group $ISO_0(3,1)$ can be constructed by the following algorithm:

\begin{enumerate}
    \item Choose the orbit $\mathcal{O}_{{p}}$ of a momentum ${p_{\mu}} \in \mathbb{R}^{3,1*}$ with mass square $m^2\in\mathbb{R}$ under the group $SL(2,\mathbb{C})$. This orbit is called the mass shell. It is a finite-dimensional submanifold inside the vector space $\mathbb{R}^{3,1*}$ of momenta, 
    \begin{equation*}
        \mathcal{O}_{{p}} \simeq \frac{SL(2,\mathbb{C})}{\ell_{{p}}},
    \end{equation*}
where $\ell_{{p}}$ is the Poincar\'e little group of the momentum $p_{\mu}$. It is the base manifold of an $\ell_{{p}}$-principal bundle over $\mathcal{O}_{{p}}$ with total space $SL(2,\mathbb{C})$.
    \item Choose a UIR of the Poincar\'e little group  $\ell_{{p}}$ on the vector space $V$, i.e. a group morphism $\rho:\ell_{{p}} \to U(V)$ from the little group  $\ell_p$ to the group $U(V)$ of unitary operators on the vector space $V$. This representation is called the Poincar\'e spin of the corresponding induced representation of $ISO_0(3,1)$.
\end{enumerate}
Then the corresponding UIR of $ISO_0(3,1)$ with mass square  $m^2$ and Poincar\'e spin $\rho$ is given by the vector space of square-integrable sections of the homogeneous vector bundle
$E\,:=\,SL(2,\mathbb{C}) \times_{\rho} V$ over the coset space $\frac{SL(2,\mathbb{C})}{\ell_{{p}}}$
or, equivalently, by $\rho$-equivariant functions $f:SL(2,\mathbb{C}) \to V$. 
\end{theorem}

The BMS algebra inherits the quadratic Casimir operator of the Poincar\'e algebra, i.e. the square of the translation generators, as was first pointed by Sachs \cite[Theorem IV.4]{sachs_asymptotic_1962}. In fact, this operator obviously commute with all Lorentz generators as well as all supertranslations generators. Its eigenvalues correspond to the square of the momenta ${p}_\mu=\pi_\mu(\P)$ associated to the supermomenta $\P$ in the orbit of the group $SL(2,\mathbb{C})$. By Schur lemma, this real number is one of the labels characterising the BMS group UIRs \cite[Theorem 6]{McCarthy_75} but there must be infinitely more labels since the BMS group is infinite-dimensional and should admit infinitely-many Casimir operators. These are unknown to the best of our knowledge \footnote{However, as emphasised in \cite{Freidel:2024jyf},  $\int dz\wedge d\bz \,\P^{2/3}(z,\bz)$ is an interesting $SL(2,\mathbb{C})-$invariant quantity. It is not polynomial in the supermomenta so, strictly speaking, not a Casimir operator. Also note that $\P$ should not be too distributional for this quantity to make sense.}.

The particular case of UIRs with zero-momentum is usually discarded on the basis that it does not correspond to genuine particles. The existence of physically relevant soft representations of the BMS group provides a motivation for briefly mentioning the following.
\begin{corollary}[Zero-momentum representations]
The UIRs of $ISO_0(3,1)$ with zero-momentum are unfaithful representations of the Poincar\'e group where all translations act trivially. Effectively, they are faithful representations of the Lorentz group $SO_0(3,1)$. In fact, the orbit of the momentum reduces to a single point (the origin of $\mathbb{R}^{3,1*}$) and the little group identifies with the whole Lorentz group. Therefore, the homogeneous vector bundle
$E$ in Theorem \ref{Wignertheo} identifies with the vector space $V$ carrying a UIR of $SO_0(3,1)$. 
\end{corollary}

\subsection{McCarthy's classification of the unitary representations of the BMS group}\label{ssection: McCarthy Classification}

Let us summarise the following result, which can be extracted from the series \cite{McCarthy_72-I,McCarthy_73-II,McCarthy_73-III,McCarthy_76-IV,McCarthy_75,McCarthy_78,McCarthy_78errata} of seminal works on the UIRs of $BMS_4$ by  McCarthy. It is a particular application of Wigner's method \cite{Wigner:1939cj} of induced representations, generalised by Mackey \cite{MacKey} (in turn generalised to infinite-dimensional Abelian normal subgroups by various authors, see e.g. \cite{Piard}). 

\begin{theorem}[Induced representations of BMS group; McCarthy, 1972-1978]\label{McCarthytheo}\mbox{}\\
All UIRs of the group $BMS_4$ can be constructed by the following algorithm:

\begin{enumerate}
    \item Choose the orbit $\mathcal{O}_{\P}$ of a supermomentum $\P \in \mathcal{E}[-3]$ under the group $SL(2,\mathbb{C})$. This orbit is a finite-dimensional submanifold inside the vector space $\mathcal{E}[-3]$ of supermomenta,
    \begin{equation*}
        \mathcal{O}_{\P} \simeq \frac{SL(2,\mathbb{C})}{\ell_{\mathcal P}},
    \end{equation*}
where $\ell_{\mathcal P}$ is the BMS little group of the supermomentum $\P$.
It is the base manifold of an $\ell_{\mathcal P}$-principal bundle over $\mathcal{O}_{\P}$ with total space $SL(2,\mathbb{C})$. 
    \item Choose a UIR of the BMS little group  $\ell_{\mathcal P}$ on the vector space $V$, i.e. a group morphism $\rho:\ell_{\mathcal P} \to U(V)$ from the BMS little group  $\ell_{\mathcal P}$ to the group $U(V)$ of unitary operators on the vector space $V$. This representation is called the BMS spin of the corresponding induced representation of $BMS_4$.
\end{enumerate}
Then the corresponding UIR of $BMS_4$ is given by the vector space of square-integrable sections of the homogeneous vector bundle
$E\,:=\,SL(2,\mathbb{C}) \times_{\rho} V$ over the coset space $\frac{SL(2,\mathbb{C})}{\ell_{\mathcal P}}$
or, equivalently, by $\rho$-equivariant functions $f:SL(2,\mathbb{C}) \to V$. 
\end{theorem}

\proof{Strictly speaking, McCarthy classified all the induced representations described above, he proved that they are irreducible and that they exhaust all UIRs which are not strictly ergodic. In this sense, the genuine complete proof that all UIRs are obtained from Wigner's method of induced representations was only later provided in \cite{Piard1977279} for the Hilbert topology (see also \cite{cattaneo1979borel}). To the best of the authors' knowledge, a similar proof is still lacking for the nuclear topology (which is physically more sound\footnote{In fact, in the Hilbert topology all BMS little groups are compact \cite{McCarthy_71,McCarthy_72-I,McCarthy_73-II}. Therefore, this topology does not allow any hard massless representation (cf. Definition \ref{Definition: hard representation}) since the latter correspond to supermomenta which are Dirac distributions (hence they have infinite norm $L^2(S^2)$\,) for which the BMS and Poincar\'e little groups are not compact (they coincide with the Euclidean group, cf. Theorem  \ref{McCarthyGPtheo}).}).
Another subtlety was the proof by McCarthy that all projective unitary representations of the BMS group lift to linear unitary representations of its universal covering \cite{McCarthy_78} (see also \cite{LauWu}).}

\begin{remark}\label{sep}
Note that the Hilbert space of any UIR of $BMS_4$ constructed in Theorem \ref{McCarthytheo} via square-integrable wave functions must be separable, since the $L^2$ space of square-integrable functions on any separable measure space (such as the orbits\footnote{The orbits $\mathcal{O}_\P$ we consider are diffeomorphic to quotients of $G=SL(2,\mathbb{C})$ by a closed subgroup $H=\ell_\P$. The Lie group $G$ is separable, so it contains a countable dense subset $S\subset G$. If we project this subset via the map $G \twoheadrightarrow G/H:g\mapsto g\cdot H$ and define the distance on $G/H$ as the infinimum of the inherited distance from $G$, the subset $S/H$ is  a countable dense subset in $G/H$.} in Theorem \ref{McCarthytheo}) is separable (see e.g. \cite[Theorem 4.13]{Brezis}). The Fock space of a separable one-particle Hilbert space is also a separable Hilbert space. Therefore, \textit{the Fock spaces of BMS particles constructed from a countable collection of one-particle UIRs of $BMS_4$ (constructed via Wigner's method of induced representations) is a separable Hilbert space}.
\end{remark}

We can now turn to the particular cases of BMS group UIRs with zero or soft supermomenta. 

\begin{corollary}[Zero-supermomentum representations]
The zero-supermomentum representations of $BMS_4$,
\begin{equation}
    \P(z,\bz) =0,
\end{equation}
are unfaithful representations of the BMS group where all supertranslations act trivially. For any choice of Poincar\'e subgroup, they coincide with the zero-momentum representations since, effectively, they are representations of the Lorentz group.
\end{corollary}
\begin{corollary}[Soft representations]
The soft UIRs of $BMS_4$,
\begin{equation}
    \P(z,\bz) = \partial_{z}^2 \partial_{\bz}^2 \N
\end{equation}
are unfaithful representations of the BMS group where all translations act trivially. The soft representations with non-zero supermomentum representations are faithful representations of the quotient group\footnote{This quotient group was called the ``Komar group'' by McCarthy in \cite{McCarthy_72-I}, as a credit to \cite{PhysRevLett.15.76}.}
\begin{equation}
    BMS_4\,/\,\mathbb{R}^{3,1}\;\simeq\; SO(3,1)\ltimes\big(\mathcal{E}[1]\big/\mathbb{R}^{3,1}\big)\,.
\end{equation}
\end{corollary}

\subsection{Some first remarks (on BMS wavefunctions in supermomentum space)}

It follows from McCarthy's theorem that UIRs of the BMS group can be realised as functionals $\Phi[\P]$ on the vector space of supermomenta supported only on a given orbit $\mathcal{O}_{\P}$. The action of an element of the group $BMS_4$, i.e. a pair\footnote{The multiplication of such elements is $( \hat{\mathcal{T}} , \hat{M}) \circ ( \mathcal{T} , M ) := ( \hat{\mathcal{T}} + \hat{M}\cdot \T , \hat{M}M)$.} $\big(\T, M \big) \in \E[1] \times SL(2,\mathbb{C})$ of a supertranslation $\T$ and of a Lorentz transformation $M$, is then given by
\begin{equation}\label{BMS action on representation}
    \Phi[\P] \quad \mapsto \quad \big(\T, M \big) \cdot \Phi[\P] :=  e^{i\, \langle \P , \T \rangle} \,M\cdot\Phi[\P]
\end{equation}
where the action of $SL(2,\mathbb{C})$ on the right-hand side is the action on sections of the associated bundle $E\,:=\,SL(2,\mathbb{C}) \times_{\rho} V$. If $\rho$ is trivial then this is just a pull back
\begin{equation}\label{BMS action on representation2}
    \Phi[\P] \quad \mapsto \quad \big(\T, M \big) \cdot \Phi[\P] :=  e^{i\, \langle \P , \T \rangle} \Phi[\P\cdot M]
\end{equation}
where the action of $SL(2,\mathbb{C})$ on supermomenta was defined in \eqref{SL2Caction}.

The inclusion $\ell_{\mathcal P} \subseteq \ell_{\pi(\P)}$ (see Proposition \ref{Proposition: BMS, Poincaré and Soft little groups}) implies that the orbit $\mathcal{O}_{\P}$ of the supermomentum is the total space of a fibre bundle whose base space is the orbit $\mathcal{O}_{\pi(\P)}$ of its associated momentum $p_{\mu} := \pi_{\mu}\left(\P\right)$ and whose typical fibre is the coset space $\frac{\ell_{\pi(\P)}}{\ell_{\mathcal{P}}}$\,.
Hence, there are three relevant fibre bundle structures, which can be summarised in the following commutative diagram
\begin{center}
\begin{tikzcd}
SL(2,\mathbb{C}) \arrow[r,twoheadrightarrow] \arrow[dr,twoheadrightarrow] 
& \mathcal{O}_{\P}\simeq \frac{SL(2,\mathbb{C})}{\ell_{\mathcal{P}}} \arrow[d,twoheadrightarrow]\\
& \qquad\mathcal{O}_{\pi(\P)}\simeq \frac{SL(2,\mathbb{C})}{\ell_{\pi(\mathcal{P})}}
\end{tikzcd}
\end{center}
The two left-to-right arrows are the projections of the principal bundles appearing in Theorems \ref{Wignertheo} and \ref{McCarthytheo}. The vertical bundle on the right therefore encapsulates the discrepancy between Poincaré and BMS representations. This is a fibre bundle over $\mathcal{O}_{\pi(\P)}$ whose typical fibre
\begin{equation}
    F \simeq \frac{\ell_{\pi(\P)}}{\ell_{\P}}\label{fibre}
\end{equation}
\emph{encodes the extra degrees of freedom} of BMS representations.
This point of view will be used in Sections \ref{Section: MasslessUIRsBMS} and \ref{Section: MassiveUIRsBMS}.

From Proposition \ref{Prop: classification of supermomenta} and Theorem \ref{McCarthytheo}, we deduce that a UIR of $BMS_4$ is given by a choice of  $SL(2,\mathbb{C})$-orbit in the set $(\mathbb{R}^{3,1})^*\times(\mathcal{E}[1]\big/\mathbb{R}^{3,1})$ of pairs of momenta and soft charges. This point of view will be used in Sections \ref{Section: BMS wavefunctions} where we will further discuss BMS wavefunctions.

\subsection{Branching rules: restricting the BMS group to a Poincar\'e subgroup}\label{branching1}

The branching rules of UIRs with respect to the restriction of the BMS group to a Poincar\'e subgroup have been studied in a series of papers \cite{McCarthy_73-II,McCarthy_73-III,McCarthy_76-IV,McCarthy_75} by McCarthy (some of which in collaboration with Crampin).  

\begin{theorem}[Branching rule of BMS group; Crampin \& McCarthy, 1973-1976]\label{branchingrulestheo}\mbox{}\\

Upon restriction to a Poincar\'e subgroup $ISO_0(3,1)\subset BMS_4$, the UIR of $BMS_4$ of mass square $m^2\in\mathbb{R}$ and of BMS spin $\rho:\ell_{\mathcal P} \to U(V)$ decomposes into a direct sum of UIRs of $ISO_0(3,1)$, all of which with identical mass square $m^2$, for all Poincar\'e spin $\rho_i:\ell_{\pi(\mathcal{P})} \to U(V_i)$  such that the BMS spin $\rho$ appears in the decomposition of $\rho_i$ upon restriction of $\ell_{\pi(\mathcal{ P})}$ to $\ell_{\mathcal{ P}}$ (i.e. $V\subset V_i$ is an $\ell_{\mathcal{ P}}$-invariant subspace of the space $V_i$ carrying the representation $\rho_i$ of $\ell_{\pi(\mathcal{ P})}$).

In particular, the  UIRs of the BMS group which are such that the BMS and Poincar\'e little groups coincide, remain irreducible after restriction to a Poincar\'e subgroup $ISO_0(3,1)\subset BMS_4$.
\end{theorem}

\proof{The first paragraph in Theorem \ref{branchingrulestheo} here is extracted from  \cite[Theorem 7]{McCarthy_75}. It is a corollary of Frobenius reciprocity theorem (recalled in \cite[Section 2]{McCarthy_73-III} for the particular case of compact groups). For concrete examples of branching rules, see \cite[Section 2]{McCarthy_73-III} for the massive case and \cite[Section 6]{McCarthy_75} for the massless and tachyonic cases.
The second paragraph in Theorem \ref{branchingrulestheo} is a direct corollary of the first one (see e.g. \cite[Theorem 14]{McCarthy_75} for the massless case).
}
This branching rule can be seen as a seesaw mechanism where the UIRs appearing in the restriction of $BMS_4\downarrow ISO_0(3,1)$ are in one-to-one correspondence with the UIRs consistent with the restriction $\ell_{\pi(\mathcal{P})}\downarrow\ell_{\mathcal{P}}$:
\begin{center}
\begin{tikzcd}
BMS_4 \arrow[dr,dash] 
& \ell_{\pi(\mathcal{P})}\\
\arrow[u,phantom, sloped, "\subset"]ISO_0(3,1) \arrow[ur,dash] & \arrow[u,phantom, sloped, "\subset"]\ell_{\mathcal{P}}
\end{tikzcd}
\end{center}

The statement of this theorem may be difficult to grasp for the moment but explicit algorithms for realising the decomposition will be provided in Sections \ref{Section: HardrepresentationsoftheBMSgroup}-\ref{Section: MassiveUIRsBMS}. This is of importance because, in physical terms, these branching rules amount to reading off the field content in terms of usual Poincaré particles -- as seen in a gravity vacuum -- of a given BMS particle. Along the way we will also provide an exhaustive discussion of the multiplicity of the Poincaré representations appearing upon branching, a point which has somewhat been left aside by McCarthy.  

\begin{remark}
Let us anticipate Sections \ref{Section: MasslessUIRsBMS}-\ref{Section: MassiveUIRsBMS} (where the branching rules of generic massless and massive UIRs are treated exhaustively) by mentioning that most UIRs of the BMS group branch into infinite sums of UIRs of any Poincar\'e subgroup. This is because BMS and Poincar\'e wavefunctions are in general sections over orbits of distinct  dimensions.
Physically, the  fact that generic UIRs of BMS group branch into infinite sums of UIRs of Poincar\'e group means that a single BMS particle corresponds, in general, to an infinite multiplet of Poincar\'e particles. Therefore, the spectrum of particles with respect to a given choice of Poincar\'e group necessarily contains an infinite tower of particle with the same mass square. Such a spectrum violates one\footnote{\label{ColemanManduala}The theorem of Coleman and Mandula assumes that there exists a finite number of particle types with mass below any given mass (cf. Assumption 2 in \cite{Coleman:1967ad} and, respectively, Assumption 1 in \cite[Appendix 24.B]{Weinberg:2000cr}).} of the crucial assumptions of Coleman-Mandula theorem \cite{Coleman:1967ad} (see also, for a pedagogical introduction, \cite[Appendix 24.B]{Weinberg:2000cr}). This could be a reason why the Coleman-Mandula theorem may not prohibit the existence of a non-trivial BMS-invariant $S$-matrix (see also Footnote \ref{CMtheorem}). 
\end{remark}

\section{Hard representations of the BMS group}\label{Section: HardrepresentationsoftheBMSgroup}

In this section, we emphasise some properties of the hard representations. These properties will justify the name of the representation and, ultimately, justify to consider the decomposition \eqref{genericdecomppsition}. First, we discuss how they fit in McCarthy's works (in terms of their little groups and their branching rules). Second, we summarise how they appear as the result of the action of the BMS group on scattering data of usual massive and massless fields.

\subsection{Supermomenta with coinciding Bondi and Poincar\'e little groups}\label{coincidingBondiandPoincarelittlegroups}

In a series of mathemathical works \cite{McCarthy_72-I,McCarthy_73-II,McCarthy_75,McCarthy_78errata,Girardello:1974sq}, McCarthy, as well as Girardello and Parravicini,  determined the form of the supermomenta stabilised by all possible subgroups of $SL(2,\mathbb{C})$. Of particular interest are those for which the BMS little group $\ell_{\mathcal P}$ coincides with the Poincar\'e little group $\ell_{\pi(\P)}$.

\begin{theorem}[McCarthy, Girardello, Parravicini, 1972-1978]\label{McCarthyGPtheo}\mbox{}\\
  Let $\P(z, \bz) \in \E[-3]$ be a supermomentum whose BMS little group $\ell_{\mathcal P}$ and Poincaré little group $\ell_{\pi(\P)}$ coincide. Then this supermomentum must take one of the following forms:
   \begin{enumerate}
       \item If $\ell_{\mathcal P} = \ell_{\pi(\P)}= SU(2)$ then $P(z,\bar z)=-\frac{m^4}{\pi}\big(p_{\mu} q^{\mu}(z,\bar z)\big)^{-3}$ for some timelike vector $p_{\mu}$.
       \item If $\ell_{\mathcal P} = \ell_{\pi(\P)}= ISO(2)$ then $P(z,\bar z)=  \omega\,\delta^{(2)} (z - \zeta,\bar z - \bar\zeta) +\frac{\sigma}{\omega^3}\partial_z^2 \partial_{\bz}^2 \delta^{(2)}(z-\zeta,\bar z - \bar\zeta)$ for $\omega\in\mathbb{R}\setminus\{0\}$, $\sigma\in\mathbb{R}{\cup\{\infty\}}$ (the limiting case where $\sigma=\infty$ is when the first of the two terms is not present) and $\zeta\in\mathbb{C}\cup\{\infty\}$.
       \item If $\ell_{\mathcal P} = \ell_{\pi(\P)}= SL(2,\mathbb{R})$ then $P(z,\bar z)=-\frac{m^4}{\pi}\big(p_{\mu} q^{\mu}(z,\bar z)\big)^{-3}$ for some spacelike vector $p_{\mu}$.
       \item If $\ell_{\mathcal P} = \ell_{\pi(\P)}= SL(2,\mathbb{C})$ then $P(z,\bar z)=0$.
   \end{enumerate}
   \end{theorem}
\noindent In other words, there exists
\begin{enumerate}
       \item two one-parameter families of three-dimensional orbits $\frac{SL(2,\mathbb{C})}{SU(2)} \simeq H^3$ whose supermomenta $\mathcal P$ are such that $\ell_{\mathcal P} = \ell_{\pi(\P)}= SU(2)$. Each family is parametrised by $m\in \mathbb{R}^+\setminus\{0\}$ and the future/past orientation of the momentum.
       \item two one-parameter families of three-dimensional  orbits $\frac{SL(2,\mathbb{C}}{ISO(2)} \simeq \mathbb{R} \times S^2$ whose supermomenta $\mathcal P$ are such that $\ell_{\mathcal P} = \ell_{\pi(\P)}= ISO(2)$. Each family is parametrised by $\sigma \in \mathbb{R}\cup\{\infty\}$ and the sign of $\omega$.
       \item a one-parameter family of three-dimensional  orbits $\frac{SL(2,\mathbb{C})}{SL(2,\mathbb{R})} \simeq dS_3$ whose supermomenta $\mathcal P$ are such that $\ell_{\mathcal P} = \ell_{\pi(\P)}= SL(2,\mathbb{R})$. This family is parametrised by $m\in \mathbb{R}^+\setminus\{0\}$.
       \item a single orbit (made of one supermomentum: the trivial one) such that $\ell_{\mathcal P} = \ell_{\pi(\P)}= SL(2,\mathbb{C})$.
   \end{enumerate}

\proof{See Table 1 in \cite{McCarthy_78errata} for the correct and final version of the result (where the notations and conventions are the ones from the paper \cite{McCarthy_75}). Several comments are in order because the result strongly depends on the functional class that one chooses to consider. 

For instance, Table 2 in \cite{McCarthy_75} included, for the massless case $\ell_{\mathcal P} = \ell_{\pi(\P)}= ISO(2)$ an additional term, which belongs to the $SL(2,\mathbb{C})$-orbit of the constant hard momentum $P(z,\bar z)=K$ for $K\in\mathbb{R}\setminus\{0\}$. However, as explained in Remark \ref{masslessremark}, this term is not admissible in the functional class that is considered. In fact, the soft massless supermomentum of the form  $\frac{1}{\omega^3}\partial_z^2 \partial_{\bz}^2 \delta^{(2)}(z-\zeta,\bar z - \bar\zeta)$ is a distributional regularisation of the singular density $\big(\omega\, q_\mu(\zeta,\zetab)\, q^{\mu}(z,\bar z)\big)^{-3}$.\footnote{This can be seen by acting with $SL(2,\mathbb{C})$ on the second example in Remark \ref{masslessremark}).}

Also note that Table 2 in \cite{McCarthy_75} and Table 1 in \cite{McCarthy_78errata} both include, for the tachyonic case $\ell_{\mathcal P} = \ell_{\pi(\P)}=SL(2,\mathbb{R})$, the further possibility of a distribution with support on a whole circle. We did not include such a term because it looks too pathological (cf. Remark \ref{Remark 2.2:}). Moreover, by analogy with the massless case (and the natural analogue of the results in \cite[Section B.2.7]{Gelfand} for the analytic case), one would expect such a distributional term to identify with the regularisation of the hard supermomentum written in Proposition \ref{McCarthyGPtheo}. For instance, via the correspondence \eqref{homogeneouscorrespondence} the hard momentum $P(z,\bar z)=\pm\,\frac{m}{\pi}(1- |z|^2)^{-3}$ is associated to the homogeneous function $\Phi(\lambda_\alpha,\bar\lambda_{\dot \alpha})=\pm\,\frac{m}{\pi}\big(|\lambda_1|^2- |\lambda_2|^2\big)^{-3}$ on $\mathbb{C}^2\setminus \{0\}$ whose regularisation in the sense of distributions would be $\tilde\Phi(\lambda_\alpha,\bar\lambda_{\dot \alpha})\,\propto\,\delta^{\prime\prime}\big(|\lambda_1|^2- |\lambda_2|^2\big)$ and corresponds to $\tilde P(z,\bar z)=\delta^{\prime\prime}(1- |z|^2)$.}

Note that the hard supermomenta $P(z,\bar z)$ appearing in the decompositions \eqref{massivedecomppsition}-\eqref{tachyondecomppsition} of generic supermomenta into hard and soft parts,  are precisely such that their Poincar\'e and BMS little groups coincide as above, as follows from \eqref{intersect}.

\begin{definition}\label{Definition: hard representation}
    We call hard representations of the BMS group, the unitary irreducible representations of $BMS_4$ with non-vanishing supermomenta for which the soft charge vanishes, i.e. $\partial_z^2 \partial_{\bar z}^2\N(z,\bar z)=0$ in \eqref{genericdecomppsition}. In other words, these are the representations whose supermomenta $\P(z,\bar z) =P(z,\bar z)$ are hard. The three possible cases are the three types of hard momenta $P(z,\bar z)$ written in \eqref{massivedecomppsition}, \eqref{masslessdecomppsition} and \eqref{tachyondecomppsition}. These hard momenta are respectively in the orbit of 
    \begin{align}\label{CanonicalMasslessHardSupermomentum}
   1.\;\; P(z,\bar z)&= \pm \,\frac{m}{\pi}(1+ |z|^2)^{-3}& 
   2.\;\;  P(z,\bar z)&= \delta^{2}(z,\bar z),&
   3.\;\; P(z,\bar z)&= \pm\,\frac{m}{\pi}(1- |z|^2)^{-3},
\end{align}
where $m>0$.
These hard supermomenta share the essential property that their stabiliser subgroups are the usual Poincar\'e little groups, i.e. $SU(2)$, $ISO(2)$ and $SL(2,\mathbb{R})$ respectively.
\end{definition}

Theorem \ref{McCarthyGPtheo} ensures that, in the massive and tachyonic cases, the hard supermomenta in \eqref{massivedecomppsition} and \eqref{tachyondecomppsition} are singled out by the property that their BMS and Poincar\'e little groups coincide. However, this property is not enough to single out massless hard supermomenta since, with respect to  
the massless hard supermomenta in \eqref{masslessdecomppsition}, there is an extra (one-parameter family of) soft charge(s) that appear in the massless supermomenta inside the second case listed in Theorem \ref{McCarthyGPtheo}.

Note that, in the massive and tachyonic cases ($m>0$) and at points where these supermomenta are not singular, they define constant-curvature metrics 
\begin{equation*}
\P^{\frac{2}{3}} dz\, d\bar z\;=\, \left(\frac{m}{\pi}\right)^\frac23\frac{4\,dz\,d\bar z}{1\pm |z|^2}
\end{equation*}
which correspond, respectively, to the metric of the round sphere $S^2$ or the hyperbolic disk $ H^2$, whose isometry groups coincide with the little groups of the supermomenta. 

\subsection{Addition of hard supermomenta}\label{Ssection: addition of supermomenta}

It is crucial to remember that the manifold of hard supermomenta $P(z,\bar z)$ is \emph{not} a vector subspace of the vector space of supermomenta. It is not stable under addition : if $p_3^{\mu} = p_1^{\mu} + p_2^{\mu}$ is the sum of two momenta $p_1^{\mu}$ and $p_2^{\mu}$, and $P_1$, $P_2$, $P_3$ are the corresponding hard momenta, then by Corollary \ref{Corollary: sum of supermomenta} one has, 
\begin{equation}\label{Hard supermomenta: sum of 2}
    P_1(z,\bz) +  P_2(z,\bz)  = P_3(z,\bz) -\frac{1}{2\pi}\partial_{z}^2\partial_{\bz}^2 \mathcal{S}(z,\bz)
\end{equation}
where $\mathcal{S} = (q \cdot p_1) \ln|q \cdot p_1| + (q \cdot p_2) \ln|q \cdot p_2| - (q \cdot p_3) \ln|q \cdot p_3|$ encodes the non-linearity.
The phenomenon is similar to the fact that orbits of usual momenta $p_\mu$ are quadrics  in $\mathbb{R}^4$ and, thus, are \textit{not} vector subspaces of the vector space of momenta. Nevertheless the subspace of hard supermomenta can be uniquely defined and, in particular, it is $SL(2,\mathbb{C})$-invariant by construction.\\

It might be important to keep in mind that the addition of two hard supermomenta \eqref{Hard supermomenta: sum of 2} gives special, non generic, type of supermomenta. As was emphasised in \cite{Chatterjee:2017zeb}, if $p_1^{\mu}$ and $p_2^{\mu}$ are two massive momenta, the sum \eqref{Hard supermomenta: sum of 2} of the corresponding hard supermomenta $P_1(z,\bz)$ and $P_2(z,\bz) $ will always have stabiliser $\ell_{p_1} \cap \ell_{p_2}\simeq SO(2)\simeq U(1)$. A similar reasoning gives the following.
\begin{proposition}\label{hard_supermomenta_2to1}
Let $p_1^{\mu}$ and $p_2^{\mu}$ be two, non-collinear 
momenta, either massive or massless (i.e. $(p_1)^2\leqslant 0$ and $(p_2)^2\leqslant 0$). Then the sum \eqref{Hard supermomenta: sum of 2} of the two corresponding hard supermomenta, as well as its soft part
\begin{equation}\label{Soft supermomentum from a 2-hard particle process}
        \partial_z^2 \partial_{\bz}^2\mathcal{S} = \partial_z^2 \partial_{\bz}^2 \Big( (q \cdot p_1) \ln|q \cdot p_1| + (q \cdot p_2) \ln|q \cdot p_2| - (q \cdot p_3) \ln|q \cdot p_3|\Big)\,,
\end{equation}
always have as non-trivial BMS little group 
\begin{equation}
    \ell_{P_1+P_2}=\ell_{\partial_z^2 \partial_{\bz}^2\mathcal{S}}=\ell_{p_1} \cap \ell_{p_2}\simeq SO(2)\simeq U(1)\,.
\end{equation}
\end{proposition}

\proof{Since $\ell_{p_1} \cap \ell_{p_2}$ stabilises both $p_1$ and $p_{2}$, this subgroup also stabilises the sum $p_3=p_1+p_2$ of these momenta, as well as the sum 
\begin{align}
    \P_3=P_1 + P_2=P_3+\Sigma_3
\end{align} of the corresponding hard supermomenta \eqref{Hard supermomenta: sum of 2} (thus here $\Sigma_3 = -\frac{1}{2\pi}\partial_{z}^2\partial_{\bz}^2 \mathcal{S}$). More precisely, we have $\ell_{p_1} \cap \ell_{p_2}\subseteq\ell_{\P_3}\subseteq \ell_{p_3}$ due to Proposition \ref{Proposition: BMS, Poincaré and Soft little groups}. 
This intersection $\ell_{p_1} \cap \ell_{p_2}$ is always non-trivial and isomorphic to $SO(2)$: to see this, recall that two non-collinear 
momenta $p_{1}$, $p_2$ generate a plane $\mathbb{R}^{1,1} \subset \mathbb{R}^{3,1}$. Then $\ell_{p_1} \cap \ell_{p_2}$ consists of rotations in the orthogonal plane $\mathbb{R}^2 = \left( \mathbb{R}^{1,1} \right)^{\perp}$.

To summarise, we have
\begin{equation}
    U(1)\simeq \ell_{p_1} \cap \ell_{p_2}\subseteq \ell_{\P_3}\subseteq\ell_{p_3},
\end{equation}
where the Poincar\'e little group $\ell_{p_3}$ is either $SU(2)$, $ISO(2)$ or $SL(2,\mathbb{R})$. It is known that there is no subgroup of $SL(2,\mathbb{C})$ sitting in between $U(1)$ and the latter subgroups (see e.g. \cite[Table 3]{McCarthy_75}). Therefore, the BMS little group $\ell_{\P_3}$ coincides either with $U(1)$ or with $\ell_{p_3}$. The equality \eqref{Hard supermomenta: sum of 2} and Theorem \ref{McCarthyGPtheo} further imply that $\ell_{\P_3}$ coincides with $U(1)$, because the soft part is non-trivial (and not of the form $\partial_{\bz}^2 \delta^{(2)}(z-\zeta,\bar z - \bar\zeta)$ in the massless case).

We can now write that
\begin{equation}
    U(1)\simeq \ell_{p_1} \cap \ell_{p_2}= \ell_{\P_3} \subseteq \ell_{\Sigma_3}
\end{equation}
since we know that $\ell_{p_2}= \ell_{\P_3}$ (due to the above paragraph) and that $\ell_{\P_3} \subseteq \ell_{\Sigma_3}$ (due to Proposition \ref{Proposition: BMS, Poincaré and Soft little groups}). 
In McCarthy's classification, the only BMS little groups including $U(1)$ as a subgroup are $SU(2)$, $ISO(2)$ and $SL(2,\mathbb{R})$ (compare \cite[Table 3]{McCarthy_75} with \cite[Table 1]{McCarthy_78errata}). However, supermomenta with such stabilisers are given in the list from Theorem \ref{McCarthyGPtheo}. Since the soft supermomentum \eqref{Soft supermomentum from a 2-hard particle process} does not belong to this list, the inclusion is in fact an equality.}

The above result can be easily generalised to three (or more) massless momenta.

\begin{proposition}\label{hard_supermomenta_nto1}
Let $p_1^{\mu},\,\ldots,\, p_n^{\mu}$ be $n\geqslant 3$ non-collinear massless momenta (i.e. $(p_i)^2\leqslant 0$, $\forall i$). Then the sum 
\begin{equation}\label{Hard supermomenta: sum of n}
\P_{n+1}(z,\bz)=    P_1(z,\bz) +  \cdots+P_n(z,\bz) 
\end{equation}
of the $n$ corresponding hard supermomenta, $P_1(z,\bz),\,\ldots,\,P_n(z,\bz)$, always has BMS little group $\ell_{\P_{n+1}}\simeq\mathbb{Z}_2$.
\end{proposition}

\proof{The hard supermomentum associated to the massless momentum $p_i^{\mu}=\omega_i\,q^\mu(z,\bz)$ is a distribution $P_i(z,\bar z)=  \omega_i\,\delta^{(2)} (z - \zeta_i,\bar z - \bar\zeta_i)$ with support at the point $z=\zeta_i\in\mathbb{C}P^1$. Therefore, the sum \eqref{Hard supermomenta: sum of n} of these hard supermomenta is a supermomentum $\P_{n+1}(z,\bz)\in\E[-3]$ whose support is the collection $\{\zeta_1,\ldots,\zeta_n\}$ of those $n$ points. The momenta $p_i$ are non-collinear, so those $n\geqslant 3$ points $\zeta_i$ are distinct. Clearly, the little group $\ell_{\P_{n+1}}$ must necessarily preserve this set of $n\geqslant 3$ points. It is well-known that a fractional linear transformation is determined uniquely by the image of three distinct points. Therefore, there are at most a finite number of fractional linear transformations preserving the set $\{\zeta_1,\ldots,\zeta_n\}\subset\mathbb{C}P^1$. This proves Proposition \ref{hard_supermomenta_nto1}.
}

To conclude this section we would like to stress that it is the non-linearity of the space of hard supermomenta which implies that hard representations cannot by themselves ensure supermomentum conservation in a scattering process.

\subsection{Absence of branching for hard representations}

The last paragraph of Theorem  \ref{branchingrulestheo} leads to the following corollary of Theorem \ref{McCarthyGPtheo}.

\begin{corollary}\label{nobranchingtheo}\mbox{}\\
The hard representations of the BMS group do not branch under restriction to a Poincar\'e subgroup.  
\end{corollary}
To flesh the meaning of this corollary, let us use  ``particle'' interchangeably with ``unitary irreducible representation'' so that ``Poincaré particle'' and ``BMS particle'' stands for UIR of the Poincaré and of the BMS group respectively. The above result means that the physical content, in terms of Poincaré particles, of a BMS hard particle is always a single Poincaré particle and, this, independently of the choice of Poincaré subgroup $ISO_0(3,1) \subset BMS_4$. In other words, \emph{for hard BMS particles, all observers agree on the field content of a given state}, even though they perhaps are in different gravity vaccua (i.e. disagree on which Poincaré subgroup to consider).

\subsection{Hard representations as scattering data}\label{scatteringdata}

The hard representations of the BMS group can in fact be realised as the (asymptotic) scattering data of (massive, massless, or tachyonic) fields on Minkowski spacetime induced at the corresponding asymptotic boundary (respectively: timelike, null, spacelike infinity). We here briefly review the corresponding results; we treat the case of a scalar field and refer to the literature for other spins.

Let $\Phi(X)$ be a solution of the Klein--Gordon equation
\begin{equation*}
    \Phi(X) \,\propto \,\int \frac{d^3 \bm{k}}{\sqrt{ \bm{k}^2 +m^2}} \bigg( a\big(p(\bm{k})\big) e^{ip_{\mu}(\bm{k}) X^{\mu}} + a\big(p(\bm{k})\big)^{\dagger} e^{-ip_{\mu}(\bm{k}) X^{\mu}} \bigg).
\end{equation*}
The procedure to lift a massless UIR of Poincaré group to a UIR of BMS group is well known and goes back to Sachs \cite{sachs_asymptotic_1962} for the case of spin zero. Since then, under the influence of the seminal works of Strominger \cite{Strominger:2013jfa,Strominger:2017zoo,He:2014laa}, it has been considerably revisited see e.g. \cite{Donnay:2022wvx,Bekaert:2022ipg,Nguyen:2023vfz}.  The procedure goes as follows.

Introduce Bondi-type  coordinates \cite{bondi_gravitational_1962,sachs_asymptotic_1962} on Minkowski spacetime $\mathbb{R}^{3,1}$, i.e. 
\begin{equation}
X^{\mu} = \Big( u+ r\big(1+|z|^2\big) , r(z+ \bar z), -ri(z-\bar z), r\big(1-|z|^2\big) \Big)    
\end{equation} 
where $u$ is retarded time and $r$ is the radius. Taking the large $r$ expansion of the fields, one obtains a field at null infinity
\begin{equation}
    \varphi(u,z,\bar z) \,\propto\, \frac{i}{2}\int d\omega\,\omega\, \Big( a(\omega,z,\bar z) e^{i\omega u} + a(\omega,z,\bar z) ^{\dagger} e^{-i\omega u} \Big) \quad  \in \C^{\infty}\left( \mathbb{R} \times S^2 \right),
\end{equation}
where here $a(\omega, z, \bz) := a\big(p_\mu=\omega\, q_{\mu}(z,\bz)\big)$. Following \cite{Donnay:2022wvx}, we will refer to the resulting space of scattering data $\varphi(u, z, \bz) \in \E[-1](\mathscr{I})$ as a Carrollian field in order to emphasise that they are intrinsically defined on a null hypersurface. Since the BMS group is realised as a subgroup of diffeomorphisms of $\mathscr{I}$ \cite{Geroch1977,Duval:2014uva}, the Carrollian fields naturally carry a UIR of the BMS group, with the action given by pullback. In particular, the action of a supertranslation $\T(z,\bz)$ is given by
\begin{equation}
     \varphi(u, z, \bz) \mapsto\varphi'(u, z, \bz) = \varphi\big(u +\T(z,\bz), z, \bz\big)
\end{equation}
which yields
\begin{equation}
    a(\omega, \zeta, \zetab)\mapsto a'(\omega, \zeta, \zetab) =e^{i\int\frac{i}{2} dz\wedge d\bz \,\P(z,\bz) \,\T(z,\bz)}a(\omega, \zeta, \zetab), \quad \text{with}\quad \P(z,\bz)= \omega\delta^{(2)}(z-\zeta).
\end{equation}
Comparing with \eqref{BMS action on representation}, we conclude that this BMS representation is a massless hard representation.

The procedure to lift a massive Poincaré particle to a BMS representation is more recent, as far as we are aware it originates from Longhi--Materassi \cite{Longhi:1997zt}. It was related to the action of asymptotic symmetries at time-like infinity in \cite{Campiglia:2015lxa,Campiglia:2015kxa} and realised as the space of Carrollian fields at $Ti$ in \cite{Have:2024dff,Borthwick:2024skd}. In this case, one introduces Beig--Schmidt-type \cite{beig_einsteins_1982} coordinates $X^{\mu} = r\left( \sqrt{1 + \bm{k}^2}, \bm{k}\right)$ on $\mathbb{R}^{3,1}$. Taking the large $r$ expansion of the bulk field, one obtains a field at $Ti\simeq \mathbb{R} \times H^3$, the extended boundary\footnote{Following \cite{Figueroa-OFarrill:2021sxz} this is meant to be pronounced ``tie" to rhyme with ``scri''.} at time-like infinity constructed in \cite{Figueroa-OFarrill:2021sxz,Borthwick:2023lye,Borthwick:2024wfn} and \cite{Borthwick:2024skd}
\begin{equation}
    \varphi(u, \bm{k} ) \, \propto\,  a(\bm{k}) e^{-im u} \quad  \in \C^{\infty}\left( \mathbb{R} \times H^3 \right),
\end{equation}
where $a(\bm{k}) := a\big(p_\mu(\bm{k})\big)$.  The resulting scattering data is once again a ``Carrollian field''. More precisely, here  $\varphi(u, \bm{k}) \in \mathcal{C}^{\infty}(Ti)$ and the null direction $u$ on $Ti$ is needed to encode the mass dependence of the field. In fact,  the ``mass'' generator is central in the Carroll algebra. The Spi group $SPI_4 \simeq  SO_0(3,1)\ltimes C^{\infty}(H^3)$ of Ashetekar--Hansen \cite{Ashtekar:1978zz,Ashtekar:1991vb} is realised as a subgroup of diffeomorphisms of Ti and thus naturally acts on Carrollian fields. In particular the action of Spi-supertranslations $\omega(\bm{k}) \in C^{\infty}(H^3)$ is given by
\begin{equation}
     \varphi(u , \bm{k})\mapsto \varphi'(u , \bm{k})= \varphi(u + \omega(\bm{k}) , \bm{k}).
\end{equation}
It then turns out \cite{Troessaert:2017jcm,Henneaux:2018hdj,Henneaux:2018cst,Henneaux:2019yax,Chakraborty:2021sbc} (see also \cite{Compere:2023qoa} for a comprehensive review and further referecences) that, under suitable assumptions at timelike infinity, there is a preferred BMS group inside the Spi group with BMS-supertranslations mapped into Spi-supertranslations via\footnote{As has been frequently emphasised in the literature, this map is in fact a bulk-to-boundary propagator for the differential operator $(\Box -3)$ on $H^3$.} $\T(z,\bz) \mapsto \omega(\bm{k}) :=  \int \frac{idz d\bz}{2} \frac{m^3\T(z,\bz)}{\pi\left(p^{\mu}(\bm{k})q_{\mu}(z,\bar z)\right)^{3}}$. This yields a representation of the BMS group on the Carrollian fields at $Ti$. In particular, supertranslations act as 
\begin{equation}
    a(\bm{k}) \mapsto a'(\bm{k}) =e^{i \int \frac{i}{2} dz\wedge d\bz \, \P(z,\bz) \T(z,\bz)}a(\bm{k}), \quad \text{with}\quad \P(z,\bz)= -\frac{m^4}{\pi}\Big(p^{\mu}(\bm{k}) q_{\mu}(z,\bar z)\Big)^{-3}
\end{equation}
and it follows from \eqref{BMS action on representation} that this representation is the massive hard representation.
The tachyonic case goes exactly along the same lines where $Ti$ is replaced with $Spi\simeq \mathbb{R}\times dS_3$.

\section{Massless unitary irreducible representations of the BMS group}\label{Section: MasslessUIRsBMS}

We will here work out explicitly aspects of massless BMS particles: in the first three subsections we describe wavefunctions, the action of the BMS group and inner products. For concreteness, a series of examples will close each of these sections. We then discuss and provide an algorithm for the branching of these UIRs of the BMS group under restriction to a Poincaré subgroup, i.e. how to decompose a given massless BMS particle in terms of usual Poincaré particles. Finally, we discuss the dependence of this decomposition on the gravity vacua, i.e. on the particular chosen Poincaré subgroup.

\begin{definition}
We call massless representations of the BMS group those UIRs for which the associated momentum $p_{\mu}$ is null $p^2=0$ and non-zero. By Proposition \ref{Prop: classification of supermomenta} it must be in the orbit of
\begin{equation}\label{BMS massless rep: Reference supermomenta1}
\K(z,\bar z) = \delta^{2}(z,\bar z) +  \partial_z^2 \partial_{\bar z}^2 \N(z,\bar z).
\end{equation}
The BMS litle group $\ell_{\K}$ is the stabiliser of this reference supermomentum. The Poincar\'e little group $\ell_k \simeq ISO(2)$ is the stabiliser of the corresponding reference momentum
\begin{equation}\label{BMS massless rep: Reference momenta1}
k^{\mu} = \pi^{\mu}(\K)=(1,0,0,1).
\end{equation}
\end{definition}
The soft little group $\ell_{\Sigma}$, defined as the stabiliser of the soft part $\Sigma =\partial_z^2 \partial_{\bar z}^2 \N$ of \eqref{BMS massless rep: Reference supermomenta1}, see Section \ref{ssection: BMS, Poincaré and soft little groups}, might a priori have any dimension between $0$ and $6$. For instance, $\ell_{\Sigma} = SL(2,\mathbb{C})$ for hard representations (since then $\partial_z^2 \partial_{\bar z}^2 \N(z,\bar z)=0$) while generically $\ell_{\Sigma}=\{I,-I\}\simeq\mathbb{Z}_2$ (since only $PSL(2,\mathbb{C})\simeq \frac{SL(2,\mathbb{C})}{\mathbb{Z}_2}$ effectively acts on $S^2$ and a generic soft supermomentum $\partial_z^2 \partial_{\bar z}^2 \N(z,\bar z)$ has no non-trivial stabiliser at all).  

The BMS little group $\ell_{\K}$ is defined as the stabiliser of \eqref{BMS massless rep: Reference supermomenta1}; it is related to the soft little group via $\ell_{\K} \simeq ISO(2) \cap \ell_{\Sigma}$ (again see Subsection \ref{ssection: BMS, Poincaré and soft little groups}). It follows from \cite{McCarthy_75} that the only possible BMS little groups for supermomenta of the form \eqref{BMS massless rep: Reference supermomenta1} (or, equivalently, \eqref{masslessdecomppsition}), are\footnote{More precisely, these are all possible little groups $\ell_{\K}$ such that the quotient $\ell_{\K}/\mathbb{Z}_2$ is connected. We leave aside the possibility of further discrete  extensions.}
\begin{equation}\label{massless reps: admissible little groups}
    \ell_{\K} \in \Big\{ \,\mathbb{Z}_2\;,\; U(1)\;, \;\mathbb{Z}_2\times\mathbb{R}\;, \;\mathbb{Z}_2\times \mathbb{R}^2\; , \; \ISO\;\Big\}\,.
\end{equation}
where $\ISO\simeq U(1)\ltimes \mathbb{C}$ stands for the double cover of the Euclidean group $ISO(2)\simeq SO(2)\ltimes\mathbb{R}^2$. The effective BMS little groups are somewhat simpler to describe, since they all are connected subgroups of the Lorentz group: $ \frac{\ell_{\K}}{\mathbb{Z}_2} \in \big\{ \,\{e\}\;,\; SO(2)\;, \;\mathbb{R}\;, \;\mathbb{R}^2\; , \; ISO(2) \,\big\}$.
As we shall work out explicitly, the quotient
\begin{equation}
    F \simeq \frac{\ISO}{\ell_{\mathcal{K}}}
\end{equation}
encodes the discrepancy between the representation considered and more usual massless hard representations.\\

\subsection{Overview}\label{Ssection: Overview}

In this subsection we give a general overview of how to realise wavefunctions, inner products and action of the BMS group. There are two steps: first, describe the orbits of the supermomenta; second, realise wavefunctions as functions on the orbits. In the following two subsections, this program will be explicitly realised by means of an explicit parametrisation of $SL(2,\mathbb{C})$.

\subsubsection{Structure of the orbit}

The orbit $\mathcal{O}_{\K}$ of a supermomentum of the form \eqref{BMS massless rep: Reference supermomenta1} is isomorphic to
\begin{equation}
    \mathcal{O}_{\K} \simeq \frac{SL(2,\mathbb{C})}{\ell_{\K}},
\end{equation}
where the BMS little group $\ell_{\K}$ satisfies $\ell_{\K} \simeq 
\big(U(1)\ltimes \mathbb{C}\big)\cap \ell_{\Sigma}$ by Proposition \ref{Proposition: BMS, Poincaré and Soft little groups} and belongs to the list \eqref{massless reps: admissible little groups}. By the definition of a massless representation, the orbit $\mathcal{O}_{k}$ of the null momentum $k_{\mu} = \pi_{\mu}\left(\K\right)$ is isomorphic to
\begin{equation}
    \mathcal{O}_{k} \simeq \frac{SL(2,\mathbb{C})}{U(1)\ltimes \mathbb{C}}\simeq \frac{SO_0(3,1)}{ISO(2)} \simeq S^2 \times \mathbb{R}^+.
\end{equation}
We thus have:
\begin{proposition}
Let $\P$ be the supermomentum of a $BMS_4$ massless representation with associated null momentum $p_{\mu}=\pi_{\mu}(\P)$. The projection from the orbit $\mathcal{O}_{\P}$ of the supermomentum onto the orbit $\mathcal{O}_{\pi(\P)}$ of its associated momentum
\begin{align}\label{projectionorbits}
    \begin{array}{ccc}
        \pi:\mathcal{O}_{\P} & \twoheadrightarrow &  \mathcal{O}_{\pi(\P)} \simeq S^2 \times \mathbb{R}^+
    \end{array}
\end{align}
is a fibre bundle of fibre $F \simeq \frac{\widetilde{ISO(2)}}{\ell_{\mathcal{P}}}$ isomorphic to $\frac{SL(2,\mathbb{C})}{\ell_{\mathcal{P}}} \twoheadrightarrow \frac{SL(2,\mathbb{C})}{\widetilde{ISO(2)}}\simeq \frac{SO_0(3,1)}{ISO(2)}$.
\end{proposition}
This bundle will in general be non-trivial. In order to see this, just note that, since the Poincaré orbit $\mathcal{O}_{\pi(\P)}$ is an homogeneous space isomorphic to the quotient $SL(2,\mathbb{C})/\widetilde{ISO(2)}$, it naturally inherits an $\widetilde{ISO(2)}$-principal bundle structure $SL(2,\mathbb{C}) \to \mathcal{O}_{\pi(\P)}$. The BMS orbit $\mathcal{O}_{\P}$ then naturally is an associated bundle $\mathcal{O}_{\P} = SL(2,\mathbb{C}) \times_{\rho} \frac{\widetilde{ISO(2)}}{\ell_{\mathcal{P}}}$ where $\rho$ is the natural action of $\widetilde{ISO(2)}$ on $\frac{\widetilde{ISO(2)}}{\ell_{\mathcal{P}}}$. Since the underlying principal bundle is non-trivial these associated bundles will in general be non-trivial as well.\\

To describe in more details the fibre bundle structure of the orbit of \eqref{BMS massless rep: Reference supermomenta1}, let us pick any supermomentum in this orbit:
\begin{equation}\label{Massless rep: fibres 1}
\P (z,\bar z)=\omega\,\delta^{(2)}(z-\zeta,\bar z- \bar{\zeta}) +  \partial_z^2 \partial_{\bar z}^2 \mathcal{M}(z,\bar z)\,\in\mathcal{O}_{\K}\,.
\end{equation}
The projection \eqref{projectionorbits} on the corresponding momentum is then given by
\begin{equation}\label{Massless rep: fibres 2}
    \P(z,\bar z)=\omega\delta^{(2)}(z-\zeta,\bar z- \bar{\zeta}) +  \partial_z^2 \partial_{\bar z}^2 {\mathcal{M}}(z,\bar z) \in \mathcal{O}_{\K} \qquad\mapsto\qquad p_{\mu}=\omega q_{\mu}(\zeta, \bar{\zeta}) \in \mathcal{O}_{k},
\end{equation}
while the fibre $F_p \simeq \frac{\ell_{p}}{\ell_{\mathcal{P}}}$ over $p_{\mu}= \omega q_{\mu}(\zeta, \bar{\zeta})$ contains all supermomenta of the form
\begin{align}\label{Massless rep: fibres 3}
     \P(z,\bar z)=\omega\delta^{(2)}(z-\zeta,\bar z- \bar{\zeta}) +  \partial_z^2 \partial_{\bar z}^2\Big((W\cdot\mathcal{M})(z,\bar z) \Big) &\in F_p, & [W] &\in \frac{\ell_{p}}{\ell_{\mathcal{P}}}.
\end{align}
In other words, all supermomenta of the form \eqref{Massless rep: fibres 3} lie in the same fibre $F_p$ as \eqref{Massless rep: fibres 1}.

\subsubsection{Wavefunctions and inner product}

Following Theorem \ref{McCarthytheo}, let us now consider a BMS wavefunction taking values in the vector space $V$ carrying the representation $\rho : \ell_{\K} \to U(V)$ of the BMS little group of \eqref{BMS massless rep: Reference supermomenta1}. More precisely, it is a section of the associated bundle $\mathcal{O}_{\K} \times_{\rho} V$ over the orbit, as we recalled in Section \ref{ssection: McCarthy Classification}. Let $\Phi[\P]$ be such field. It can equivalently be rewritten as a field $\Phi[M]$ on $SL(2,\mathbb{C})$,
\begin{equation}\label{Massless reps: wavefunction of M}
\Phi[M]:= \Big(M^{-1}\cdot\Phi\Big)\left[\K\right]\,,
\end{equation}
with equivariance 
\begin{equation}\label{equivariance}
    \Phi[M B] = \rho(B^{-1})\cdot\Phi[M]
\end{equation} for any element $B\in\ell_{\K}$ of the BMS little group. In a neighbourhood of the identity in $SL(2,\mathbb{C})$, one can write the matrix elements in a factorised form\footnote{The notation is meant to parallel the notation $L(p)W$ of Weinberg's classical textbook \cite{Weinberg:1995mt}, where $W$ stands for Wigner little group; accordingly here $B$ here stands for an element of BMS
little group.}
\begin{equation}
    M= L(\P)\,B\label{MLPB}
\end{equation}
where $B\in \ell_{\K}$ and $L(\P): \mathcal{O}_{\K} \to SL(2,\mathbb{C})$ is a map such that 
\begin{equation}\label{mapKtoP}
 L(\P)\cdot \K = \P.   
\end{equation}
Then, denoting the BMS wavefunction by $\psi(\P) :=\Phi[L(\P)]$, one has the relation
\begin{equation}\label{Massless reps: wavefunction gauged fix}
    \Phi\big[L(\P)\,B\big] 
    = \rho(B^{-1}) \cdot \psi(\P).
\end{equation}
In this expression, $B$ parametrises all the possible BMS little group elements acting on the wavefunction $\psi(\P)$.

 One can further factorise elements of $SL(2,\mathbb{C})$ in order to obtain coordinates adapted to the fibre bundle structure \eqref{Massless rep: fibres 2} of the obit: 
\begin{equation}
    L(\P)= L(p)\,W(\gamma)
\end{equation}
where $L(p): \mathcal{O}_{k} \to SL(2,\mathbb{C})$ is a map such that $L(p)\cdot k_{\mu} = p_{\mu}$ and $W(\gamma): \frac{\ell_{k}}{\ell_{\K}} \to \ell_{k} \simeq \widetilde{ISO(2)}$ parametrises the fibre
\begin{align}\label{Massless rep: fibres 4}
     \delta^{(2)}(z,\bar z) +  \partial_z^2 \partial_{\bar z}^2\Big((W\cdot \N)(z,\bar z) \Big) &\in F_k, & [W] &\in \frac{\ell_{k}}{\ell_{\K}}.
\end{align}
This allows to write
\begin{equation}
\psi(p,\gamma) :=  \Phi\Big(L(p)\,W(\gamma)\Big).
\end{equation}
where $p_{\mu} = \omega q_{\mu}(\zeta,\zetab)$ is a null momentum parametrising the base manifold \eqref{Massless rep: fibres 2}, while $\gamma$ collectively denotes local coordinates along the fibre \eqref{Massless rep: fibres 3}.

Since the representation $\rho$ of the group $\ell_\K$ on the space $V$ is unitary, there exists a positive-definite  inner product
 $(\,.\,,.)$ on $V$. The corresponding inner product on the states can then be written as
\begin{equation}
    \langle \Phi_1 | \Phi_2 \rangle = \int\limits_{\mathbb{R}^+ } 
    \omega d\omega \int\limits_{S^2} \frac{i}{2}d \zeta \wedge  d\bar{\zeta} \;\int\limits_{F_k} 
    d\gamma
    \;  (\psi_1,\psi_2)
\end{equation}
where $d\gamma
$ is a natural invariant volume form on $F_{k} \simeq {\widetilde{ISO(2)}}/{\ell_{\K}}$, which is constructed case by case by McCarthy \cite{McCarthy_75}. In practice, as detailed below, the volume form is rather obvious.

\subsubsection{Examples}

It might be useful to have in mind some examples in order to keep our feet on the ground.\\

\noindent $\bullet$ First, for hard representations, i.e. $\N=0$, and assuming a trivial representation of the litle group $\widetilde{ISO(2)}$, wavefunctions are functions on $\mathcal{O}_{k} \simeq SL(2,\mathbb{C}) / \ISO = \mathbb{R}^+ \times S^2$, i.e. the wavefunctions are of the form $\psi(\omega,\zeta,\bar\zeta)$.\\

\noindent $\bullet$  An interesting non-hard massless representation is given by orbits of the supermomentum of little group\footnote{These are examples of supermomenta in \cite[Table 2]{McCarthy_75} denoted there as $B \delta^{(2,0)} +\bar{B} \delta^{(0,2)}+ C\delta$, with stabiliser denoted $\Sigma$.} isomorphic to the additive group $\mathbb{Z}_2\times\mathbb{R}^2$:
\begin{equation}\label{BMS massless rep: Reference supermomenta2}
\hat{\K}(z,\bar z) = \delta^{(2)}(z,\bar z) +  \sigma \,\Big( \partial_z^2 + \partial_{\bz}^2  \Big) \delta^{(2)}(z,\bar{z})\,,
\end{equation}
where $\sigma\neq 0$ is a non-vanishing real number. Note that the second term on the right-hand side of \eqref{BMS massless rep: Reference supermomenta2}, i.e. the summand proportional to $\sigma$, is a soft supermomentum, since it clearly annihilates any translation $\T\in\text{Im}\,q$.
The supermomentum $\hat{\K}$ has BMS little group $\ell_{\hat{\K}}\simeq \mathbb{Z}_2\times \mathbb{R}^2$ and Poincaré little group  $\ell_{k} \simeq \ISO$.
As a result, the orbit $\mathcal{O}_{\hat{\K}} \simeq \frac{SL(2,\mathbb{C})}{\mathbb{Z}_2\times \mathbb{R}^2}$
of the supermomentum is the total space of a fibre bundle over the base $\mathcal{O}_p\simeq\mathbb{R}^+ \times S^2$ with fibre $F \simeq \frac{\ISO}{\mathbb{Z}_2\times\mathbb{R}^2} \simeq SO(2)$.  
Wavefunctions will be of the form $\psi(\omega, \zeta,\zetab, \alpha)$ with the angle $\alpha$ being a coordinate on the fibre. \\

\noindent $\bullet$ A second interesting non-hard example, now with trivial little group, is obtained by spoiling the alignment of the two delta functions, e.g.
\begin{equation}\label{BMS massless rep: Reference supermomenta3}
\tilde{\mathcal{K}}(z,\bar z) = \delta^{(2)}(z,\bar z) + \sigma \Big(  \partial_z^2 + \partial_{\bz}^2  \Big) \delta^{(2)}(z -\infty,\bar{z} - \infty) .
\end{equation}
This supermomentum has BMS little group $\ell_{\tilde{\mathcal{K}}}\simeq\mathbb{Z}_2$ and Poincaré little group  $\ell_{k}\simeq \ISO$. As a result, the fibre bundle $\mathcal{O}_{\tilde{\mathcal{K}}}\simeq \frac{SL(2,\mathbb{C})}{\mathbb{Z}_2} \to  \mathcal{O}_k\simeq\mathbb{R}^+ \times S^2$ has fibre $F \simeq \frac{\ISO}{\mathbb{Z}_2}\simeq ISO(2)$ and wavefunctions will be of the form $\psi(\omega, \zeta,\zetab,\alpha, \beta ,\betab)$ where $\left(\alpha,\beta,\betab \right)$ are coordinates on the fibre.\\

\subsubsection{Action of the BMS group on the wavefunction}

We here describe the action of the BMS group in general terms. The presentation is meant to parallel that of \cite{Weinberg:1995mt} for Poincaré particles. 

In terms of the wavefunctions $\Phi[M]$ defined by \eqref{Massless reps: wavefunction of M}, the action of the BMS group \eqref{BMS action on representation} becomes
\begin{align}\label{Masseless BMS rep: action}
    \Big(\big( \T, \Lambda \big) \cdot \Phi \Big) [M] &= e^{i\langle M \cdot \K , \T \rangle} \Phi[\Lambda^{-1}M] = e^{i\langle \P, \T \rangle} \Phi[\Lambda^{-1}L(\P)\,B]\,,
\end{align}
where we used \eqref{MLPB}-\eqref{mapKtoP} in the second equality.
Let us introduce 
\begin{equation}\label{Btilde}
    \tilde{B}(\Lambda,\P)\; := \; L^{-1}(\P) \;\Lambda\; L(\Lambda^{-1}\cdot\P)\,,
\end{equation} which crucially is an element of the BMS little group $\ell_\K\,$, i.e. $\tilde{B}(\Lambda,\P)\cdot \K = \K$ due to \eqref{mapKtoP}. Making use of the definition \eqref{Btilde} and of the equivariance \eqref{equivariance} of the wavefunction, one as
\begin{align*}
    \Big(\big( \T, \Lambda \big) \cdot \Phi \Big) [M]     &= e^{i\langle \P, \T \rangle} \, \Phi\big[L(\Lambda^{-1}\cdot\P) \tilde{B}^{-1}(\Lambda,\P) B\big]\\
    &=  e^{i\langle \P, \T \rangle} \,  \rho\Big(B^{-1} \tilde{B}(\Lambda,\P)\Big) \cdot \Phi\Big(L(\Lambda^{-1}\cdot\P)\Big)\,.
\end{align*}
Therefore, the action of the BMS group on the wavefunction $\psi(\P)$, defined by \eqref{Massless reps: wavefunction gauged fix}, is
\begin{equation}
    \Big(\big( \T, \Lambda \big) \cdot \psi \Big) (\P) = e^{i\langle \P, \T \rangle} \,\rho\Big(\tilde{B}(\Lambda,\P)\Big) \cdot \psi(\Lambda^{-1}\cdot\P).
\end{equation}

\subsection{The supermomentum orbit}

In this subsection, we realise explicitly the first part of the program that has been highlighted in Subsection \ref{Ssection: Overview}, i.e. we give an explicit description of the orbit of the supermomenta under consideration.

\subsubsection{Local coordinates on the orbit}

Despite the fact that the $\ISO$-principal bundle $SL(2,\mathbb{C})$ over  the coset space $SL(2,\mathbb{C})\big/\ISO\simeq \mathbb{R}^+\times S^2$ is non trivial, there exists local charts which covers it, up to a set of zero measure. The Gauss decomposition of elements of $SL(2,\mathbb{C})$ as products
\begin{equation}\label{BMS massless rep: SL(2,C) factorisation}
    L(\omega, \zeta,\bar\zeta) W(\alpha, \beta,\bar\beta) = \begin{pmatrix}
        \omega^{-\tfrac12} & \omega^{\tfrac12}\, \zeta \\ 0 & \omega^{\tfrac12} 
    \end{pmatrix}\begin{pmatrix}
        e^{i\tfrac{\alpha}{2}}& 0 \\[0.2em] \beta \,e^{-i\tfrac{\alpha}{2}} & e^{-i\tfrac{\alpha}{2}}
    \end{pmatrix},
\end{equation}
where $\omega>0$, $\zeta\in\mathbb{C}$, $\alpha\in\mathbb R$, $\beta\in\mathbb{C}$, will be especially convenient. The matrices $W(\alpha, \beta,\bar\beta)$ above parametrise the Poincaré little group $\ISO$ while the matrices $L(\omega, \zeta,\bar\zeta)$ parametrise the coset space $SL(2,\mathbb{C})\big/\ISO\simeq \mathbb{R}^+\times S^2$. The collection $(\omega, \zeta,\bar\zeta,\alpha, \beta,\bar\beta)$ provide local coordinates on $SL(2,\mathbb{C})$ which cover the group, up to matrices of the form
\begin{equation*}
\begin{pmatrix}
        0& \omega^{\tfrac12} \\ \omega^{-\tfrac12} & 0
    \end{pmatrix}\begin{pmatrix}
        e^{i\tfrac{\alpha}{2}}& 0 \\[0.2em] \beta \,e^{-i\tfrac{\alpha}{2}} & e^{-i\tfrac{\alpha}{2}}
    \end{pmatrix}
\end{equation*}
which form a set of measure zero. In particular, this means that one can integrate in this chart without any further global complication.

We wish to write down an explicit expression for the orbit of the reference supermomentum  \eqref{BMS massless rep: Reference supermomenta1} under the action of \eqref{BMS massless rep: SL(2,C) factorisation} i.e.
\begin{equation}
    \mathcal{P}\left[ \omega, \zeta, \bar{\zeta}  , \alpha, \beta, \bar{\beta} \right](z,\bar{z}) :=  \Big(L(\omega, \zeta,\bar\zeta) W(\alpha, \beta,\bar\beta) \cdot \K \Big)\big(z,\bar{z}\big) .
\end{equation}
We will note 
\begin{align}\label{BMS massless rep: z' expression}
    z'(z) &= \frac{\omega(z - \zeta)}{e^{i\alpha} - \beta\, \omega (z-\zeta)}, &\longleftrightarrow && z(z') &= \frac{ (e^{i\alpha}+ \omega \zeta \beta) \,z' + \omega\zeta}{\omega( 1 + \beta z')} 
\end{align}
the M\"obius transformation corresponding to the left action given by $z\mapsto\begin{pmatrix}a&b\\ c& d \end{pmatrix} \cdot \,z = \frac{az+b}{cz+d}$.
of, respectively, the matrices
 \begin{align}\label{BMS massless rep: NM expressions}
     \left(L W\right)^{-1} &= e^{-i\tfrac{\alpha}{2}}\omega^{-\tfrac12} \begin{pmatrix}
         \omega  & - \omega \zeta  \\ -\beta\omega   & \quad \beta\omega\, \zeta  + e^{i\alpha} 
     \end{pmatrix}, & LW &= e^{-i\tfrac{\alpha}{2}}\omega^{-\tfrac12} \begin{pmatrix}
         \beta\omega\, \zeta  + e^{i\alpha}   & \quad \omega \zeta  \\ \beta\omega   &\quad \omega
     \end{pmatrix}
 \end{align}
 acting on $z$ and $z'$. For later use, we also note that \eqref{BMS massless rep: z' expression} implies
\begin{align}
     \frac{\partial z'}{\partial z} & = \frac{\omega  e^{i\alpha}}{\big(e^{i\alpha}  - \beta \omega  (z-\zeta)\big)^2}, &  \frac{\partial z}{\partial z'} & = \frac{ \omega e^{i\alpha}}{\big(\omega (1 + \beta z')\big)^2}.\label{dz/dz'}
\end{align}
 Such Möbius transformations act on supermomenta $\P \in \mathcal{E}[-3]$ as
 \begin{equation}\label{transfolawcalP}
\left(LW\right) \cdot \P \;=\; \P \cdot\left(LW\right)^{-1} \;=\; \left|\frac{\partial z'}{\partial z}\right|^{3} \P(z',\bar z') ,
\end{equation} 
from which we can obtain the following proposition.

\begin{proposition}\label{Proposition: orbit of supermomenta}
The orbit of \eqref{BMS massless rep: Reference supermomenta1} under $SL(2,\mathbb{C})$, as parametrised by \eqref{BMS massless rep: SL(2,C) factorisation}, is given by elements of the form
    \begin{align}\label{BMS massless rep: generic massless supermomenta}
    \begin{alignedat}{2}
    \mathcal{P}\left[ \omega, \zeta, \bar{\zeta}  , \alpha, \beta, \bar{\beta} \right](z,\bar{z}) = \; &  \omega\,\delta^{(2)}(z-\zeta,\bar z-\bar\zeta) + \left|\frac{\partial z'}{\partial z}\right|^{3} \partial_{z'}^2 \partial_{\bar z'}^2 \N(z',\bar z')  \\[0.3em]
    =\; &  \omega\,\delta^{(2)}(z-\zeta,\bar z-\bar\zeta) +  \partial_z^2 \partial_{\bar z}^2\left(\, \left|\frac{\partial z'}{\partial z}\right|^{-1} \N(z',\bar z')\right) 
    \end{alignedat}
    \end{align}
    where $z'$ is given by \eqref{BMS massless rep: z' expression}.
\end{proposition}

\proof{The first line in \eqref{BMS massless rep: generic massless supermomenta} is merely the consequence of the transformation law \eqref{transfolawcalP}. The expression for the hard part of the supermomenta of \eqref{BMS massless rep: generic massless supermomenta}  can be obtained as follows.
As above, we note $z \mapsto z' = \frac{az+ b}{cz +d}$ the Möbius transformation corresponding to $(NM)^{-1}$ then
\begin{align*}
    \left|\frac{\partial z'}{\partial z}\right|^{3}\delta^{(2)}(z',\bar z')
    &= \left|\frac{\partial z'}{\partial z}\right|\delta^{(2)}(z -z_0,\bar z-\bar z_0)
\end{align*}
where $z_0$ is the solution of $z'(z_0)=0$, i.e. $z_0=-b/a$. Therefore,
\begin{align*}
    \left|\frac{\partial z'}{\partial z}\right|^{3}\delta^{(2)}(z',\bar z')
    &= |cz+d|^{-2}\delta^{(2)}\left(z +\tfrac{b}{a},\bar z +\tfrac{\bar b}{\bar a}\right)= |a|^{2}\delta^{(2)}\left(z +\tfrac{b}{a},\bar z +\tfrac{\bar b}{\bar a}\right)
\end{align*}
since \eqref{JacobianSL2C} and $ad-bc=1$.
Making use of \eqref{BMS massless rep: NM expressions} one obtains
\begin{align*}
    \left|\frac{\partial z'}{\partial z}\right|^{3}\delta^{(2)}(z',\bar z')\,=\, \omega\, \delta^{(2)}(z-\zeta,\bar z-\bar\zeta).
\end{align*}
Finally, the expression for the soft part of the supermomenta in the second line of \eqref{BMS massless rep: generic massless supermomenta}  is obtained by making use of the equivariance of the Paneitz operator (cf. Proposition \ref{Paneitz operator Lemma}). 
}

\subsubsection{Examples}  

\noindent $\bullet$ The orbit of $\delta(z,\bz)$ under $SL(2,\mathbb{C})$, as parametrised by \eqref{BMS massless rep: SL(2,C) factorisation}, is $\omega \delta^{(2)}(z-\zeta, \bz -\zetab)$.\\

\noindent $\bullet$ The orbit of \eqref{BMS massless rep: Reference supermomenta2} under $SL(2,\mathbb{C})$, as parametrised by \eqref{BMS massless rep: SL(2,C) factorisation}, is given by elements of the form
    \begin{align}\label{BMS massless rep: Reference supermomenta2 orbit}
    \hat{P}\left[ \omega, \zeta, \bar{\zeta}  , \alpha \right](z,\bar{z}) =\; & \omega\,\delta^{(2)}(z-\zeta,\bar z-\bar\zeta) +  \sigma \left (\frac{e^{2i\alpha}}{\omega}\,\partial_{z}^2 \delta^{(2)}(z-\zeta,\bar z-\bar\zeta) + c.c. \right)\,.
    \end{align}
In particular, from the disappearance of the complex coordinate $\beta$, one recovers that this orbit has little group isomorphic to the additive group $\mathbb{C}\simeq\mathbb{R}^2$.\\

\noindent $\bullet$ The orbit of \eqref{BMS massless rep: Reference supermomenta3} under $SL(2,\mathbb{C})$, as parametrised by \eqref{BMS massless rep: SL(2,C) factorisation}, is given by elements of the form
    \begin{align}\label{BMS massless rep: Reference supermomenta3 orbit}
&    \tilde{\P}\left[ \omega, \zeta, \bar{\zeta}  , \alpha ,\beta , \bar{\beta}\right](z,\bar{z}) \\ &= \omega\delta^{(2)}(z-\zeta,\bar z-\bar\zeta)  
    +\sigma\left(\frac{ e^{2i\alpha}}{\omega}  \;\left(\frac{\bar \beta}{\beta}\right)^2\; \partial_{z}^2 \delta^{(2)}\left(z- \zeta - \frac{e^{i\alpha}}{\omega \beta},\bar z- \bar\zeta - \frac{e^{-i\alpha}}{\omega \bar\beta}\right) + c.c. \right).\nonumber
    \end{align}
All $SL(2,\mathbb{C})$ coordinates appear on the right-hand side, from which one recovers that the little group is trivial. This orbit is formally related to the previous one by taking the limit $\beta \to \infty$. However, they are in fact distinct; note in particular that their stabilisers are different. \\

\subsection{Wavefunctions}\label{ssection: Wavefunctions}

In this subsection, we realise explicitly the second part of the program that has been highlighted in Subsection \ref{Ssection: Overview}, i.e. we give give an explicit description of the BMS wavefunctions.

\subsubsection{Wavefunctions and inner product}

Depending on each particular possible BMS little group $\ell_{\K}$, the explicit form of the wavefunction and inner product will vary.

\paragraph{Generic orbits and Bosonic versus Fermionic representations:} We will first treat the most common case where the BMS little group is effectively trivial
\begin{equation}\label{Z2little}
    \ell_{\K} =\{I,-I\}\simeq\mathbb{Z}_2\,.
\end{equation}
Other cases will be discussed below. There are two representations of $\mathbb{Z}_2$, parametrised by the quantity $F\in\{0,1\}$ : they respectively correspond to bosonic ($F=0$) or fermionic ($F=1$) BMS particles. Therefore $V\simeq\mathbb{C}$ and the fields carrying the corresponding UIR of $BMS_4$ will be given by complex functions on  
$SL(2,\mathbb{C})\simeq Spin(3,1)$ parametrised as \eqref{BMS massless rep: SL(2,C) factorisation}:
\begin{align}\label{BMS massless rep: fields}
 \Phi(\omega,\zeta,\bar\zeta,\alpha, \beta,\bar\beta)
\end{align}
and satisfying the parity condition 
\begin{equation}\label{paritymassless}
    \Phi(\omega,\zeta,\bar\zeta,\alpha+2\pi, \beta,\bar\beta)=(-1)^F\Phi(\omega,\zeta,\bar\zeta,\alpha, \beta,\bar\beta)\,.
\end{equation} 
which realises the needed $\mathbb{Z}_2$-equivariance.

Since $\mathbb{Z}_2$ is always part of the little group this distinction between bosonic and fermionic representations will be common to all BMS representations, for massless representations it is always related to the condition \eqref{paritymassless} and, in order to lighten the presentation, we will most of the time leave it implicit in what follows.

 The inner product is given by
\begin{equation}\label{BMS massless rep: inner product}
    \langle \Phi_1 | \Phi_2 \rangle\; =\; \int\limits_{\mathbb{R}^+}  \omega\, d\omega\int\limits_{S^2}  \frac{i}{2}  d\zeta \wedge d\bar{\zeta} \;\int\limits_{\ISO} dW\;  \bar\Phi_1\Phi_2
\end{equation}
where $dW = \frac{i}{2}\,d \alpha\wedge d\beta\wedge d\bar{\beta}$ is the Haar measure on $\ISO$. The overall integral above is taken over $SL(2,\mathbb{C})$ and the corresponding volume form is the Haar measure on $SL(2,\mathbb{C})$. 
\begin{proposition}
    Let $(\omega, \zeta,\bar{\zeta}, \alpha , \beta, \bar{\beta})$ be local coordinates on $SL(2,\mathbb{C})$ given by \eqref{BMS massless rep: SL(2,C) factorisation}. Then the Haar mesure is given by
    \begin{equation}
        d\mu \,=\, \tfrac12\omega \,d\omega\, d\zeta_1\, d\zeta_2\, d \alpha\, d\beta_1\, d\beta_2\,,
    \end{equation}
where $\zeta=\zeta_1+i\,\zeta_2$ and $\beta=\beta_1+i\,\beta_2$.
\end{proposition}
\proof{
This is the volume form of the Killing metric on $SL(2,\mathbb{C})$ computed in the coordinate system \eqref{BMS massless rep: SL(2,C) factorisation}. See appendix \ref{Appendix: Haar mesure} for details.
}

\paragraph{Orbits with non trivial little groups:}
If the little group $\ell_{\K}$ is effectively non-trivial and the spin of the representation is $\rho : \ell_{\K} \to U(V)$, then wavefunctions take values in $V$ and must be of the form \eqref{BMS massless rep: fields}, but now with a special dependence in the  coordinates of $\ISO$ in order to realise the equivariance
\begin{equation}\label{Massless reps: equivariance}
    \Phi[M B] = \rho(B^{-1})\cdot \Phi[M]
\end{equation}
 for any element $B\in\ell_{\K}$. To realise this explicitly, one should recall that the coordinates $(\omega,\zeta,\bar\zeta,\alpha, \beta,\bar\beta)$ parametrise elements of $SL(2,\mathbb{C})$ via $M=L(\omega, \zeta,\bar\zeta) W(\alpha, \beta,\bar\beta)$, see \eqref{BMS massless rep: SL(2,C) factorisation}. Thus if $B= W(\theta, b,\bar{b})$, equivariance \eqref{Massless reps: equivariance} of the wavefunction reads
 \begin{equation}
      \Phi(\omega,\zeta,\bar\zeta,\alpha', \beta',\bar\beta') = \rho\big(B^{-1}\big)\cdot \Phi(\omega,\zeta,\bar\zeta,\alpha, \beta,\bar\beta) 
 \end{equation}
with
\begin{equation}\label{Massless reps: equivariance, transfo ISO(2)}
    \left(\alpha', \beta',\bar{\beta}'\right) =\left(\alpha + \theta, \beta e^{i\theta} +b, \bar{\beta} e^{-i \theta} +\bar{b}\right).
\end{equation}
The form of the wavefunction is then fixed by this expression and the particular representation $\rho$ chosen. The resulting wavefunctions, for every possible admissible little groups \eqref{massless reps: admissible little groups} are listed below. The choices of representation $\rho$ of those little groups will perhaps seem \textit{ad hoc} below, but their justification should become more clear in Subsection \ref{Ssection: branching}. \\

\noindent$\bullet$ $\ell_{\K} = U(1)$ : The subgroup $U(1) \subset \ISO$ identifies with the subset of matrices
\begin{equation}\label{Massless reps: U(1) matrices}
B(\theta)=W\left(\theta,0,0\right)=
\begin{pmatrix}
e^{+i\,\tfrac{\theta}{2}}& 0 \\[0.2em] 0 & e^{-i\,\tfrac{\theta}{2}}
\end{pmatrix}\,  .
\end{equation}
The UIRs of this Abelian group are labeled by a number $\lambda=\frac{n}2\in\frac12\mathbb{Z}$ and, by equivariance, wavefunctions must be of the form
\begin{equation}\label{Wavefunction: U(1) equivariance}
    \Phi(\omega,\zeta,\bar\zeta,\alpha, \beta,\bar\beta) \,=\, e^{i\lambda\alpha}\;\psi(\omega, \zeta,\zetab,\beta, \betab)\,,
\end{equation}
so $\lambda$ corresponds to the helicity.
The inner product is obtained from \eqref{BMS massless rep: inner product} by replacing the last integral by $\int\limits_{\mathbb{R}^2} \frac{i}{2}d\beta \wedge d\bar{\beta}\; \bar{\psi}_1 \psi_2$.\\

\noindent$\bullet$ $\ell_{\K} = \mathbb{Z}_2\times\mathbb{R}$ : The subgroup $\mathbb{R}\subset \ISO$  identifies with the subset of matrices
\begin{equation}\label{Massless reps: R matrices}
B(b) = W\left(0,b,\bar b\right)=
\begin{pmatrix}
\,1\,& \,0\,\, \\[0.2em] \,b\, & \,1\,\,
\end{pmatrix}\, \qquad b\in \mathbb{R}.
\end{equation} The UIRs of this Abelian group are labeled by a real number $\stackrel{\circ}{\pi}_1\,\in\mathbb{R}$. By equivariance, the wavefunction take the form 
\begin{align}\label{Wavefunction: R equivariance}
     \Phi(\omega,\zeta,\bar\zeta,\alpha, \beta,\bar\beta) 
     &= \int\limits_{\mathbb{R}} d\pi_2 \; e^{i\,  R_{\alpha}(\vec{\pi})\,\cdot\,\vec{\beta}} \;\psi(\omega,\zeta,\bar\zeta,\alpha, \pi_2  )
\end{align}
where $\vec{\pi} = (\stackrel{\circ}{\pi}_1, \pi_2)$ and $\vec\beta=(\beta_1,\beta_2)$ with $\beta=\beta_1+i\beta_2$, while $R_{\theta}$ denotes a rotation of angle $\theta$. The inner product is obtained from \eqref{BMS massless rep: inner product} by replacing the last integral by $\int\limits_{\mathbb{R}} d\pi_2 \, \int\limits_{S^1} d\alpha \; \bar{\psi}_1 \psi_2$.\\

\noindent$\bullet$ $\ell_{\K} = \mathbb{Z}_2\times\mathbb{R}^2$ : The subgroup $\mathbb{R}^2\subset \ISO$  identifies with the subset of matrices
\begin{equation}\label{Massless reps: R^2 matrices}
B(b,\bar{b}) = W\left(0,b,\bar{b}\right)=
\begin{pmatrix}
1& 0 \\[0.2em] b & 1
\end{pmatrix}\, \qquad b\in \mathbb{C}.
\end{equation}
The UIRs of this Abelian group are labeled by a plane vector $\vec\pi_0$. By equivariance, wavefunctions must be of the form
\begin{equation}\label{Wavefunction: R2 equivariance}
    \Phi(\omega,\zeta,\bar\zeta,\alpha, \beta,\bar\beta) 
     = e^{i\, R_{\alpha}(\vec{\pi}_0)\,\cdot\, \vec{\beta}} \;\psi(\omega,\zeta,\bar\zeta,\alpha)\,.
\end{equation}
The inner product is obtained from \eqref{BMS massless rep: inner product} by replacing the last integral by $\int\limits_{S^1} d\alpha\; \bar{\psi}_1 \psi_2$.\\

\noindent$\bullet$ $\ell_{\K} = \ISO$ : This corresponds to hard massless BMS representations. Usual massless fields with helicity $\lambda$ must be of the form 
\begin{equation}\label{Wavefunction: ISO(2) equivariance 1}
    \Phi(\omega,\zeta,\bar\zeta,\alpha, \beta,\bar\beta) = e^{i\lambda\alpha}\; \psi(\omega, \zeta,\zetab)
\end{equation}
in particular, the  coordinates $(\beta, \betab)$ drop out. These coordinates will however appear for continuous-spin representations.\footnote{Since continuous-spin particles, despite being part of Wigner's classification, are not so familiar, they are reviewed, for the benefit of the reader, in Appendix \ref{Appendix: CSP}, see e.g. \cite{Bekaert:2006py,Bekaert:2017khg} and refs therein for more details.} The wavefunctions for the corresponding representations can be realised in various equivalent ways, which will both be natural from the point of view of generic BMS particles. Firstly, one realisation is 
\begin{equation}\label{Wavefunction: ISO(2) equivariance 2}
    \Phi(\omega,\zeta,\bar\zeta,\alpha, \beta,\bar\beta)
     = e^{i\, R_{\alpha}(\vec{\pi}_0)\,\cdot\, \vec{\beta}} \;\psi(\omega,\zeta,\bar\zeta,\alpha)\,.
\end{equation}
where $\vec\pi_0$ is a fixed plane vector of radius $\mu=|\vec{\pi}_0| > 0$. Secondly, other equivalent representations for continuous spin are
\begin{equation}\label{Wavefunction: ISO(2) equivariance 3}
    \Phi(\omega,\zeta,\bar\zeta,\alpha, \beta,\bar\beta)
     = e^{i\lambda \alpha}\int_{S^1} d\theta\, e^{i\,R_{\theta}(\vec{\pi}_0) \cdot \vec{\beta}} \;\psi(\omega,\zeta,\bar\zeta,\theta)\,.
\end{equation}
for any fixed $\lambda \in \tfrac12\mathbb{Z}$.
The inner product is the usual inner product obtained from \eqref{BMS massless rep: inner product} by dropping out the last integral (for usual massless fields) or by replacing it with an integral over the circle (for continuous-spin fields).\\

\subsubsection{Action of supertranslations}

Irrespectively of the BMS little group, one has the following.

\begin{proposition}\label{Action of supertranslations_prop}
A supertranslation $\T(z,\bar z)$ acts on the BMS wavefunction $\Phi[M]$ as
\begin{align}\label{BMS massless rep: supertranslation shift}
     \Phi(\omega,\zeta,\bar\zeta,\alpha, \beta,\bar\beta) \;&\mapsto\;e^{i \omega \T(\zeta,\bar \zeta) +i \varphi(\omega,\zeta,\bar\zeta,\alpha, \beta,\bar\beta)}\; \Phi(\omega,\zeta,\bar\zeta,\alpha, \beta,\bar\beta)\,.
\end{align}
with the phase $\varphi$ given by 
    \begin{align}\label{BMS massless rep: supertranslation shift phaseB}
    \varphi(\omega,\zeta,\bar\zeta,\alpha, \beta,\bar\beta)&= \frac{e^{2i\alpha}}{\omega} \int dz'^2 \;\left(\frac{1+ \bar \beta \bar z'}{1+\beta z'}\right)^2\; \partial_{z}^2 \T(z,\bar z)  \; \partial_{\bar z'}^2 \N(z',\bar z')
\end{align}
where the explicit expression of the functions $z'(z)$ were given in \eqref{BMS massless rep: z' expression}. 
\end{proposition}
Other, alternative and useful, expressions for the phase are
    \begin{align}
    \varphi(\omega,\zeta,\bar\zeta,\alpha, \beta,\bar\beta)&= \omega \int dz'^2 \; \T(z,\bar z)\; \left| 1 + \beta z'\right|^{2} \partial_{z'}^2 \partial_{\bar z'}^2 \N(z',\bar z'),\\
 &=  \omega^{-3}\int dz'^2 \;\partial_{z}^2 \partial_{\bar z}^2 \T(z,\bar z)\; \left|1 + \beta z'\right|^{-6} \N(z',\bar z') , \nonumber
\end{align}
Here again, the expression of the functions $z'(z)$ and $z(z')$ are given by \eqref{BMS massless rep: z' expression}. 

\proof{See Appendix \ref{Appendix: transformedsupermomentum}.}

\subsubsection{Examples}

\noindent $\bullet$ Hard representations have wavefunctions of the form \eqref{Wavefunction: ISO(2) equivariance 1}, \eqref{Wavefunction: ISO(2) equivariance 2} or \eqref{Wavefunction: ISO(2) equivariance 3}, and transform under the action of a supertranslation $\T(z\bz) \in \E[1]$, as 
\begin{equation}
    \psi(\omega,\zeta,\bar\zeta)\mapsto   e^{i \omega \T(\zeta,\bar \zeta)}\;\psi(\omega,\zeta,\bar\zeta).
\end{equation}

\noindent $\bullet$ The wavefunctions on the orbit \eqref{BMS massless rep: Reference supermomenta2 orbit} must be complex functions of the form \eqref{Wavefunction: R2 equivariance} with action of supertranslations given by
\begin{align}
\psi(\omega,\zeta,\bar\zeta,\alpha)\mapsto  e^{i \omega \T(\zeta,\bar \zeta) + i\sigma \left( \frac{e^{2i\alpha}}{\omega}\partial_{z}^2 \T(\zeta,\zeta) + c.c. \right) }\;\psi(\omega,\zeta,\bar\zeta,\alpha)\,.
\end{align}

\noindent $\bullet$  Wavefunctions on the orbit \eqref{BMS massless rep: Reference supermomenta3 orbit} must be of the form $\Phi(\omega,\zeta,\bar\zeta,\alpha,\beta, \bar{\beta})$ with action of supertranslations given by 
\begin{align}\label{Massless reps, example l=1. Supertranslations}
\Phi(\omega,\zeta,\bar\zeta,\alpha,\beta,\bar{\beta})\mapsto e^{i \omega \T(\zeta,\bar \zeta) + i\sigma \left( \frac{e^{2i\alpha}}{\omega}  \;\left(\frac{\bar \beta}{\beta}\right)^2\; \partial_{z}^2 \T\left(\zeta + \frac{e^{i\alpha}}{\omega \beta}, \bar \zeta + \frac{e^{-i\bar\alpha}}{\omega \bar\beta}\right) + c.c.  \right) }\;  \Phi(\omega,\zeta,\bar\zeta,\alpha,\beta,\bar{\beta})\,.
\end{align}\\

\subsection{Branching to a Poincaré subgroup}\label{Ssection: branching}

We will now choose a Poincar\'e subgroup $ISO_0(3,1) \subset BMS_4$ and decompose a massless representation of BMS group into irreducible representations of this particular Poincaré group. We will first discuss the generic case where the little group is effectively trivial. All other cases will be worked out in a second time.

\subsubsection{Decomposition of BMS particles in sum of Poincaré particles}

As mentioned at the end of Theorem \ref{branchingrulestheo}, a BMS particle decomposes generically into an infinite collection of Poincar\'e particles. In the massless case, this collection will in general involve an infinite tower of so-called continuous-spin particles.\footnote{\label{CMtheorem}Incidentally, note that a (somewhat implicit) assumption of the theorem of Coleman and Mandula is the absence of infinite-component UIRs of the Poincar\'e group (such as the continuous-spin representations) in the spectrum (this is implicit in the assumption that the matrices $b_\alpha(p)$ are \textit{finite} Hermitian matrices in \cite[p.14]{Weinberg:2000cr}). This can be another way out of the Coleman-Mandula theorem that would naively preclude the existence of a non-trivial BMS-invariant $S$-matrix between massless particles (see also Footnote \ref{ColemanManduala}).}

Let $\Phi(\omega,\zeta,\bar\zeta,\alpha, \beta,\bar\beta)$ be a state vector in a generic UIR of the BMS group, i.e. for which $\ell_{\K}=\{I,-I\}\simeq\mathbb{Z}_2$, see \eqref{BMS massless rep: fields}. This wavefunction transforms as a scalar under $SL(2,\mathbb{C})$ transformations (right multiplication) and, under the action of a supertranslation $\T(z,\bar z)$, according to \eqref{BMS massless rep: supertranslation shift} where the phase $\varphi$ is given by \eqref{BMS massless rep: supertranslation shift phaseB}. Since, in order to realise concretely the BMS group, we are here writing elements of the group as pairs $(\T, M) \in \E[1] \times SL(2,\mathbb{C})$, this implies that a Poincar\'e subgroup $ISO_0(3,1) \subset BMS_4$ has been singled out.\footnote{This Poincar\'e subgroup reads $(T^{\mu}q_{\mu}(z,\bz), M) \in \E[1] \times SL(2,\mathbb{C})$.}. We would like to decompose $\Phi(\omega,\zeta,\bar\zeta,\alpha, \beta,\bar\beta)$ into irreducible representations of this particular Poincaré subgroup.\footnote{Note, however, that there is nothing special about this particular Poincaré subgroup; branching with respect to others will be discussed in Section \ref{Ssection: Branching in different gravity vacua}.} 

\paragraph{Fourier transform over the massless little group}\mbox{}

\noindent The idea is to decompose the $\ISO$ dependence (i.e. the dependence in the variables $\alpha,\beta,\bar\beta$) of $\Phi$ in continuous Fourier modes
\begin{align}
     \Phi(\omega,\zeta,\bar\zeta,\alpha, \beta,\bar\beta) 
     &=  \int\limits_{\mathbb{R}^2} d\pi_1d\pi_2 \; e^{i\, \vec{\pi}\cdot \vec{\beta}} \,\widetilde{\Phi}
     (\omega,\zeta,\bar\zeta,\alpha\,;\vec\pi)\,,\label{decompalphabeta}
\end{align}
where $\vec\pi=(\pi_1,\pi_2)$ and $\vec\beta=(\beta_1,\beta_2)$ with $\beta=\beta_1+i\beta_2$. 
We can also further decompose in discrete Fourier modes
\begin{align}     
 \widetilde{\Phi}
     (\omega,\zeta,\bar\zeta,\alpha\,;\vec\pi) = \sum\limits_{n\in\mathbb Z}\, e^{i\frac{n}{2}\alpha}\, \widetilde{\Phi}_n
     (\omega,\zeta,\bar\zeta\,;\vec\pi) \,.\label{decompalphabeta2}
\end{align}

Let us recall that the collection $(\omega,\zeta,\bar\zeta)$ are coordinates on the null cone $\mathcal{O}_k\simeq\mathbb{R}\times S^2$, thus, the mode $\widetilde{\Phi}_n(\omega,\zeta,\bar\zeta\,;\vec\pi)$ is a field defined on the mass shell $p^2=0$, indexed by a discrete label $n\in\mathbb Z$ together with a vector $\vec\pi\in\mathbb{R}^2$ in a plane (which can somewhat be thought as the transverse plane). 

Physically, the label $n$ corresponds to the helicity $\lambda= \frac{n}{2}$ of the field $\widetilde{\Phi}_n$. 
While coordinates $(\alpha, \beta,\bar\beta)$ parametrise the Poincar\'e little group $\ISO$ as in \eqref{BMS massless rep: SL(2,C) factorisation},  the subgroup $U(1)\subset\ISO$ of rotations $R_\theta$ of angle $\theta$ in the transverse plane identifies with $W(\theta,0,0)$, see \eqref{Massless reps: U(1) matrices}. 
From \eqref{Massless reps: equivariance, transfo ISO(2)} one sees that its acts by right multiplication as the shift $\alpha\to \alpha+\theta$ and the phase transformation $\beta\to e^{i\theta}\beta$. Accordingly, the field $\widetilde{\Phi}_n$ transforms as $\widetilde{\Phi}_n(\omega,\zeta,\bar\zeta\,;\vec\pi)\to e^{i\frac{n}2\theta}\widetilde{\Phi}_n
(\omega,\zeta,\bar\zeta\,;\vec\pi')$ where $\vec\pi'=R_\theta(\vec\pi)$. This proves that $\lambda =\tfrac{n}{2}$ is the helicity of the field $\widetilde{\Phi}_n$. The parity condition \eqref{paritymassless} imposes the helicities to be either all integer ($F=1$) or all half-integer ($F=-1$). 

Note that when $\vec\pi=\vec 0$ each mode $\widetilde{\Phi}_n(\omega,\zeta,\bar\zeta\,;\vec 0)$ spans an irreducible representation of $\ISO$. However, when $\vec{\pi}\neq0$, the action of the subgroup $\mathbb{R}^2\subset \ISO$ will mix helicities. 
The action of this subgroup identifies with $W(0,b,\bar{b})$ in \eqref{Massless reps: R^2 matrices}, and one obtains from \eqref{Massless reps: equivariance, transfo ISO(2)} that its acts by right multiplication as the shift $\beta\to\beta+b$. Accordingly, the field $\widetilde{\Phi}$ transforms as $\widetilde{\Phi}(\omega,\zeta,\bar\zeta,\alpha\,;\vec\pi)\to e^{i\vec\pi\cdot\vec b}\,\widetilde{\Phi}
(\omega,\zeta,\bar\zeta,\alpha\,;\vec\pi)$ with $\vec b=(b_1,b_2)$ and $b=b_1+ib_2$. Therefore, when $\vec\pi=\vec 0$ each mode $\widetilde{\Phi}_n(\omega,\zeta,\bar\zeta\,;\vec 0)$ spans an irreducible representation of $\ISO$ where $\mathbb{R}^2$ acts trivially.
The physical interpretation of the transverse vector when $\vec\pi\neq\vec 0$ in $\widetilde{\Phi}_n(\omega,\zeta,\bar\zeta\,;\vec\pi)$ is exotic because it corresponds to continuous-spin degrees of freedom. They are usually discarded but they are unavoidable in the case of massless BMS particles with generic momenta (see \cite{Bekaert:2006py,Bekaert:2017khg} for reviews and see also Appendix \ref{Appendix: CSP}).\\

\paragraph{Decomposition of BMS particles in sum of Poincaré particles}\mbox{}

\noindent When considering the decomposition \eqref{decompalphabeta} of a generic field $\Phi(\omega,\zeta,\bar\zeta,\alpha, \beta,\bar\beta)$ on $SL(2,\mathbb{C})$ in Fourier modes $\widetilde{\Phi}_n (\omega,\zeta,\bar\zeta\,;\vec\pi)$ with respect to $\ISO$, there are two particular modes which stand out:
\begin{enumerate}
    \item The mode $\widetilde{\Phi}_n(\omega,\zeta,\bar\zeta\,;\vec 0)$ labeled by the helicity $\lambda = n/2$ and with trivial vector (i.e. $\vec\pi=\vec 0$) describes a spin-$\frac{n}{2}$ massless field. 
    \item When $\vec\pi\neq\vec 0$, the mode $\widetilde{\Phi}_0(\omega,\zeta,\bar\zeta\,;\vec \pi)$ manifestly describes an infinite sum of different continuous-spin representations. More precisely, there is an infinite tower of continuous-spin representations corresponding to the infinite range of the parameter $|\vec{\pi}|=\mu\in\mathbb{R}^+$. Concretely, a given UIR of Poincar\'e is spanned by fields $\widetilde{\Phi}_0(\omega,\zeta,\bar\zeta\,;\vec \pi)$ with support on the circle of vectors $\vec\pi$ with a fixed radius $\mu$. 
\end{enumerate}
More generally  all the modes $\widetilde{\Phi}_n(\omega,\zeta,\bar\zeta\,;\vec \pi)$ with $\vec\pi\neq\vec 0$ describe an infinite tower of continuous-spin representations, one for each value of the parameter $|\vec{\pi}|=\mu\in\mathbb{R}^+$. They all are equivalent (for fixed $\mu$), irrespectively of the value for $n$ (cf. Remark \ref{helicityremark}).
This shows that 

\begin{proposition}\label{massless_e_branching}
A massless UIR of $BMS_4$ with effectively trivial little group of supermomenta decomposes, with respect to any Poincar\'e subgroup, into a direct sum of massless representations over all possible helicities (either all integers, $\lambda\in\mathbb{Z}$, or all half-integers, $\lambda\in\mathbb{Z}+\tfrac12$) with multiplicity one, plus a direct integral of continuous-spin representations over all values of the parameter $\mu\in\mathbb{R}^+$ with infinite multiplicity.
\end{proposition}

\begin{remark}
The direct integral (i.e. the continuous orthogonal sum) of Poincar\'e UIRs is not a bug, it a feature of generic BMS particles. In fact, one may expect the appearance of a direct integral (in the branching rule from BMS group to Poincar\'e subgroup) for all UIRs of $BMS_4$ for which the typical fibre \eqref{fibre} of the fibre bundle $\mathcal{O}_{\P}\twoheadrightarrow\mathcal{O}_{\pi(\P)}$ is not compact.
\end{remark}

\subsubsection{Decomposition of BMS particles with non trivial little groups}\label{sssection: DecompositioninsumofPoincaréparticles2}

Having established the decomposition of generic BMS particles in terms of Poincaré particles, we are ready to work out branching rules for non trivial little groups.

\paragraph{$\bullet$ Orbits with little group $\ell_{\K} = U(1)$}\mbox{}
\vspace{0.2cm}

The UIRs of this Abelian group $U(1)$ are labeled by the helicity $\lambda=\frac{n}2\in\frac12\mathbb{Z}$ and the corresponding wavefunctions are of the form \eqref{Wavefunction: U(1) equivariance}. This implies, in the massless case, that only the modes $\widetilde{\Phi}_n(\omega,\zeta,\bar\zeta\,;\vec \pi)$ are present in the Fourier decomposition \eqref{decompalphabeta} of $\Phi(\omega,\zeta,\bar\zeta,\alpha,\beta,\bar\beta)$. A similar discussion to the one above shows the following branching rule.

\begin{proposition}\label{massless_U(1)_branching}
    A massless UIR of $BMS_4$ with BMS little group $U(1)$ is labeled by the helicity $\lambda\in\frac12\mathbb{Z}$ and decomposes, with respect to any Poincar\'e subgroup, into the direct sum of one usual massless representation with helicity $\lambda$ plus a direct integral of continuous-spin representations over all values of the parameter $\mu\in\mathbb{R}^+$ with multiplicity one.
\end{proposition}

\proof{This proposition agrees with Theorem \ref{branchingrulestheo} since 1) the unfaithful representation of $ISO(2)$ of helicity $\lambda$ is effectively the faithful representation of $SO(2)$ of helicity $\lambda$, and 2)  all faithful UIRs of $ISO(2)$ contain the UIR of $SO(2)$ of helicity $\lambda$ with multiplicity one. Proposition \ref{massless_U(1)_branching} can also be obtained as the direct consequence of the decomposition \eqref{decompL2(R2)}.}

\begin{remark}
The direct integral of Poincar\'e UIRs in the above propositions might suggest that the Hilbert space of a massless BMS particle with such supermomenta is \textit{not} separable (see e.g. \cite{Prabhu:2022zcr,Prabhu:2024zwl} for discussions of the separability in a related context) but this would be an incorrect conclusion (cf. Remark \ref{separableISO(2)}). In fact, the Hilbert space of any UIR of $BMS_4$ constructed in Theorem \ref{McCarthytheo} via square-integrable wave functions is separable (cf. Remark \ref{sep}).
\end{remark}
\mbox{}

\paragraph{$\bullet$ Orbits with little group $\ell_{\K} = \mathbb{Z}_2\times\mathbb{R}^2$}\mbox{}
\vspace{0.2cm}

Consider now the case of massless BMS representations with effective little group isomorphic to $\mathbb{R}^2$. From \cite{McCarthy_75} it must be in the orbit of \eqref{BMS massless rep: Reference supermomenta2 orbit}. The UIRs of this Abelian group are labeled by the plane vector $\vec\pi_0$ and the corresponding wavefunctions are of the form \eqref{Wavefunction: R2 equivariance}.
Therefore, when $\vec \pi_0=\vec 0$,
only the modes $\widetilde{\Phi}_n(\omega,\zeta,\bar\zeta)$ are present in the Fourier decomposition \eqref{decompalphabeta} of $\Phi$, that is to say
\begin{equation}\label{PhiomegaR2}
    \widetilde{\Phi}
     (\omega,\zeta,\bar\zeta,\alpha\,;\vec 0)
     = \sum\limits_{n\in\mathbb Z}\, e^{i\frac{n}{2}\alpha}\; \widetilde{\Phi}_n(\omega,\zeta,\bar\zeta)\,.
\end{equation} 
Again, there is a parity condition restricting the helicities in the above sum to either all integer or all half-integer.
So, when $\vec \pi_0=\vec 0$, one can directly read off the wave function of an infinite tower of massless Poincar\'e UIRs of all possible, integer or half-integer, helicities. This is somewhat reminiscent of the Kaluza-Klein mechanism since the fibre is also a circle, although this is quite distinct since the fibre is internal and corresponds to a BMS extra spin degree of freedom here.

When $\vec\pi_0\neq \vec 0$, the internal dependence of the wavefunction \eqref{Wavefunction: R2 equivariance} should be thought as a wave function on the circle of radius $\mu = |\vec{\pi}_0|$ parametrised by $\alpha$, hence the wavefunction \eqref{Wavefunction: R2 equivariance}  describes a single continuous-spin particle labeled by $\mu$, see Remark \ref{Branching: Rm, second realisation of continuous spin} and \eqref{Wavefunction: ISO(2) equivariance 2}. This can be summarised in the following.

\begin{proposition}\label{massless_R2_branching}
    A massless UIR of $BMS_4$ with effective BMS little group $\mathbb{R}^2$ is labeled by two real numbers $(\pi_1,\pi_2)\in\mathbb{R}^2$.
    When $(\pi_1,\pi_2)=(0,0)$, the massless UIR of $BMS_4$ decomposes, with respect to any Poincar\'e subgroup, into a direct sum of usual massless representation over all possible helicities (either all integers, $\lambda\in\mathbb{Z}$, or all half-integers, $\lambda\in\mathbb{Z}+\tfrac12$) with multiplicity one. However, when $(\pi_1,\pi_2)\neq (0,0)$, the massless UIR of $BMS_4$ remains irreducible under restriction to any Poincar\'e subgroup, and identifies with a single continuous-spin representation of parameter $\mu=\sqrt{(\pi_1)^2+(\pi_2)^2\,}$.
\end{proposition}

\proof{This is in agreement with Theorem \ref{branchingrulestheo} since, for instance, only the UIR of $\ISO$ with parameter $\mu=\sqrt{(\pi_1)^2+(\pi_2)^2}$ contains the mode $e^{i\,\vec\pi\cdot\vec\beta}$ for a given $\vec\pi=(\pi_1,\pi_2)$.}
\mbox{}

\paragraph{$\bullet$ Orbits with little group $\ell_{\K} = \mathbb{Z}_2\times\mathbb{R}$}\mbox{}
\vspace{0.2cm}

To conclude, one considers the last possible case for a massless BMS representation: an effective little group isomorphic to $\mathbb{R}$.
The UIRs of this Abelian group $\mathbb R$ are labeled by a real number $\stackrel{\circ}{\pi}_1\,\in\mathbb{R}$ and the corresponding wavefunctions are of the form \eqref{Wavefunction: R equivariance}.

This is a continuous sum of $\mathbb{R}^2$-equivariant wavefunctions, i.e. continuous spin representations when $\mu = |\vec{\pi}|\neq0$ and an extra infinite sum of massless fields with usual helicity when $\mu=0$.  It is clear that, for any given $\pi_2\in\mathbb R$,
\begin{equation}
    \mu=\sqrt{\big(\stackrel{\circ}{\pi}_1\big)^2+\big(\pi_2\big)^2\,}\;\geqslant\; \big|\,\stackrel{\circ}{\pi}_1\big|\,.
\end{equation}
Therefore, Theorem \ref{branchingrulestheo} implies the following branching rule.

\begin{proposition}\label{massless_R_branching}
    A massless UIR of $BMS_4$ with effective BMS little group $\mathbb R$ is labeled by a real number $\pi\in\mathbb R$ and decomposes, with respect to any Poincar\'e subgroup, into a direct integral of continuous-spin representations over all values of the parameter $\mu\geqslant|\pi|$ (when $\pi\neq 0$) with multiplicity one. In the particlar case $\pi=0$, the direct integral is over all values $\mu>0$ and must be supplemented by a direct sum of usual massless representations over all possible helicities (either all integers, $\lambda\in\mathbb{Z}$, or all half-integers, $\lambda\in\mathbb{Z}+\tfrac12$) with multiplicity one.
\end{proposition}

\begin{remark}
    Let us  note that the direct integral of continuous-spin representations in Propositions \ref{massless_e_branching} and \ref{massless_U(1)_branching} is mentioned in \cite[Appendix]{McCarthy_73-III} but the direct sum of usual massless representations is not mentioned explicity. Furthermore, the specific cases of branching in Propositions \ref{massless_R2_branching} and \ref{massless_R_branching} seem not have been analysed at all by McCarthy in his series of papers.
\end{remark}

\subsection{Massless Poincaré particles in another gravity vacuum}\label{Ssection: Branching in different gravity vacua}

\subsubsection{Supertranslation of a Poincar\'e particle}

From the results of this section on sees that, for the vast majority of massless BMS representations, one can consider states of the form
\begin{equation}\label{Massless rep: 1-particle state}
    \Phi(\omega,\zeta,\zetab, \alpha, \beta,\betab) = e^{i\lambda\alpha} a(\omega,\zeta,\zetab)\,.
\end{equation}
Then, as we saw, under the branching $ISO_0(3,1) \subset BMS_4$ given by
\begin{equation}\label{Massless reps: Poincarré group1}
    \Big(T^{\mu}q_{\mu}(z,\bz)  , M\Big) \subset \Big(\T(z,\bz), M\Big),
\end{equation}
the state \eqref{Massless rep: 1-particle state} describes a usual massless Poincaré particle of helicity $\lambda\in \frac{1}{2}\mathbb{Z}$ and momentum $p_\mu=\omega \,q_\mu(\zeta,\zetab)$.

Now there is nothing unique about the Poincaré group \eqref{Massless reps: Poincarré group1}. Conjugation by a supertranslation $\C(z,\bz) \in \E[1]$ gives
\begin{equation}\label{Massless reps: Poincarré group2}
    \Big(\,T^{\mu}q_{\mu}(z,\bz)+ (M\cdot\C)(z,\bz) - \C(z,\bz) \,,\, M\,\Big) \subset \Big(\T(z,\bz), M\Big)\,
\end{equation}
which, when $\partial_z^2 \C \neq 0$, defines\footnote{It might here be useful to remind the reader that we take the multiplication of elements of the BMS group to be $\left( \hat{\mathcal{T}} , \hat{M} \right) \cdot \Big( \mathcal{T} , M \Big) := \left( \hat{\mathcal{T}} + \hat{M}\cdot \T , \hat{M}M\right)$ with $\hat{M}\cdot \T := \T \cdot \hat{M}^{-1}$.} another gravity vacuum, i.e. another choice of Poincaré subgroup of $BMS_4$. With respect to the Poincaré group \eqref{Massless reps: Poincarré group2}, the state corresponding to a usual massless Poincaré particles of helicity $\lambda$ are of the form
\begin{equation}\label{Massless rep: 1-particle state2}
    \Phi(\omega,\zeta,\zetab, \alpha, \beta,\betab) = e^{i\lambda\alpha} a(\omega,\zeta,\zetab)\;e^{i\omega\C(\zeta,\bar\zeta)}\; e^{i\, \langle \,\partial^2 \bar{\partial}^2 (\N')  \,,\, \C\,\rangle}\,.
\end{equation}
This is just a supertranslated version of the state \eqref{Massless rep: 1-particle state}. From Proposition \ref{Proposition: orbit of supermomenta}, this can be rewritten as
\begin{align}\label{Massless rep: 1-particle state3}
     \Phi(\omega,\zeta,\zetab,\alpha, \beta, \betab) = e^{i \lambda \alpha}a(\omega,\zeta,\zetab)\;e^{i\omega\C(\zeta,\bar\zeta)}\; e^{i \varphi(\omega,\zeta,\bar\zeta,\alpha, \beta,\bar\beta)},
\end{align}
where
\begin{align}
    \varphi(\omega,\zeta,\alpha, \beta) &= \frac{e^{2i\alpha}}{\omega} \int d^2z' \;\left(\frac{1+ \bar \beta \bar z'}{1+\beta z'}\right)^2\; \partial_{z}^2 \C(z,\bar z)  \; \partial_{\bar z'}^2 \N(z',\bar z'), & z&=\frac{ (e^{i\alpha}+ \omega \zeta \beta) \,z' + \omega\zeta}{\omega( 1 + \beta z')}.
\end{align}

The main point here is that the interpretation of a given BMS state in terms of Poincaré particles \emph{depends on the gravity vacuum}, i.e. on the Poincaré subgroup $ISO_0(3,1) \subset BMS_4$. For example, \eqref{Massless rep: 1-particle state} is a Poincaré particle in the vacua \eqref{Massless reps: Poincarré group1}, while \eqref{Massless rep: 1-particle state3}  is the (same) Poincaré particle but placed in the vacua \eqref{Massless reps: Poincarré group2}. Furthermore, it follows from the previous branching rules that the latter state \eqref{Massless rep: 1-particle state3} is, on the one hand, when seen in the vacuum \eqref{Massless reps: Poincarré group2}, the state of a usual particle and, on the other hand, when seen from the initial vacua \eqref{Massless reps: Poincarré group1}, \emph{consists in general of an infinite superposition of massless particles of all (including continuous) spins}.\footnote{\label{nogo}More precisely, massless or massive UIRs of $BMS_4$ whose Poincar\'e and BMS little groups do not coincide contain at least one of the following two exotic representations:
 a continuous-spin representation (in the massless case) and/or
 an infinite tower of UIRs with identical mass but unbounded spin.
If this unusual particle content of generic BMS multiplets is rejected, then it only leaves the hard and soft BMS particles.}

\subsubsection{Examples}

\noindent$\bullet$ As a first concrete example, let us consider bosonic BMS particles with supermomenta of the form \eqref{BMS massless rep: Reference supermomenta2 orbit}. These massless BMS particles have BMS little group $\mathbb{R}^2$ and a generic state must be of the form \eqref{Wavefunction: R2 equivariance}.
Consider the case $\vec\pi_0=\vec 0$, hence, the wavefunction is of the form 
\begin{equation}\label{expansionR2littlegroup}
    \psi(\omega,\zeta,\bar\zeta,\alpha) 
     =  \sum\limits_{s\in\mathbb Z}\, e^{is\alpha} \,
     a_s(\omega,\zeta,\bar\zeta)\,.
\end{equation}
For such BMS particles, a generic state is a superposition of massless Poincar\'e particles with all integer helicities $s\in \mathbb Z$ (cf. Proposition \ref{massless_R2_branching}), each of them characterised by a wavefunction $a_s(\omega, \zeta, \bar{\zeta})$.

Let us consider the very simple state :
\begin{equation}
    \psi(\omega,\zeta,\bar\zeta,\alpha) 
     =  a(\omega,\zeta,\bar\zeta)\,.
\end{equation}
comparing with \eqref{expansionR2littlegroup} one sees that, in the gravity vacuum \eqref{Massless reps: Poincarré group1} this is just a usual scalar field. Now, the corresponding supertranslated state
\begin{align}
    \psi'(\omega,\zeta,\bar{\zeta},\alpha) &= a(\omega, \zeta, \bar{\zeta})\,e^{i \langle \mathcal{K}, \C \rangle}\\
    &= a(\omega, \zeta, \bar{\zeta})\, e^{i\Big(\omega \C(\zeta, \bar{\zeta}) +\frac{\sigma}{\omega}\big(e^{i 2\alpha} \partial_\zeta^2 \C(\zeta, \bar{\zeta}) + e^{-i2\alpha} \partial_{\bar\zeta}^2 \C(\zeta, \bar{\zeta})\big)\Big)}.    
\end{align}
is the same scalar field but now placed in a different gravity vacuum \eqref{Massless reps: Poincarré group2}. To read off its meaning in terms if the initial vacuum one only needs to  expand in powers of $e^{i\tfrac{\alpha}{2}}$ and compare this Fourier series with \eqref{expansionR2littlegroup}. This yields an infinite quantum superposition of massless particles of all integer spins. We will further study this phenomenon in Section \ref{Section: memoryeffect}.\\

\noindent$\bullet$ As a second example, let us consider bosonic BMS particles with supermomenta in the orbit of \eqref{BMS massless rep: Reference supermomenta3 orbit}. These massless BMS particles have trivial effective BMS little group and a generic state must be of the form \eqref{BMS massless rep: fields} and decompose into Poincaré particles via \eqref{decompalphabeta}. In particular one can consider the simple state 
\begin{equation}
    \Phi(\omega,\zeta,\zetab, \alpha, \beta,\betab) = e^{i\lambda\alpha} a(\omega,\zeta,\zetab)\,,
\end{equation}
describing a usual massless particle of helicity $\lambda$. This state is not normalisable as a BMS particle, because of the integral over $\vec{\beta} \in \mathbb{R}^2$ in \eqref{BMS massless rep: inner product}, it can however be regularized by considering states of the form
\begin{equation}\label{Massless reps, example: Poincarré particle regularization}
    \Psi(\omega ,\zeta, \zetab,\alpha,\beta,\betab)= e^{i\lambda\alpha} \int d^2\vec{\pi} \,e^{i\vec{\pi}\cdot \vec{\beta}} \psi(\omega, \zeta,\zetab) \, \rho^{-2}e^{-\frac{|\vec{\pi}|^2}{2 \,\rho^2\,\omega^2}} 
\;\propto\;
    e^{i\lambda\alpha}\, \psi(\omega, \zeta,\zetab)\, \omega^2 e^{-\rho^2 \frac{\,\omega^2 |\vec{\beta}|^2}{2}}.
    \end{equation}
    where $\rho\ll 1$ is some dimensionless parameter playing the role of regulator. Comparing with \eqref{decompalphabeta} and the previous branching rules, one sees that this state is the sum of our initial state, for $|\vec{\pi}|=0$, with infinitely many continuous spin particles, for $|\vec{\pi}|>0$. The Gaussian means that the continuous spins are here all localised in the ultraviolet: 
    The continuous-spin particles with fixed parameter $\mu=|\vec{\pi}|\neq0$  will be exponentially suppressed at energies $\omega \ll \Lambda$ with $\Lambda:=\mu/\rho$ playing the role of an infrared cutoff.
    In this sort regime it has been argued (see \cite{Schuster:2023xqa} and refs therein) that continuous-spin particles do not differ significantly from usual massless particles, so it is plausible that the regularised state has physical properties very similar to the initial one.

Note that, after making the Gaussian integral in $\vec{\beta}$, the norm of \eqref{Massless reps, example: Poincarré particle regularization} effectively becomes
\begin{equation}\label{Massless reps, example: Poincarré particle regularization, norm}
    \langle \Psi | \Psi \rangle\; \propto \; \rho^{-2}\int\limits_{\mathbb{R}^+}  \omega^3\, d\omega\int\limits_{S^2}  \frac{i}{2}  d\zeta \wedge d\bar{\zeta}\, |\psi(\omega, \zeta,\zetab)|^2,
\end{equation}
and therefore weakens the constraint on the infrared behaviour of $\psi(\omega, \zeta,\zetab)$: states with poles are now normalisable (as BMS particles).

\section{Massive unitary irreducible representations of the BMS group}\label{Section: MassiveUIRsBMS}

We will here discuss some aspects of massive BMS particles. We will be more brief than in the case of massless BMS particles, since both cases shares some similarities and, when they diverge, the results in the massive case are somewhat more intuitive.
\begin{definition}
We call massive representations of the BMS group those UIRs for which the associated momentum $p_{\mu}$ is time-like, $p^2=-m^2<0$, and non-zero. By Proposition \ref{Prop: classification of supermomenta}, it must be in the orbit of
\begin{equation}\label{BMS massive rep: Reference supermomenta}
\K(z,\bar z) = \pm\frac{m}{\pi}\big(1+|z|^2\big)^{-3} +  \partial_z^2 \partial_{\bar z}^2 \N(z,\bar z).
\end{equation}
The BMS litle group $\ell_{\K}$ is the stabiliser of this reference supermomentum. The Poincar\'e little group, now a maximal compact subgroup $\ell_k \simeq SU(2)\subset SL(2,\mathbb{C})$, is the stabiliser of the corresponding reference momentum
\begin{equation}\label{BMS massive rep: Reference momenta1}
k^{\mu} = \pi^{\mu}(\K)=(\pm1,0,0,1).
\end{equation}
\end{definition}

Here again, the soft little group $\ell_{\Sigma}$ might a priori have any dimension between $0$ and $6$. However it follows from \cite{McCarthy_75} that the only possible little groups $\ell_{\K} = SU(2) \cap \ell_{\Sigma}$ of supermomenta $\K$ of the form \eqref{BMS massive rep: Reference supermomenta} (or, equivalently, \eqref{massivedecomppsition}), are
\begin{equation}\label{massive reps: admissible little groups}
    \ell_{\K} \in \Big\{ \;\mathbb{Z}_2\;,\; U(1)\;, \; SU(2) \;\Big\}\,,
\end{equation}
which correspond to the effective little groups $   \frac{\ell_{\K}}{\mathbb{Z}_2} \in \Big\{ \;\{e\}\;,\; SO(2)\;, \; SO(3) \;\Big\}$.
Here again, the quotient
\begin{equation}
    F \simeq \frac{SU(2)}{\ell_{\K}}
\end{equation}
encodes the discrepancy between the representation considered and more usual massive hard representations.

\subsection{Massive representations of the BMS group}\label{ssection: Massive representations of the BMS group}

\subsubsection{Supermomentum orbit}

The orbit $\mathcal{O}_{\K}$ of a supermomentum of the form \eqref{BMS massive rep: Reference supermomenta} is isomorphic to
\begin{equation}
    \mathcal{O}_{\K} \simeq \frac{SL(2,\mathbb{C})}{\ell_{\K}},
\end{equation}
where the BMS little group $\ell_{\mathcal{K}}$  belongs to the list \eqref{massive reps: admissible little groups}. By the definition of a massive representation, the orbit $\mathcal{O}_{k}$ of the massive momentum \eqref{BMS massive rep: Reference momenta1} is isomorphic to 
\begin{equation}
    \mathcal{O}_{k} \simeq \frac{SL(2,\mathbb{C})}{SU(2)} \simeq H^3.
\end{equation}

We thus have:
\begin{proposition}
Let $\P$ be the supermomentum of a $BMS_4$ massive representation with associated  time-like momentum $p_{\mu}=\pi_{\mu}(\P)$. The projection from the orbit $\mathcal{O}_{\P}$ of the supermomentum onto the orbit $\mathcal{O}_{\pi(\P)}$ of its associated momentum
\begin{align}
    \begin{array}{ccc}
        \pi:\mathcal{O}_{\P} & \twoheadrightarrow &  \mathcal{O}_{\pi(\P)} \simeq H^3
    \end{array}
\end{align}
is a fibre bundle of fibre $F = \frac{SU(2)}{\ell_{\mathcal{P}}}$ isomorphic to $\frac{SL(2,\mathbb{C})}{\ell_{\mathcal{P}}} \twoheadrightarrow \frac{SL(2,\mathbb{C})}{SU(2)}$.
\end{proposition}

We will use the Iwasawa decomposition of $SL(2,\mathbb{C})$ since it relies on the maximal compact subgroup $SU(2)$:
\begin{eqnarray}
&&G=NA\,K\,,\quad\text{with}\quad G=SL(2,\mathbb{C})\,,\quad\text{and}\\
\quad&&K=SU(2)\,,\quad  A=\mathbb{R}^+=
    \left\{\,\left( {\begin{array}{cc}
   a & 0 \\
   0 & a^{\scriptscriptstyle -1}\\
  \end{array} } \right)\,\mid\,a\in\mathbb{R}\, \right\}\,,\quad N=\mathbb{C}=
    \left\{\left( {\begin{array}{cc}
   1 & b \\
   0 & 1\\
  \end{array} } \right)\,\mid\,b\in\mathbb{C} \right\}\,.
\end{eqnarray}
The coset space $G/K\simeq NA$, i.e. $SL(2,\mathbb{C})\,/\,SU(2)\simeq \mathbb{R}^+\times\mathbb{R}^2$ identifies with the one-sheeted hyperboloid $H^3\cong SO_0(3,1)\,/\,SO(3)$ parametrise by the coordinates $a>0$ and $b \in\mathbb{C}$. The maximal compact subgroup $K(\alpha,\theta,\phi)$ can be parametrised locally by the Euler angles $\alpha,\theta,\phi$. Recall that $SU(2) \simeq S^3$ can be realised as the total space of a $U(1)$-principal bundle over $S^2$ (by the Hopf fibration). In order to compare with Section \ref{Section: MasslessUIRsBMS}, it will often be convenient to also use the stereographic coordinates $(\beta,\betab)$ on the base $S^2\simeq \mathbb{CP}^1$. Note that $\beta=|\beta|e^{i\phi}$ and that $\alpha$ is a coordinate on the fibre $U(1)$.

\subsubsection{Wavefunctions}

Let $\Phi$ be a state vector for a UIR of the BMS group given by the orbit of the massive supermomentum \eqref{BMS massive rep: Reference supermomenta} with little group $\ell_{\K}$.\\

\noindent$\bullet$ $\ell_{\K} = \mathbb{Z}_2$ :  By assumption, the orbit is then $\mathcal{O}_{\P}\simeq \frac{SL(2,\mathbb{C})}{\mathbb{Z}_2}\simeq SO_0(3,1)$. There are only two UIRs of $\mathbb{Z}_2\simeq\{1,-1\}$ : the trivial representation $\mathbb{Z}_2\to1$ and the identity map $\mathbb{Z}_2\to\mathbb{Z}_2$. Therefore, Theorem \ref{McCarthytheo} implies that the wave function is a complex function $\Phi(a,b,\bar b,\alpha,\beta,\betab)$ on $SL(2,\mathbb{C})$ which is square-integrable with respect to the Haar measure on $SL(2,\mathbb{C})$ and which obeys the parity condition (equivalently, $\mathbb{Z}_2$-equivariance)  
\begin{equation}\label{paritymassive}
\Phi(a,b,\bar b,\alpha+2\pi,\theta,\phi+2\pi)=(-1)^F\Phi(a,b,\bar b,\alpha,\theta,\phi)
\end{equation}
where $F=\pm 1$ in the bosonic/fermionic case. 
\\

\noindent$\bullet$ $\ell_{\K} = U(1)$ : By assumption, the orbit is $\mathcal{O}_{\P}\simeq SL(2,\mathbb{C})/U(1)$. The UIRs of this $U(1)$ subgroup are given by a choice of ``helicity'' $\lambda \in \tfrac12\mathbb{Z}$. Therefore, Theorem \ref{McCarthytheo} implies that the wave function is a complex function on $SL(2,\mathbb{C})$ of the form 
\begin{equation}\label{U1massive}
    \Phi(a,b,\bar b,\alpha,\beta,\betab) = e^{i\lambda\alpha} \psi(a,b,\bar b,\beta,\betab)
\end{equation}
where $\psi$ is square-integrable.\\

\noindent$\bullet$ $\ell_{\K} =SU(2)$ :  This corresponds to hard massive BMS representations. The orbit is $\mathcal{O}_{\K}\simeq SL(2,\mathbb{C})/SU(2)\simeq H^3$. The UIRs of $SU(2)$ are characterised by a choice of spin $j\in\tfrac12\mathbb N$. The wavefunction is a complex function on $SL(2,\mathbb{C})$. Various equivalent forms exist, depending on the choice of the $U(1)$ subgroup chosen for the helicity. 
For integer $j\in\mathbb N$, the wavefunction of a hard massive BMS particle can be of the form 
\begin{equation}
    \Phi(a,b,\bar b,\alpha,\theta,\phi) = \sum\limits_{m=-j}^j Y_j^m(\theta,\phi) \,\psi_m(a,b,\bar b)
\end{equation}
if the $U(1)$ helicity subgroup corresponds to shifts of the angle $\phi$, or of the form
\begin{equation}
    \Phi(a,b,\bar b,\alpha,\beta,\betab) = \sum\limits_{m=-j}^j Y_j^m(\beta,\betab) \,\psi_m(a,b,\bar b)\,.
\end{equation}
if the $U(1)$ helicity subgroup instead corresponds to phase shifts of $\beta$.
Both realisations define equivalent UIRs of the Poincar\'e group (and there are infinitely-many others).
The above modes $\psi_m(a,b,\bar b)$ and $\psi_m(a,b,\bar b)$ must be square-integrable functions on $SL(2,\mathbb{C})\,/\,SU(2)$.

\subsection{Branching to a Poincar\'e subgroup}\label{branching3}

\paragraph{$\bullet$ Generic orbits $\ell_{\K} = \mathbb{Z}_2$}\mbox{}

\vspace{1mm}

For any compact Lie group $G$, the Peter-Weyl theorem gives the orthogonal decomposition of the space $L^2(G)$ of square integrable functions (for the Haar measure) on $G$ as the direct sum of all finite-dimensional UIRs of $G$ with multiplicity equal to the corresponding (finite) dimension.
In particular,
\begin{equation}
L^2\Big(SU(2)\Big)=\bigoplus\limits_{2j\in\mathbb N}\,(2j+1)\,D_j\,,    
\end{equation}
where $D_j$ is the spin-$j$ representation of $SU(2)$. 
In the above decomposition, one can think of $SU(2)$ as the hypersphere $S^3$. In this interpretation, the term $(2j+1)\,D_j$ corresponds to the UIR $D_j\otimes D_j$ of $Spin(4)\cong SU(2)\times SU(2)$ realised as the vector space of scalar spherical harmonics on $S^3$ (see e.g. \cite[Section 4]{Jantzen} and \cite[Section 5]{Nashabeh_2024}). To be more precise, a basis of scalar spherical harmonics on $S^3$ is provided by the Wigner D-matrix elements $D^j_{m,m'}(\alpha,\theta,\phi)=\langle j,m|R(\alpha,\theta,\phi)|j,m'\rangle$ with $j\in\frac12\mathbb N$ and $m,m'=-j,-j+1,\ldots,j-1,j$ (see e.g.  \cite[Section II.C]{RevModPhys.38.330}). For integer $j=\ell\in\mathbb N$, the Wigner D-matrix elements are proportional to the spin-weighted spherical harmonics (see e.g. \cite{Eastwood_Tod_1982} for the definition). More precisely, the relation is (see e.g. \cite[Appendix B]{Shiraishi:2012bh} for this relation):
\begin{equation}
   \overline{D^\ell_{m,m'}(\alpha,\theta,\phi)}\;\;\propto\;\; {}_{m'}Y_\ell^{m}(\theta,\phi)\,e^{im'\alpha}\,.
\end{equation}
We also mention the relation between Wigner's D-matrix and $d$-matrix elements, $D^j_{m,m'}$ and $d^j_{m,m'}$ respectively:
\begin{equation}\label{smalld}
D^j_{m,m'}(\alpha,\theta,\phi)\;=\;  d^j_{m,m'}(\theta)\,e^{-im\phi}\,e^{-im'\alpha}\,,
\end{equation}
which underlines the similar role played by the angles $\alpha$ and $\phi$.

Performing the above decomposition for the $SU(2)$ factor in the Iwasawa decomposition of $SL(2,\mathbb{C})$ suggests that the square-integrable function $\Phi$ on $SL(2,\mathbb{C})$ can be decomposed into a discrete sum of modes 
\begin{equation}\label{decompmassive}
\Phi(a,b,\bar b,\alpha,\beta,\betab)=\sum\limits_{j\in\frac12\mathbb N}\;\sum\limits_{m=-j}^j\;\sum\limits_{m'=-j}^j    \tilde\Phi_{j}^{m,m'} (a,b,\bar b)\;\overline{D^j_{m,m'}(\alpha,\beta,\betab)}\,,
\end{equation}
which is the  massive analogue of the decomposition \eqref{decompalphabeta}-\eqref{decompalphabeta2} in the massless case.
As one can see, the modes $\tilde\Phi_{j}^{m,m'}(a,b,\bar b)$ are, for fixed $j$ and fixed $m'$, a field on $SL(2,\mathbb{C})\,/\,SU(2)\simeq H^3$ taking values in the spin-$j$ representation $D_j$ of $SU(2)$ (since $m=-j,-j+1,\ldots,j-1,j$) and picking a phase $e^{im'\theta'}$ with respect to the $U(1)$ transformation $\alpha\to\alpha+\theta'$. The UIR of $ISO_0(3,1)$ corresponding to the modes $\tilde\Phi_{j}^{m,m'}(a,b,\bar b)$ with fixed $j$ and $m'$ is thus a massive representation of spin $j$, since it is induced from a spin-$j$ representation $D_j$ of the Poincar\'e little group $SU(2)$. 

By making use of \eqref{smalld}, the decomposition \eqref{decompmassive} can also be written in the Euler angle coordinates as
\begin{equation}\label{massivedecomp2}
\Phi(a,b,\bar b,\alpha,\theta,\phi)=\sum\limits_{j\in\frac12\mathbb N}\;\sum\limits_{m=-j}^j\;\sum\limits_{m'=-j}^j    \tilde\Phi_{j}^{m,m'} (a,b,\bar b)\;\overline{d^j_{m,m'}(\theta)}\,e^{im\phi}\,e^{im'\alpha}\,.
\end{equation}
Equivalently, above one could have considered the modes $\tilde\Phi_{j}^{m,m'}(a,b,\bar b)$ for fixed $j$ and fixed $m$: they define again a field on $SL(2,\mathbb{C})\,/\,SU(2)\simeq H^3$ taking values in the spin-$j$ representation $D_j$ of $SU(2)$ (since $m'=-j,-j+1,\ldots,j-1,j$) but now picking a phase $e^{im\tilde\theta}$ with respect to the $U(1)$ transformation $\beta\to e^{i\tilde\theta}\beta$ or, equivalently, $\phi\to \phi+\tilde\theta$. Therefore, the UIR of $ISO_0(3,1)$ corresponding to the modes $\tilde\Phi_{j}^{n,n'}(a,b,\bar b)$ with fixed $j$ and $n$ is thus a again massive representation of spin $j$. 

Consequently, in both ways one recovers the following branching rule.

\begin{proposition}[McCarthy \& Crampin, 1973]\label{massivebranchingBtoP}\mbox{}\\
A massive UIR of $BMS_4$ of mass $m$ with trivial effective little group of supermomenta decomposes, with respect to any Poincar\'e subgroup $ISO_0(3,1)\subset BMS_4$, into a direct sum of massive UIRs of $ISO_0(3,1)$ of the same mass $m$ and of all possible spins (either all integers, $j\in\frac12\mathbb N$, or all half-integers, $j\in\mathbb{N}+\frac12$) with multiplicity $2j+1$.
\end{proposition}

\proof{The above argument based on the Iwasawa decomposition of $SL(2,\mathbb{C})$ provides an independent proof of  Proposition \ref{massivebranchingBtoP}. But the latter proposition can also be deduced from the results in \cite[Section 2]{McCarthy_73-III} for the case of the cyclic group $C_2$ (the so-called ``$j$-towers'' for all $j\in\frac12\mathbb N$ imply the multiplicites given in Proposition \ref{massivebranchingBtoP}).}

\paragraph{$\bullet$ Orbits with little group $\ell_{\P} = U(1)$}\mbox{}

\vspace{1mm}

The UIRs of the Abelian subgroup $U(1)$ are labeled by the helicity $\lambda=\tfrac{n}{2}\in\frac12\mathbb{Z}$. Therefore, Theorem \ref{McCarthytheo} implies that the wave function can be taken as an equivariant $\mathbb C$-valued function on $SL(2,\mathbb{C})$ of the form \eqref{U1massive} if one takes as $U(1)$ subgroup the one parametrised by the angle $\alpha$. This implies that only the modes $\widetilde{\Phi}_j^{n,n'}$ (with $n$ fixed) are present in the Fourier decomposition \eqref{massivedecomp2} of $\Phi$. Note that necessarily $j\geqslant|\lambda|$. This shows 

\begin{proposition}[McCarthy \& Crampin, 1973]\label{massivebranchingbranchrule}\mbox{}\\
A massive UIR of $BMS_4$ with BMS little group $U(1)$ is labeled by the helicity $\lambda\in\frac12\mathbb{Z}$ and decomposes, with respect to any Poincar\'e subgroup $ISO_0(3,1)\subset BMS_4$, into a direct sum of massive representations with spin $j=|\lambda|,\,|\lambda|+1,\,|\lambda|+2,\,\ldots$    
\end{proposition}

\proof{The above argument in terms of the mode decomposition in spherical harmonics on $S^3$ provides a concrete realisation of this proposition. This result agrees with the statement in \cite[Section 2]{McCarthy_73-III}. It can also be derived from Theorem \ref{branchingrulestheo} by noting that the UIRs of $SU(2)$ whose branching to a subgroup $U(1)$ contains the eigenmode $e^{i\lambda\alpha}$ are the spin-$j$ representations $D_j$ with $j=|\lambda|,\,|\lambda|+1,\,|\lambda|+2,\,\ldots$}

\begin{remark}
The appearance of an infinite tower of particles with  unbounded spin\footnote{For an analogy, observe that infinite sums of Poincar\'e UIRs appears for instance in the fusion rules of $ISO_0(3,1)$ giving the decomposition of the tensor product of two UIRs in terms of individual ones \cite[p.553]{Barut:1986dd} where, in a sense, the appearance of such a direct sum merely signals that it is not very natural to think about two-particle states as superpositions of single-particle states.} in the branching rules of UIRs of $BMS_4$ restricted to $ISO_0(3,1)$ may appear surprising, if not pathological. However, it merely signals that the usual notion of Poincar\'e particle is not adapted for describing generic BMS particles. In fact, it is simply unnatural to describe BMS particles in terms of a fixed gravity vacua. If one absolutely want to do that, then the price to pay is the presence of infinite tower of Poincar\'e particles in general. In more mundane terms, one can say that the usual Wigner basis of momentum eigenstates for one-particle states is simply not well-suited in the presence of supertranslations but should rather be replaced with supermomentum eigenstates  \cite{Bekaert:2024jxs}, as briefly explained in the next section.     
\end{remark}

\section{BMS wavefunctions}\label{Section: BMS wavefunctions}

In \cite{Bekaert:2024jxs}, some (infinite-dimensional) Fourier transforms of BMS wavefunctions were introduced. It resulted in three natural choices of picture (supermomentum space, BMS space, mixed space) for the wavefunctions of BMS particles and it was shown that the last two choices also allow to interpret BMS particles as quantum superposition of usual Poincaré particles. We here come back on results of \cite{Bekaert:2024jxs} in view of the representation-theoretic tools developed in the present article.

This section intends to make contact with physics literature and will accordingly be sloppy with respect to formal integration over infinite-dimensional spaces.

\subsection{In supermomentum space}\label{Insupermomentumspace}

A general state $|\Psi\rangle$ of a BMS particle is given by a wavefunction $\Psi(\P)$, i.e. a functional on the infinite-dimensional space $\E[-3]$ of supermomenta $\P$. 
Supertranslations $\T \in \E[1]$ act on these BMS wavefunctions in supermomentum space as mere phase shifts $\Psi(\P)\mapsto e^{\,i\,\langle \,\P\,,\,\T\,\rangle}\Psi(\P)$.
As previously recalled, the UIRs of $BMS_4$ are in one-to-one correspondence with spaces of fields (taking values in a UIR of the BMS little group) which only have support on a fixed orbit of a supermomentum $\mathcal{P}$ under the action of $SL(2,\mathbb{C})$. Such orbits $\mathcal{O}_{\mathcal{P}}$ directly generalise the mass-shell of Poincaré representations to our infinite-dimensional context.

As was discussed in Proposition \ref{Prop: classification of supermomenta}, a supermomentum $\P(z,\bar z)$ can be uniquely written as the sum $ \P=P+ \Sigma $ of  a hard supermomentum $P(z,\bar z)$ associated to a four-momentum $p_{\mu}$ and a soft charge $\Sigma(z,\bar z)=\partial^2_z\partial^2_{\bz} \N(z,\bar z)$. We can thus denote the wave function as
\begin{align}
 \Psi\Big(p,\partial^2_z \N \Big).
\end{align}

\subsection{In BMS space}

Minkowski spacetime $\mathbb{M}^{3,1} \simeq \frac{ISO_0(3,1)}{SO_0(3,1)}$ is an affine space modeled on the vector space of translations $\mathbb{R}^{3,1}$. By close analogy, we will define the space of cuts at null infinity (called here ``BMS space'') and the space of gravity vacua, as homogeneous spaces of the BMS group.
\begin{definition}
The BMS space is the homogeneous space
\begin{equation}
    \mathbb{B}\simeq \frac{BMS_4}{ SO_0(3,1)} = \frac{SO_0(3,1)\ltimes\E[1]}{ SO_0(3,1)}\,.
\end{equation}
It is an affine space modeled on the vector space of supertranslations $\E[1]$. Its elements can be realised geometrically as sections of null infinity $\mathscr{I}$ (seen as a line bundle over $S^2$) called cuts, hence $\mathbb{B}\simeq\Gamma(\mathscr{I})$.
\end{definition}

Here, following Ashtekar \cite{Ashtekar:1987tt,Ashtekar:2014zsa}, we define the space of gravity vacua as a homogeneous space of the BMS group, see also \cite{Herfray:2020rvq,Herfray:2021xyp,Herfray:2021qmp} for realisations in terms of Cartan geometry.
\begin{definition}\label{gravacua}
The space of gravity vacua is the homogeneous space 
\begin{equation}
   \mathbb{V}\simeq \frac{BMS_4}{ ISO_0(3,1)} = \frac{SO_0(3,1)\ltimes\E[1]}{SO_0(3,1)\ltimes \mathbb{R}^{3,1} }\,.
\end{equation} It is an affine space modeled on the quotient vector space $\E[1]/\mathbb{R}^{3,1}$. For this reason, we referred to  elements of $\E[1]/\mathbb{R}^{3,1}$ as shifts of vacua (cf. Definition \ref{quotientgravityshifts}).
\end{definition}

A corollary of Proposition \ref{Proposition: Injection translation -> super} is that the translation group $\mathbb{R}^{3,1} \subset BMS_4$ forms a canonical subgroup of the BMS group. This implies that the BMS space $\mathbb{B}$ is naturally an $\mathbb{R}^{3,1}$-principal bundle over the space $\mathbb{V}$ of gravity vacua. In fact, this is modelled on the property that the supertranslation group $\E[1]$ is naturally a $\mathbb{R}^{3,1}$-principal bundle over the quotient group $\E[1]/\mathbb{R}^{3,1}$ of vacua shifts:
\begin{center}
\begin{tikzcd}
\mathbb{R}^{3,1} &\lefttorightarrow&
 \mathbb{B} 
 \arrow[d,twoheadrightarrow]&\qquad&\mathbb{R}^{3,1}\arrow[r,hookrightarrow]&\E[1]\arrow[d,twoheadrightarrow]\\
&& \mathbb{V} &  &&\E[1]/\mathbb{R}^{3,1}
\end{tikzcd}
\end{center}
In practice, BMS space $\mathbb{B}$ is foliated by affine leaves which are gravity vacua, each diffeomorphic to Minkowkski spacetime. More concretely, above each element  $\partial^2\C$ of the base $\mathbb V$ sits, as fibre, the corresponding Minkowski spacetime $\mathbb{M}_{\C}^{3,1}$. 

An important property is that there exists a quantitative notion of ``distance'' between gravity vacua.\footnote{Physically, this purely mathematical notion is presumably related to the relative magnitude of the displacement (almost literally a ``distance'') in the memory effect.} This follows as a straightforward corollary of Proposition \ref{innerproductprop}.

\begin{corollary}\label{distancebetweenvacua}
The affine $\mathbb{V}$ space of gravity vacua is a metric space. More precisely, the Lorentz-invariant distance between two gravity vacua $\mathcal{V}_1,\mathcal{V}_2\in\mathbb{V}$ related by a shift $[\mathcal{T}]\in\E[1]/\mathbb{R}^{3,1}$ is defined by the norm of this vacua shift, i.e.
$$d\big(\mathcal{V}_1,\mathcal{V}_2\big)=\sqrt{\llangle\mathcal{T},\mathcal{T}\rrangle}$$
where 
\begin{equation} \label{Proof, main lemma: hermitian product}
    \llangle\T  , \T \rrangle = \langle \,  \eth^2 \bar{\eth}^2 \T , \T\, \rangle=\int_{S^2} \frac{i}{2}
    dz\wedge d\bar z \; \T\, \partial_{z}^2\partial_{\bar z}^2
    \T   
    =  \int_{S^2} \frac{i}{2}
    dz\wedge d\bar z \; |\partial_z^2 \T |^2.
\end{equation}
\end{corollary}
\noindent Note that the last equality in \eqref{Proof, main lemma: hermitian product} holds since $\T$ is real.

\vspace{2mm}

Since the space of supermomenta $\P \in \E[-3]$ is the dual space of supertranslations $\T \in \E[1]$, we can make use of an infinite-dimensional version of the Fourier transform to define the BMS wavefunction \cite{Bekaert:2024jxs}
\begin{align}
    \Psi(\T) &= \int \mathcal{D}\P \;e^{i
    \langle\mathcal{P},\mathcal{T}\rangle}\;  \Psi(\P)& \Longleftrightarrow && \Psi(\P) &= \int \mathcal{D}\P \;e^{-i
    \langle\mathcal{P},\mathcal{T}\rangle}\;  \Psi(\T).
\end{align}

A supertranslation
$\T = X^{\mu}\q_{\mu} +\C$ can be represented\footnote{For example by making use of spherical harmonics. We recall that, while such decomposition is not unique and breaks Lorentz invariance on $S^2$, the projection of a supertranslation on a vacua shift $\T \mapsto \partial_z^2 \C =\partial_z^2 \T$ is invariant.} by a pair $\left(X^{\mu},\partial_z^2 \C\right)$ of a spacetime point $X^{\mu}$ and a shift of vacua $\partial_z^2 \C = \partial_z^2 \T$. We can make use of this decomposition and
\begin{align}
\langle \,\P\,,\,  \T\, \rangle &= p\cdot X + \langle P , \C \rangle +  \langle\partial^2_z\N,\partial^2_z\C\rangle, \label{pairingcomputed1}
\end{align} 
where 
$\langle\partial^2_z\N,\partial^2_z\C\rangle:=\llangle\N,\C\rrangle=\int_{S^2} \frac{i}{2}
    dz\wedge d\bar z \; \partial_{z}^2\N\,\partial_{\bar z}^2
    \C =\int_{S^2} \frac{i}{2}
    dz\wedge d\bar z \; \partial_{z}^2\N\,\widebar{\partial_{z}^2
    \C}$, to rewrite the BMS wavefunction as
\begin{align}\label{BMS wavefunction as superposition of fields in different vaccua}
    \Psi\Big(X,\partial^2_z \C\Big) &:= \int d^4p \; e^{i\, p\cdot X }\; e^{i \langle P , \C \rangle} \;\int \mathcal{D}\N \;  e^{i\langle\partial^2_z\N,\partial^2_z\C\rangle}\; \Psi\Big(p,\partial^2_z \N\Big).
\end{align}

Since the projection $\T \mapsto \partial_z^2 \T$ of supertranslations onto vacua shifts is a Lorentz-invariant operation (cf. Proposition \ref{Proposition: Injection translation -> super}), it makes sense to consider the restriction of the above wavefunction to any fixed gravity vacuum $\partial_z^2 \C_0$. The result $\psi_{\C_0}(X) := \Psi\big(X,\partial^2_z \C_0\big)$ then becomes the usual wavefunction of a field $\psi_{\C_0}$ propagating on the corresponding background Minkowski spacetime $\mathbb{M}^{3,1}_0 \subset \mathbb{B}$ defined by $\partial_z^2 \C_0$. Therefore BMS states can be understood as sums of usual spacetime states, each of them propagating on a given gravity vacuum.

Let us stress the following. The expression for BMS wavefunctions \eqref{BMS wavefunction as superposition of fields in different vaccua} is written as a path integral, and as such is a complicated infinite-dimensional functional. However, the UIRs of the BMS group are always spanned by fields supported on a finite-dimensional manifold (``BMS shell'') of dimension $\leqslant 6$ inside the space of supermomenta and, as such, are much simpler objects than the above discussion might suggest (in more physical words, ``on-shell'' BMS particles are infinitely simpler than ``off-shell'' ones).

\subsection{In momenta and vacua space}\label{Inmomentaandvacuaspace}

Due to the existence of the Lorentz-invariant projections $\P(z,\bar z) \mapsto p_{\mu}$ of supermomenta onto momenta  and $\T \mapsto \partial_z^2 \C$ of supertranslations onto vacua shifts, one can perform only ``half'' of the above Fourier transform \eqref{BMS wavefunction as superposition of fields in different vaccua} to obtain \cite{Bekaert:2024jxs}
\begin{align}
    \Psi\Big(p,\partial^2_z \C\Big) &= \int \mathcal{D} \N \;e^{i\langle\partial^2_z\N,\partial^2_z\C\rangle} \;\Psi\Big(p,\partial^2_z \N\Big)  =  e^{-i\langle P,\, \C\rangle} \int d^4 X \;e^{-i\, p\,\cdot X} \; \Psi\Big(X,\partial^2_z \C\Big).
\end{align}
Supertranslations then act on these BMS wavefunction as
\begin{align*}
    \Psi\Big(p,\partial^2_z \C
    \Big) \qquad \to \qquad  e^{i \langle P , \T \rangle}\;\Psi\Big(p,\partial^2_z (\C+ \T)
    \Big).
\end{align*}
These BMS wavefunctions factorise in a natural way because they depend on momenta and gravity vacua, both of which are elements of Lorentz-invariant (quotient) spaces, respectively $\E[-3]\,/\,\text{Ker}\,\pi$ and $\E[1]\,/\,\text{Im}\,q$. Let us once again stress that, for irreducible representations, the above path integrals over the soft charge actually localise to finite-dimensional (generically, three-dimensional) integrals. 

It is convenient to denote $\psi_{\C_0}\left(p\right) := \Psi\big(p\,,\,\partial^2_z \C_0\big)$ the Poincaré state obtained by restricting the above wavefunction to a fixed gravity vacuum $\partial^2_z \C_0$. In general, a BMS particle can again be interpreted as the superposition of different usual Poincaré particles $\psi_{\C_0}\left(p\right)$ each of them propagating on a different gravity vacuum $\partial_z^2\C_0$. This picture seems very reminiscent to us to the ideas described in \cite{Kapec:2021eug,Kapec:2022axw,Kapec:2022hih}.

\subsection{Relation to hard representations}\label{relationhardrepresentations}

Let $\psi(p)$ be the wavefunction of a usual massless spin-$s$ field. To which BMS wave function $\psi_{\C}(p)$ does this correspond? There is in fact a huge ambiguity corresponding to all the possible ways to lift a Poincaré representation to a BMS representation.

The most obvious choice is to lift the Poincaré representation to a hard BMS representation by declaring that the BMS wavefunction does not depend on the vacuum
\begin{equation}\label{BMS wavefunction: Relation to hard representations, hard}
    \psi_{\C}\left(p\right) = \psi(p)\,.
\end{equation}
In supermomentum space, this means that
\begin{equation}
    \Psi\Big(p,\partial^2_z \N\Big) = \psi\left(p\right)\,\delta(\N)
\end{equation}
and this indeed coincides with the hard representation of a massless field.

As an alternative, one might declare that the Poincar\'e wavefunction $\psi(p)$ corresponds to a state localised in a given gravity vacuum
\begin{equation}\label{BMS wavefunction: vacua eigenstate}
    \psi_{\C}\left(p\right) = \psi(p)\,\delta(\C - \C_0)\,,
\end{equation}
i.e it is only defined in a particular vacuum and is zero in all others. 
This state is the closest thing to the classical picture of working in a fixed particular background. In supermomentum space, this however means that
\begin{equation}
    \Psi\Big(p,\partial^2_z \N\Big) = \psi\left(p\right)\;e^{i \langle\partial^2_z\N,\partial^2_z\C_0\rangle}.
\end{equation}
and thus it is not localised on a single orbit of supermomenta (because there is no delta function to restrict $\partial^2_z\N$ to an orbit of the Lorentz group): therefore it does not belong to an irreducible representation. This parallels position eigenstates in quantum mechanics which are not very good physical states.

Much more interesting states are the ones associated to a BMS particle in the sense that $\Psi(p,\partial_z^2 \N)= \psi(p)$, hence
\begin{equation}\label{statelambda}
    \psi_{\C}\left(p\right)  = \int\limits_{\ell_{p}/\ell_{\P}}  d^m
    W \;e^{\,i\,\langle\,\partial^2_z(W\cdot\N ),\,\partial^2_z(\C-\C_0)\,\rangle}\;  \psi(p),
\end{equation}
where $\ell_{p} \simeq ISO(2)$ is the Poincaré little group which stabilises $p_{\mu}$, while $\ell_{\P}$ is the BMS little group stabilising $\P = P + \partial_z^2 \partial_{\bz}^2 
\N$ and $m=3-\dim(\ell_{\P})$. As mentioned in Section \ref{Ssection: Branching in different gravity vacua}, such a state is a usual state for a Poincar\'e particle propagating in the Minkowski spacetime $\mathbb{M}_0^{3,1}$ defined by the gravity vacuum $\C_{0}$. In this distinct sense, the state \eqref{statelambda} is also ``localised'' in a gravity vacuum $\C_0$. However, as opposed to \eqref{BMS wavefunction: vacua eigenstate}, it belongs to a definite UIR of BMS group. Although the state \eqref{statelambda} might not have a finite norm if the quotient $\ell_{p}/\ell_{\P}$ is not compact\footnote{This is not a problem for massive representations. For massless representations this will happen whenever $\mathbb{R}^2$ is not in the BMS little group, see \eqref{Massless reps, example: Poincarré particle regularization} for a natural regularisation in such cases.}, it is much more well-behaved. Note that if the representation is hard, $\partial_z^2\N=0$, then this state coincides with \eqref{BMS wavefunction: Relation to hard representations, hard} and cannot discriminate a particular gravity vacuum.

\subsection{Fock space of BMS particles}\label{Fock}

We can now mention the second quantisation of BMS particles proposed in \cite{Bekaert:2024jxs} by sketching the corresponding Fock space of multi-particle states. 
If $\mathcal{H}$ denotes the Hilbert space of a single BMS particle, which we will assume bosonic for the sake of simplicity, then the symmetric tensor product algebra $\odot(\mathcal{H})$ is the corresponding Fock space. 

Let us denote by $\mid 0\,\rangle$ the BMS vacuum, which is annihilated by all BMS generators. Considering the supermomentum representation of the wavefunction (cf. Section \ref{Insupermomentumspace}), let us denote by $\mid\P\,\rangle=| p,\partial^2_z \N\rangle$ the supermomentum eigenstates of a single BMS particle. The corresponding multi-particle states 
 are defined as the symmetric tensor products 
\begin{equation}
    \mid\P_1,\ldots,\P_n\,\rangle\,:=\,\mid\P_1\,\rangle \odot\cdots\odot\mid\P_n\,\rangle
\end{equation}
Equivalently, they can be denoted 
\begin{equation}\label{Fockstates}
    | p_1,\ldots,p_n,\partial^2_z \N_1,\ldots,\partial^2_z \N_1 \rangle
\,:=\,|p_1,\partial^2_z \N_1\rangle\odot\cdots\odot| p_n,\partial^2_z \N_n\rangle
\end{equation}

\begin{remark}
The present Fock space $\odot(\mathcal{H})$ is distinct from previous proposals such as \cite{Prabhu:2022zcr,Prabhu:2024zwl,Prabhu:2024lmg}. In fact, our construction must be contrasted with another possible construction (somewhat analogous to the one in \cite{Prabhu:2024lmg}) where multi-particle supermomentum eigenstates would have been symmetric tensor products $|p_1,0\rangle\,\odot\,\cdots\odot\,| p_n,0\rangle\,\odot\,|0,\partial^2_z \N_n\rangle$ of $n$ hard BMS particles and one soft BMS particle. Such state could arguably be denoted $| p_1,\ldots,p_n,\partial^2_z \N\rangle$. From the point of view of the present paper, such states belong to the tensor product of Fock spaces for distinct UIRs (hard and soft). The present paper focuses on generic BMS representations but we do not exclude the possibility that the Hilbert space suitable for BMS-invariant scattering theory only makes use of hard and soft representations (cf. Footnote \ref{nogo}). Nevertheless, we believe that BMS representation theory should be the ultimate guideline for building the proper Hilbert space.
\end{remark}

Instead of the supermomentum eigenbasis, one can also consider the mixed (i.e. momentum \& vacuum) eigenbasis of on-shell BMS particles. Let us denote by $| p,\partial^2_z \mathcal{C}\rangle$ the momentum eigenstate in a given gravity vacuum (cf. Section \ref{Inmomentaandvacuaspace}). The corresponding multi-particle states would read 
\begin{equation}
    | p_1,\ldots,p_n,\partial^2_z \mathcal{C}_1,\ldots,\partial^2_z \mathcal{C}_n \rangle:=|p_1,\partial^2_z \mathcal{C}_1\rangle\odot\cdots\odot|p_n,\partial^2_z \mathcal{C}_n\rangle\,.
\end{equation}
Let us stress that the latter multi-particles states ($n\geqslant 2$) are labeled by several distinct gravitational vacua. The multi-particle state $| p_1,\ldots,p_n,\partial^2_z \mathcal{C}_1,\ldots,\partial^2_z \mathcal{C}_n \rangle$ is to be contrasted with a state made of $n$ particles in a \textit{single} gravitational vacuum (which would have been denoted  $| p_1,\ldots,p_n,\partial^2_z \mathcal{C} \rangle$). Although the latter state may appear intuitively more natural (along the standard picture of hard particles propagating in the \textit{same} gravitational vacuum), it is actually not as natural from a group-theoretical perspective (where one would try construct a BMS-invariant QFT \`a la Weinberg, by replacing everywhere in \cite{Weinberg:1995mt} the Poincar\'e group by the BMS group). In fact, the terminology ``gravitational vacua'' may be misleading in this respect since the present work shows that it is logically consistent for each BMS particle to somehow feel its own ``gravitational vacuum''. This last point of view is very natural if one considers the situation in terms of the supermomentum representation, where each BMS particle has its own soft charge, cf. \eqref{Fockstates}.

\section{On the relation to the memory effect}\label{Section: memoryeffect}

The goal of this section is to present, on a very particular example, how generic BMS particles evade in a subtle way the no-go result on the absence memory effect for hard BMS particles. Accordingly, the style of this section (like the previous one) slightly departs from the rest of the paper. We refer to reviews such as \cite{Strominger:2017zoo,Ashtekar:2018lor} for basic definitions and results on this topic.

We would like to stress that there is no memory effect contribution to the average supermomenta of hard particules.
To see this, let us consider one-particle states and loosely use the language of first-quantisation. If $a(\omega, \zeta, \bar{\zeta})$ is the wave function (in the momentum representation) of a hard particle, say a graviton, then $|a(\omega, \zeta, \bar{\zeta})|^2$ is the probability that the particle has supermomentum $\P(z,\bar z)=\omega\,\delta^{(2)}(z-\zeta,\bar z-\bar\zeta)$. The average supermomentum is thus
$$    \langle \P(z,\bz) \rangle = \int \omega\, d \omega \int \frac{i}{2}d\zeta \wedge d\bar{\zeta} \; |a(\omega, \zeta, \bar{\zeta})|^2 \, \omega\,\delta^{(2)}(z-\zeta)
    = \int \omega^2 d \omega\, \big|a(\omega, z, \bar{z})\big|^2 
    = \int du\, |\partial_u C_{zz}(u, z, \bz)|^2
$$
where $C_{zz}(u, z, \bz)=\int d\omega\,e^{i\omega u}a(\omega, z, \bar{z})$ is the shear. 
The flux balance equation, $\partial_u m= |N_{zz}|^2 + \partial^2_z N_{\bz\bz}$ where $N_{\bz\bz}=\partial_u C_{\bz \bz}$ is the news, implies that
\begin{align}
    &m(u=\infty, z, \bz) - m(u=-\infty, z, \bz) = \int du\, \partial_u m(u,z, \bz)
    = \int du\, \big|\partial_u C_{zz}(u, z, \bz)\big|^2 
    +\int du \, \partial^2_z \N_{\bz\bz}(u, z, \bz)\nonumber\\
    & \qquad\qquad= \int \omega^2 d \omega\, \big|a(\omega, z, \bar{z})\big|^2  + \lim\limits_{\omega\to 0}\,[\,\omega\, \partial^2_z a(\omega, z,\bz)\,]
\end{align}
As Ashtekar pointed out \cite[Section II.C.3]{Ashtekar:1987tt} (see also \cite[Section IV.B]{Ashtekar:2014zsa}), for a state inside the hard representation to be normalisable, we need\footnote{This necessary condition for the norm finiteness is also sufficient if the news belongs to the Schwarz space of fastly decreasing functions \cite[Section II.C.3]{Ashtekar:1987tt}.} to have $ \lim\limits_{\omega\to 0}\,[\,\omega\,\partial^2_z a(\omega, z,\bz)\,] =0$ or, equivalently, $C_{zz}(u=\infty, z, \bz)\,=\,C_{zz}(u=-\infty, z, \bz)=0$. In other words, there is no memory effect for a hard graviton state.
In this way, the average supermomentum of a hard graviton reproduces the standard expression 
\begin{equation}
\int du \,\big|\partial_u C_{zz}(u, z, \bz)\big|^2\;=\;  m(u=\infty, z, \bz) - m(u=-\infty, z, \bz)
\end{equation}
for the hard flux at null infinity of a graviton. 

Can we produce example of BMS states with non-zero memory? Yes: let us for instance\footnote{Another type of interesting example is the state \eqref{Massless reps, example: Poincarré particle regularization} whose effective norm \eqref{Massless reps, example: Poincarré particle regularization, norm} is less stringent then that of usual hard massless fields, due to the continuous spins spectrum which has been integrated out, and, in particular, allows for memory.} consider bosonic BMS particles with supermomenta of the form \eqref{BMS massless rep: Reference supermomenta2 orbit}. These massless BMS particles have effective BMS little group $\mathbb{R}^2$ and a generic state must be of the form \eqref{Wavefunction: R2 equivariance}. 
Consider the case $\vec\pi_0=\vec 0$, then the wavefunction is of the form \eqref{expansionR2littlegroup}.
For such BMS particles, a generic state is a superposition of massless Poincar\'e particles with all possible helicities $s\in\mathbb Z$ (cf. Proposition \ref{massless_R2_branching}), each of them characterised by a wavefunction $a_s(\omega, \zeta, \bar{\zeta})$. Let us now consider, as particular state, the quantum superposition of a scalar field of wavefunction $a(\omega, \zeta, \bar{\zeta})$ and the same field but prepared in an other gravity vacua; the first being obtained from the second by the action of a supertranslation $\C(z,\bar z)$:
\begin{align}\label{State with memory}
    \Phi(\omega,\zeta,\bar{\zeta},\alpha) &= \frac{1}{\sqrt 2}\left(a(\omega, \zeta, \bar{\zeta})\,e^{i \langle \mathcal{K}, \C \rangle}+ a(\omega, \zeta, \bar{\zeta})\right) \\
    &= \frac{1}{\sqrt 2}\,a(\omega, \zeta, \bar{\zeta})\left[e^{i\Big(\omega \C(\zeta, \bar{\zeta}) +\frac{\sigma}{\omega}\big(e^{i2\alpha} \partial_\zeta^2 \C(\zeta, \bar{\zeta}) + e^{-i 2\alpha} \partial_{\bar\zeta}^2 \C(\zeta, \bar{\zeta})\big)\Big)} +1\right],  
\end{align}
where the second line follows from the evaluation of the pairing \eqref{pairing} with the supermomentum \eqref{BMS massless rep: Reference supermomenta2 orbit}.
Since this state interpolates between two gravity vacua, it has a chance to encode memory. 

Let us introduce the uniform (aka $L^\infty$) norm 
\begin{equation}
||\partial_z^2\C||_\infty:= \textrm{Max}_{_{S^2}} \big|\partial_z^2\C(z,\bar z)\big|    \,,
\end{equation}
i.e. the maximum of the modulus $|\partial_z^2\C|$ of the spin-weighted density $\partial_z^2\C\in\mathcal{O}(-3,1)$ on the sphere. This uniform norm effectively sets an energy scale above which the summands in the decomposition \eqref{expansionR2littlegroup} behave like hard states. More precisely, when the dimensionless ratio\footnote{This limit had a Lorentz-invariant meaning. In fact, the dimensionless ratio $\omega/||\partial_z^2 \C||_\infty$ of the energy by the uniform norm is $SL(2,\mathbb{C})$-invariant. This can be seen by observing that the pairing $\langle \mathcal{K}, \C \rangle=\omega \C(\zeta, \bar{\zeta}) +\frac{\sigma}{\omega}\big(e^{i2 \alpha} \partial_\zeta^2 \C(\zeta, \bar{\zeta})+c.c.$ is $SL(2,\mathbb{C})$-invariant. So are each summand on the right-hand side.
Therefore, its maximum on the whole sphere $|\partial_\zeta^2 \C(\zeta,\bar\zeta)|/\omega$ is $SL(2,\mathbb{C})$-covariant, so that $||\partial_z^2\C(z,\bar z)||_\infty\,/\omega$ is $SL(2,\mathbb{C})$-invariant.} of the energy by the uniform norm is very large, $\frac{\omega\;}{||\partial_z^2 \C||_\infty}\gg 1$, one has the following expansion
\begin{align}\label{expansionPhi}
    e^{-i\omega\C(\zeta,\bar \zeta)}\Phi(\omega,\zeta,\bar{\zeta},\alpha)= a_0(\omega, \zeta, \bar{\zeta}) +  e^{i2\alpha} a_2(\omega, \zeta, \bar{\zeta}) + e^{-i2\alpha} a_{-2}(\omega, \zeta, \bar{\zeta}) + \mathcal{O}\left(\left(\tfrac{||\partial_z^2 \C||_\infty}{\omega}\right)^2 \right)
\end{align}
which effectively looks like the quantum superposition of a massless spin-0 particle
\begin{equation}
    a_0(\omega, \zeta, \bar{\zeta}):= \frac{1 + e^{-i\omega \C(\zeta, \bar{\zeta})}}{\sqrt 2}\;a(\omega, \zeta, \bar{\zeta})  + \mathcal{O}\left(\left(\tfrac{||\partial_z^2 \C||_\infty}{\omega}\right)^2 \right)
\end{equation}
complemented by two spin-2 particles of opposite helicities whose amplitudes are suppressed when $\omega\gg||\partial_z^2 \C||_{\infty}$
\begin{align}
    a_2(\omega, \zeta, \bar{\zeta})& := \frac{i\,\sigma}{\sqrt{2}} \;  \frac{\partial^2_{\zeta} \C(\zeta,\bar{\zeta})}{\omega}\;a(\omega, \zeta, \bar{\zeta})  + \mathcal{O}\left(\left(\tfrac{||\partial^2 \C||_\infty}{\omega}\right)^3 \right)\,,\\
    a_{-2}(\omega, \zeta, \bar{\zeta}) &:= \frac{i\,\sigma}{\sqrt{2}} \;  \frac{\partial^2_{\bar\zeta} \C(\zeta,\bar{\zeta})}{\omega}\;a(\omega, \zeta, \bar{\zeta})  + \mathcal{O}\left(\left(\tfrac{||\partial^2 \C||_\infty}{\omega}\right)^3 \right)\,.
\end{align}
More generally, when the energy $\omega$ is high compared to the uniform norm $||\partial_z^2 \C||_\infty$, one has
\begin{align}
    a_s(\omega, \zeta, \bar{\zeta}) = \frac{1}{\left(\frac{s}{2}\right)!\,\sqrt 2} \left(\frac{i \sigma \partial^2_\zeta \C(\zeta,\bar\zeta)}{\omega}\right)^{\frac{s}{2}}a(\omega, \zeta, \bar{\zeta}) + \mathcal{O}\left(\left(\tfrac{||\partial^2 \C||_\infty}{\omega}\right)^{\frac{s}{2}+2} \right).
\end{align}

Thus the state \eqref{State with memory} effectively appears as a spin-$0$ state at high energy, but is filled with higher-spin states in the infrared. In particular, the leading correction are spin-$2$ states which superficially seem to have a pole in the energy $\omega$\,: this is suggestive that they might carry some memory. However, if the full spin-2 state $a_2(\omega,\zeta,\bar{\zeta})$ really had a pole, these states would not be normalisable by themselves which would imply that the state \eqref{State with memory} would also be of infinite norm. The norm of a state of the form \eqref{expansionR2littlegroup} indeed reads
 \begin{equation}
     \langle\Phi , \Phi \rangle := \int \omega\, d \omega \int \frac{i}{2}d\zeta \wedge d\bar{\zeta} \; 
     \int \frac{d\alpha}{2\pi} \left| \Phi(\omega,\zeta,\bar{\zeta},\alpha) \right|^2 = \sum\limits_{s\in\mathbb Z}  \int \omega\, d \omega \int \frac{i}{2}d\zeta \wedge d\bar{\zeta} \;  \left| a_s(\omega,\zeta,\bar{\zeta}) \right|^2.
 \end{equation}
Fortunately, this problem does not arise: the state \eqref{State with memory} is a perfectly well-behaved state inside the UIR of $BMS_4$. This is because if $a(\omega, \zeta, \bar{\zeta})$ has finite norm then so does its supertranslated counterpart $a(\omega, \zeta, \bar{\zeta})e^{i \langle \mathcal{K}, \P \rangle}$. Therefore, since the states of a BMS particle form a Hilbert space, the sum of these state has finite norm as well.

We are now left with the question of the average supermomentum of the state:
\begin{align}
    \langle \P(z,\bz) \rangle &= \int \omega d\omega d \zeta d\bar{\zeta} \int \frac{d\alpha}{2\pi}\, \P(\zeta,\bar{\zeta})\,  \left| \Phi(\omega,\zeta,\bar{\zeta},\alpha) \right|^2 \nonumber\\
    &= \int \omega d\omega  \int \frac{d\alpha}{2\pi}\, \Bigg( \omega +  \frac{\sigma}{\omega}\left( e^{i2\alpha} \partial_z^2  + e^{-i2\alpha} \partial_{\bz}^2\right) \Bigg) \left| \Phi(\omega,z,\bar{z},\alpha) \right|^2.
\end{align}
We will suppose that $a(\omega, \zeta, \bar{\zeta})$ is a hard state in the sense that it is only supported at energies $\omega > \Lambda$ above a cut off $\Lambda\gg||\partial_z^2 \C||_{_\infty}$, so that the average momentum is only influenced by the leading-order behaviour of the modulus square of \eqref{expansionPhi}
\begin{align}
    \big| \Phi(\omega,z,\bz,\alpha) \big|^2 & = \big| a_0(\omega,z,\bz) \big|^2 + \Big[e^{2i\alpha }\Big(\bar{a}_0(\omega,z,\bz) a_2(\omega,z,\bz) + a_0(\omega,z,\bz)\bar{a}_{-2}(\omega,z,\bz)  \Big)+c.c.\Big]\nonumber\\
    &\hspace{5cm}+\mathcal{O}\left(\left(\frac{||\partial_z^2 \C||_\infty}{\omega}\right)^2 \right)\,.
\end{align}
In order to ensure the finiteness of the scalar field energy, we have at least $\omega^2 |a(\omega,z,\bz)|^2 = \mathcal{O}\big(\frac{1}{\omega}\big)$.
Therefore, the average supermomentum is the sum
\begin{align*}
    \langle \P(z,\bz) \rangle  &= \langle P(z,\bz) \rangle + \partial_z^2 \partial_{\bz}^2 \, \langle \N(z,\bz) \rangle
\end{align*}
of a hard contribution which, at high energy, coincides with the average supermomentum of the hard state $a_0(\omega, z, \bar{z})$,
\begin{align}
    \langle P(z,\bz) \rangle  := \int \omega^2 d\omega  \int \frac{d\alpha}{2\pi}\,\left| \Phi(\omega,z,\bar{z},\alpha) \right|^2 =
    \int_{\Lambda}^{\infty} \omega^2 d\omega\, \big| a_0(\omega, z, \bar{z}) \big|^2  +\mathcal{O}\left( \left(\frac{||\partial_z^2 \C||_\infty}{\Lambda} \right)^2 \right)
\end{align}
complemented by an effective memory effect
\begin{align*}
    \partial_{z}^2\partial_{\bz}^2\langle \N(z,\bz)\rangle &:= \int \omega d\omega  \int \frac{d\alpha}{2\pi}\, \Bigg(\frac{\sigma}{\omega}\left( e^{i2\alpha} \partial_z^2  + e^{-i2\alpha} \partial_{\bz}^2\right) \Bigg) \left| \Phi(\omega,z,\bar{z},\alpha) \right|^2\\
    =&\,   \partial_{\bz}^2 \left( \sigma \int d\omega  \int \frac{d\alpha}{2\pi}\, e^{-i2\alpha}\left| \Phi(\omega,z,\bar{z},\alpha) \right|^2 \right) + c.c.\\
    =&\, \partial_{\bz}^2 \left(\sigma\!\int_{\Lambda}^{\infty}\! d\omega\Big( \bar{a}_0(\omega,z,\bz) a_2(\omega,z,\bz) + a_0(\omega,z,\bz)\bar{a}_{-2}(\omega,z,\bz)\Big) \right)+ c.c.+ \mathcal{O}\left( \left(\tfrac{||\partial_z^2 \C||_\infty}{\Lambda}\right)^4 \right)\\
    =&\, \partial_{\bz}^2 \left(-\sigma\,\partial_{z}^2 \C(z,\bz) \int_{\Lambda}^{\infty}\! d\omega\,\big|a(\omega,z,\bz)\big|^2  \, \frac{\sin\big(\omega\,\C(z,\bz)\big)}{\omega}\right)\, + c.c.\,+\,\mathcal{O}\left( \left(\tfrac{||\partial_z^2 \C||_\infty}{\Lambda}\right)^4 \right)\,.
\end{align*}

Let us conclude this section by sketching a possible generalisation of the previous case to supermomenta $\P = P[\omega,\zeta,\bar\zeta]+\partial_z^2 \partial_{\bar z}^2 \N[\omega,\zeta,\bar\zeta,\alpha, \beta,\bar\beta]$ in a generic $SL(2,\mathbb{C})$-orbit, where the hard piece is $P\big[\omega,\zeta,\bar\zeta\big](z,\bar z)= \omega \,\delta^{(2)}(z-\zeta,\bar z-\bar\zeta)$ as usual. For such a massless BMS particle, consider a particular state in the decomposition \eqref{decompalphabeta}
\begin{align}\label{partstate}
     \Phi(\omega,\zeta,\bar\zeta,\alpha, \beta,\bar\beta) 
     =   e^{i\lambda\alpha} \, \widetilde{\Phi}_n
     (\omega,\zeta,\bar\zeta\,;\vec\pi)
\end{align}
which can be interpreted as the state of a massless Poincar\'e particle of helicity $\lambda$ and momentum $p_\mu=\omega\, q_\mu(\zeta,\bar\zeta)$ in a given gravitational vacuum. The corresponding BMS state in another gravitational vacuum is obtained by acting with the corresponding shift $\C$. Concreteley, the state \eqref{partstate} picks a phase 
\begin{align}
\langle \,\P\,,\,  \T\, \rangle &= \omega\,\C(\zeta,\bar\zeta) +  \big\langle\partial^2_z\,\N[\omega,\zeta,\bar\zeta,\alpha, \beta,\bar\beta]\,,\,\partial^2_z\C\,\big\rangle
\end{align}
due to \eqref{pairingcomputed1}. The first term on the right-hand side does not depend on the $ISO(2)$-ccordinates so it does not change the particle content but the second term does. In the Hilbert space topology, the space $\mathcal{O}(-3,1)$ of spin-weighted densities is a Hilbert space and the Cauchy-Schwarz inequality implies that the second term can be bounded as follows
\begin{align}
\Big\lvert\big\langle\partial^2_z\,\N[\omega,\zeta,\bar\zeta,\alpha, \beta,\bar\beta]\,,\,\partial^2_z\C\,\big\rangle\Big\rvert\;\leqslant\; 
\Big\lVert\,\partial^2_z\,\N[\omega,\zeta,\bar\zeta,\alpha, \beta,\bar\beta]\,\Big\rVert_2\;\Big\lVert\,\partial^2_z\C\,\Big\rVert_2 \,,
\end{align}
for the $L^2$-norm $\lVert\cdot\rVert_{_2}$ on the sphere.\footnote{The analogue of such a norm is a delicate issue in the nuclear topology, so it will not be investigated here.} For a shift $[\mathcal{C}]\in\E[1]/\mathbb{R}^{3,1}$ of gravitational vacua, the latter norm coincide with \eqref{Proof, main lemma: hermitian product}: $\llangle\C  , \C \rrangle = \lVert\partial^2_z\C\rVert_{2}$\,.
By analogy with the previous case, one would expect the superposition of states in distinct vacua to be controlled by the distance between these gravitational vacua (in the sense of Corollary \ref{distancebetweenvacua}).

\section{Conclusion and summary}\label{Section: concl}

Let us conclude by summarising our main results.

In Section \ref{Section: Hard/Softdecompositionofsupermomenta}, we introduced a Lorentz-invariant decomposition of supermomenta into hard and soft pieces (Proposition \ref{Prop: classification of supermomenta}). This decomposition is based on the following important properties: i) a supermomentum is soft iff it projects onto a vanishing momentum (Proposition \ref{Paneitz operator Lemma}); ii) the hard representations of the BMS group coincide with usual Poincaré representations, in particular,  they do not branch under restriction to a Poincar\'e subgroup (Corollary \ref{nobranchingtheo}) since the little group of a hard supermomentum is identical to the little group of its associated momentum (Subsection \ref{coincidingBondiandPoincarelittlegroups}). Some conceptual and technical subtleties related to the nonlinearity of the hard vs soft decomposition (Propositions \ref{Prop: massless hard supermomenta identity}-\ref{Prop: final hard supermomenta identity}) and their relations with Weinberg soft factors, were discussed in Subsection \ref{remarksaboutthedecomposition}.

We emphasised in Section \ref{Section: HardrepresentationsoftheBMSgroup} that the non-linearity of the decomposition means that hard states alone cannot fulfil conservation of supermomenta (Propositions \ref{hard_supermomenta_2to1}-\ref{hard_supermomenta_nto1}); therefore, in order to enforce this conservation law, other type of BMS particles need to be introduced. While hard BMS particles coincide with Poincar\'e particles, realised in terms of their natural scattering data (Subsection \ref{scatteringdata}), in all other cases, a single BMS particle is in general a BMS multiplet of usual Poincar\'e particles.

In Sections \ref{Section: MasslessUIRsBMS} and \ref{Section: MassiveUIRsBMS}, we provided the explicit form of the wavefunctions of generic BMS particles (Subsection \ref{ssection: Wavefunctions} and \ref{ssection: Massive representations of the BMS group}).
An explicit algorithm for the decomposition of these wavefunctions in suitable bases of functions (on the quotient of the Poincar\'e little group by the BMS little group) was provided for the decomposition of the BMS particle in terms of Poincar\'e particles.\footnote{Propositions \ref{massless_e_branching} and \ref{massless_U(1)_branching} were mentioned in \cite{McCarthy_73-III} but the direct sum of usual massless representations was not mentioned explicity. Furthermore, Propositions  \ref{massless_R2_branching}-\ref{massless_R_branching} are new and complete the analysis of all possible cases of branching in the massless case. Finally, Propositions \ref{massivebranchingBtoP}-\ref{massivebranchingbranchrule} of McCarthy were recovered in an explicit way.} In Section \ref{Ssection: Branching in different gravity vacua}, we stressed that this interpretation of generic BMS particles in terms of superposition of usual Poincaré particles crucially depend on the choice of gravity vacuum, that is on the choice of a Poincaré subgroup $ISO_0(3,1) \subset BMS_4$ for the branching. For instance, a state appearing as a simple scalar field in one gravity vacuum will be interpreted, in another gravity vacuum, as a complicated quantum superposition of particles of all possible spins (including, typically, continuous spins, in the massless case).

In Section \ref{Section: BMS wavefunctions}, the wavefunctions of BMS particle introduced in \cite{Bekaert:2024jxs} were discussed in terms of the homogenous spaces appearing in the asymptotic quantisation program (the space of cuts at null infinity, the space of gravitation vacua, etc). We stressed that there exists a Lorentz-invariant notion of distance between gravitational vacua (Corollary \ref{distancebetweenvacua}). We also pointed out that the Fock space of any countable collection of BMS particles is a separable Hilbert space (Subsection \ref{Fock}).

In Section \ref{Section: memoryeffect}, we exhibited, on a particular example of state, a concrete high-energy limit in which one can control the relative weight of the tower of higher-spin particles in the BMS multiplet and which furthermore allowed for an average supermomentum including memory. It remains to be seen if such types of limits may help to obtain a correspondence principle where usual S-matrix results are recovered, in some suitable regime, from a BMS-invariant S-matrix. At the end of this section, we briefly argued that the distance between gravitational vacua could play a role in such a mechanism. We also constructed examples of normalisable $BMS$ states allowing for a pole in the energy, with the resulting $L^2$ norm being soften by the presence of continuous-spin particles in the spectrum. The physical properties of such examples need to be explored further.

Another important direction to explore in the future would be to perform a systematic investigation of soft BMS particles (i.e. UIRs of BMS group with vanishing momentum) along the same lines as what we did for generic BMS particles. Another work in progress is the classification of the UIRs of the asymptotic symmetry group of quantum electrodynamics (QED) with massless charged fields, i.e. the group $SO(3,1)\ltimes\big(\mathbb{R}^{3,1}\times \E[0]\big)$. This could provide an alternative path for defining a Hilbert space of asymptotic multi-particle states suitable for an infrared-finite scattering theory of massless QFTs, such as QED with massless charged fields or Yang-Mills theory. This different road could perhaps circumvent some of the undesirable features mentioned e.g. in \cite{Prabhu:2022zcr}.

\section*{Acknowledgements}

We are grateful to G.~Barnich, R.~Gicquaud, B.~Oblak, E.~Skvortsov, B.~Valsesia for discusions. We also would like to acknowledge decisive discussions with L.~Donnay on the physical motivations underlying this work; in particular the present fresh perspective on BMS representations would not have been pursued without her persistent insistence that BMS representations ought to capture infrared physics. 

This research was partially completed during the thematic programme ‘Carrollian Physics and Holography’ at the Erwin Schr\"odinger Institute of Vienna,

\newpage

\appendix

\appendixpage

\section{Paneitz operator: image and kernel}\label{Appendix: proofsPaneitz}

In Appendix \ref{proofPaneitz}, we provide a proof that Proposition \ref{Paneitz operator Lemma} works for smooth densities. We prove that it extends to distributions in Appendix \ref{proofPaneitz:distrib}. 
Finally, in Appendix \ref{heuristic} we provide a proof of Proposition \ref{Paneitz operator Lemma} in the Hilbert topology. 
We include the latter proof because this reasoning turns out to be instructive and closer to usual computations (in terms of Hermitian forms and Hilbert spaces).

Let us assume that $\E[w]$ is the vector space of smooth conformal densities of weight $w$ on the celestial sphere. The topological  dual $\E[w]^\prime$ is spanned by the corresponding distributions (i.e. continuous linear forms on $\E[w]$). In particular, $\E[-2-w]\subset \E[w]^\prime$ since smooth conformal densities of weight $-2-w$ can be thought as distributions acting on smooth conformal densities of weight $w$. In particular,
\begin{align}
    \E[1] &\subset \E[-3]^\prime,& \E[-3] &\subset \E[1]^\prime.
\end{align}
where $\E[1]^\prime$ is the vector space of distributions corresponding to supermomenta. 
Let us denote $Ann(\mathbb{R}^{3,1}) \subset \E[-3]$ the space of smooth conformal densities  of weight $-3$ annihilating the translation subspace $\mathbb{R}^{3,1} \subset \E[-1]$ and $\big(Ann(\mathbb{R}^{3,1})\big)^\prime$ the corresponding dual space. One has
\begin{align}
  \E[1]/\mathbb{R}^{3,1} &\subset \big(Ann(\mathbb{R}^{3,1})\big)^\prime\,,& Ann(\mathbb{R}^{3,1})  &\subset  \big(\E[1]/\mathbb{R}^{3,1}\big)^\prime\,.
\end{align}

\subsection{Proof of Proposition \ref{Paneitz operator Lemma} in the smooth case}\label{proofPaneitz}

For smooth densities, Proposition \ref{Paneitz operator Lemma} can be expressed as the following

\begin{proposition}\label{Paneitzisom}
The Paneitz operator
\begin{equation}
     \eth^2 \bar{\eth}^2 : \E[1]/\mathbb{R}^{3,1} \stackrel{\sim}{\to} Ann(\mathbb{R}^{3,1})
\end{equation}
is an isomorphism in the smooth category.
\end{proposition}

\proof{This can be checked very concretely by performing some computations in terms of the representative $f\in C^\infty(S^2)$, for which one can use spherical harmonics $Y_{\ell}^m$ as a basis. From \cite[Eq. (3.24)]{Newman:1966ub} one has
\begin{equation}
     \eth^2\bar{\eth}^2  Y_{\ell}^m  = \gamma^{-3} \ell(\ell^2-1)(\ell+2) \, Y_{\ell}^m.
\end{equation}
Therefore the kernel of the Paneitz operator is spanned precisely by the representatives with only spherical harmonics $\ell=0$ and $\ell=1$ in their decomposition. In other words, the kernel coincides with translations $\mathbb{R}^{3,1} \subset \E[1]$.  Furthermore, the image of the Paneitz operator coincides with representatives whose spherical harmonics $\ell=0$ and $\ell=1$ vanish. Put differently, its image coincides with $Ann(\mathbb{R}^{3,1}) \subset \E[-3]$. This proves Proposition \ref{Paneitz operator Lemma} for smooth densities.}

\begin{remark}
Proposition \ref{Paneitzisom} can also be extracted from \cite[Sections III.3.3 and III.5.2]{Gelfand2}. More precisely, there exist  two short exact sequences of $SL(2,\mathbb{C})$ representations (cf. \cite[Fig.4 on p.155]{Gelfand2}):
\begin{equation}\label{B1}
\begin{tikzcd}
&&&\mathcal{O}(-3,1)\arrow[dr]&\\
0 \arrow[r] & \mathbb{C}^{3,1} \arrow[r,hookrightarrow,"q"]& \E_{\mathbb{C}}[1] \arrow[ur,twoheadrightarrow,"\eth^2"]\arrow[dr,twoheadrightarrow,"\bar{\eth}^2"]
& &0\\
&&&\mathcal{O}(1,-3)\arrow[ur]&
\end{tikzcd}
\end{equation}
and
\begin{equation}\label{B2}
\begin{tikzcd}
&\mathcal{O}(-3,1)\arrow[dr,hookrightarrow,"\bar{\eth}^2"]&&&\\
0 \arrow[ur]\arrow[dr] &&\E_{\mathbb{C}}[-3]\arrow[r,twoheadrightarrow,"\pi"]& \mathbb{C}^{3,1*} \arrow[r]
 &0\\
&\mathcal{O}(1,-3)\arrow[ur,hookrightarrow,"\eth^2"]&&&
\end{tikzcd}
\end{equation}
It is well-known that two short exact sequences of this form can be spliced together and are equivalent to a single long exact sequence (see e.g. \cite[Exercise 13 on p.180]{hungerford}). Applying this splicing procedure to \eqref{B1}-\eqref{B2} gives a long sequence which, when restricting to real conformal densities, yields \eqref{longexact}.
\end{remark}

\subsection{Proof of Proposition \ref{Paneitz operator Lemma} in the case of distributions}\label{proofPaneitz:distrib}

By duality, we deduce from Proposition \ref{Paneitzisom} the following
\begin{proposition}\label{Paneitzlemmafordistributions}
The Paneitz operator extends to an isomorphism
\begin{equation}
     \eth^2 \bar{\eth}^2 : \big(Ann(\mathbb{R}^{3,1})\big)^\prime \stackrel{\sim}{\to} \big(\E[1]/\mathbb{R}^{3,1}\big)^\prime\,.
\end{equation}
between spaces of distributions.
\end{proposition}

\proof{
Let us first prove surjectivity. If $\beta \in  \big(\E[1]/\mathbb{R}^{3,1}\big)^\prime$ we can define a distribution $\alpha \in \big(Ann(\mathbb{R}^{3,1})\big)^\prime$ via
\begin{align}
    \langle \alpha , \psi \rangle &:= \langle \beta , \chi \rangle\,,& \text{since}&& \forall \psi\in Ann(\mathbb{R}^{3,1})\,,\exists!\,\chi\in\E[1]/\mathbb{R}^{3,1}:\psi=\eth^2 \bar{\eth}^2\chi\,,
\end{align}
where we used Proposition \ref{Paneitzisom}.
Therefore,
\begin{align}
    \langle \beta , \chi \rangle &= \langle \alpha , \eth^2 \bar{\eth}^2\chi \rangle = \langle \eth^2 \bar{\eth}^2 \alpha , \chi \rangle\,,& \forall\chi \in \E[1]/\mathbb{R}^{3,1}\,.
\end{align}
Hence, $\beta = \eth^2 \bar{\eth}^2 \alpha$.

Let us now prove injectivity. If $\alpha \in \big(Ann(\mathbb{R}^{3,1})\big)^\prime$ satisfies $\eth^2 \bar{\eth}^2\alpha=0$, then
\begin{align}
    \langle \eth^2 \bar{\eth}^2\alpha , \chi \rangle=\langle \alpha , \eth^2 \bar{\eth}^2 \chi \rangle =0\,, \quad \forall \chi \in\E[1]/\mathbb{R}^{3,1}. 
\end{align}
Since any $\psi \in Ann(\mathbb{R}^{3,1})$ can always be written as $\psi = \eth^2 \bar{\eth}^2 \chi$ for some $\chi \in\E[1]/\mathbb{R}^{3,1}$, we deduce that $\alpha=0$.
}

From Proposition \ref{Paneitzlemmafordistributions}, we can deduce that soft supermomenta always belong to the image of the Paneitz operator.
\begin{corollary}
Any element $\P \in \E[-1]^\prime$ which annihilates the translation subspace $\mathbb{R}^{3,1} \subset \E[-1]$ takes the form $\P = \eth^2 \bar{\eth}^2 \N$ where $\N \in \big(Ann(\mathbb{R}^{3,1})\big)^\prime$. 
\end{corollary}

\proof{Since $\P$ annihilates translations by assumption, it defines a distribution on $\E[-1]/\mathbb{R}^{3,1}$ that we will denote with the same letter: $\P \in (\E[1]/\mathbb{R}^{3,1})^\prime$. By Proposition \ref{Paneitzlemmafordistributions}, this means that there exists $\N \in \big(Ann(\mathbb{R}^{3,1})\big)^\prime$ such that $\P = \eth^2 \bar{\eth}^2 \N$.
}

\subsection{Proof of Proposition \ref{Paneitz operator Lemma} in the Hilbert topology}\label{heuristic}

Let $\T \in \mathcal{E}[1]$ be a supertranslation and let us define the positive semi-definite quadratic form $\llangle.\, ,\, .\rrangle$ on $\mathcal{E}[1]$ given explicitly by the expression \eqref{Proof, main lemma: hermitian product}.
 In order for $\llangle\T  , \T\rrangle$ to exist, it is enough to assume that the representatives of supertranslations belong to the Sobolev space $H^2(S^2)$, i.e. $\mathcal{T}\in L^2(S^2)$ as well as all its derivatives up to second order. In particular, this works for smooth densities on the celestial sphere (and clearly does not work for distributions since their norm can be infinite).

The comments in the proof of Proposition \ref{innerproductprop} already ensure the exactness on the left in \eqref{longexact}: $\text{Im}(q)=\text{Ker}(\eth^2 \bar{\eth}^2)$. So one should prove the exactness on the right:
$\text{Im}(\eth^2 \bar{\eth}^2)=\text{Ker}(\pi)$. 
The inner product \eqref{innerprod} implies that $\text{Im}(\eth^2 \bar{\eth}^2)\subseteq\text{Ker}(\pi)$ since any supermomentum 
of the form $\eth^2 \bar{\eth}^2 \T_1$ annihilates all translations $\T_2$, that is to say $\langle   \eth^2 \bar{\eth}^2 \T_1 , \T_2 \rangle=0$
for any $\T_1 \in \mathcal{E}[1]$ and $\T_2\in \text{Im}(q)\simeq \mathbb{R}^{3,1}$. Since the scalar product \eqref{innerprod} is non-degenerate on the quotient $\mathcal{E}[1]\big/\mathbb{R}^{3,1}$, we have that $\text{Im}(\eth^2 \bar{\eth}^2) \simeq \mathcal{E}[1]/\mathbb{R}^{3,1}$ is a Hilbert space \cite[Section III.6.5]{Gelfand2} and, by Fr\'echet-Riesz theorem, 
\begin{equation}
    \begin{array}{ccc}
         \mathcal{E}[1]/\mathbb{R}^{3,1} &  \to &  (\mathcal{E}[1]\big/\mathbb{R}^{3,1})^\prime =Ann(\mathbb{R}^{3,1})\\[0.4em]
          \T & \mapsto & \llangle\,\T\, ,\,\, .\, \,\rrangle =  \langle \,  \eth^2 \bar{\eth}^2 \T\, , \,\,.\,\,\, \rangle
    \end{array}
\end{equation}
is an isomorphism. 
This concludes the proof in the Hilbert topology. 

\section{Distributional identity for hard supermomenta}\label{Appendix: Distributional identity for hard supermomenta}

In Appendices \ref{proof: identity for massless hard supermomenta} and \ref{proof: identity for massive hard supermomenta} below, we provide proofs for Proposition \ref{Prop: massless hard supermomenta identity} and Proposition \ref{Prop: massive hard supermomenta identity}, respectively.

\subsection{Proof of Proposition \ref{Prop: massless hard supermomenta identity}: identity for massless hard supermomenta}\label{proof: identity for massless hard supermomenta}

For completeness, let us explain the origin of the identity \eqref{delta identity}. It is based on the Cauchy–Pompeiu formula 
\begin{equation}\label{oint1}
    \T(\zeta,\bar\zeta)=\frac1{2\pi i}\oint_{\partial D}\frac{\T(z,\bar z)}{z-\zeta}\,dz+\frac1{2\pi i}\int_{D}\!\!\!dz\wedge d\bar z\; \frac{\partial_{\bar z}\T(z,\bar z)}{z-\zeta}
\end{equation}
where $D\subset\mathbb C$ is an open domain (topologically a disk) inside the complex plane and $\zeta\in D$. Making the replacement $\T \mapsto (\bar z-\bar\zeta)\partial_{\bar z}\T(z,\bar z)$ in \eqref{oint1} one obtains
\begin{align}\label{oint2}
0&= \frac1{2\pi i}\oint \frac{\bar z - \bar \zeta}{z-\zeta}\,\partial_{\bar z}\T(z,\bar z )\,dz  + \frac1{2\pi i}\int\!\!dz\wedge d\bar z\;\frac{\partial_{\bar z}[(\bar z-\bar\zeta)\partial_{\bar z}\T(z,\bar z)]}{z-\zeta}\\
&= \frac1{2\pi i}\oint \frac{\bar z - \bar \zeta}{z-\zeta}\,\partial_{\bar z}\T(z,\bar z )\,dz  + \frac1{2\pi i}\int\!\!dz\wedge d\bar z\; \frac{\partial_{\bar z}\T(z,\bar z)}{z-\zeta}+ \frac1{2\pi i}\int\!\!dz\wedge d\bar z\;\frac{\bar z-\bar\zeta}{z-\zeta}\partial_{\bar z}^2\T(z,\bar z)\nonumber
\end{align}
where the domain $D$ and the contour $\partial D$ have been and will be left implicit from now on.
Subtracting  \eqref{oint2} to \eqref{oint1}, one obtains
\begin{align}\label{Massles unicity proof: boundary term general expression}
    &\int\frac{i}{2}dz\wedge d\bar z\; \left( \T(z,\bar z) \,\delta^{(2)}(z-\zeta,\bar z - \bar \zeta) -\frac{1}{\pi}\partial_{\bar z}^2 \T(z,\bar z ) \frac{\bar z - \bar \zeta}{z-\zeta} \right) \\   &\; =\; \frac{1}{2\pi i}\oint \left(\frac{\T(z,\bar z ) }{z-\zeta} - \partial_{\bar z}\T(z,\bar z ) \frac{\bar z - \bar \zeta}{z-\zeta}\right)dz \,,\nonumber
\end{align}
which prove \eqref{delta identity}.

 Since $\T(z,\bar z ) = |\hat z|^{-2}\hat{\T}(\hat z, \bar{ \hat{z}})$ is a smooth globally-defined weight-one density on the celestial sphere, we have
\begin{align}\label{Massles unicity proof: asymptotic behavior of T}
    &\T(z,\bar z ) = |\hat z|^{-2}\hat{\T}(\hat z, \bar{ \hat{z}})\\
    &= |\hat z|^{-2}\left(\hat{\T}(0, 0) + \hat z\,\partial_{\hat{z}}\hat{\T}(0, 0) + \bar{\hat z}\,\partial_{\bar{\hat{z}}}\hat{\T}(0, 0) + \hat z \bar{\hat z}\,\partial_{\hat{z}}\partial_{\bar{\hat{z}}}\hat{\T}(0, 0) +\frac{\bar{\hat z}^2}{2}\partial_{\bar{\hat{z}}}^2\hat{\T}(0, 0)+\frac{\hat z^2}{2}\partial_{\hat{z}}^2\hat{\T}(0, 0)+ \mathcal{O}\big(|\hat z|^3\big)\right)  \nonumber\\
    &= z\bar{z}\,\hat{\T}(0, 0) +  \bar{ z}\,\partial_{\hat{z}}\hat{\T}(0, 0) +   z\,\partial_{\bar{\hat{z}}}\hat{\T}(0, 0) +\partial_{\hat{z}}\partial_{\bar{\hat{z}}}\hat{\T}(0, 0)  +\frac{z}{2\bar{z}}\partial_{\bar{\hat{z}}}^2\hat{\T}(0, 0)+\frac{\bar z}{2z}\partial_{\hat{z}}^2\hat{\T}(0, 0)+ \mathcal{O}\big(|z|^{-1}\big).\nonumber
\end{align} 
Making use of this asymptotic behavior, the boundary term on the right-hand side of \eqref{Massles unicity proof: boundary term general expression} can then be explicitly computed and gives a non zero contribution
\begin{align*}\label{nonzerocontribution}
 \frac1{2\pi i}  \oint \left(\frac{\T(z,\bar z ) }{z-\zeta} - \partial_{\bar z}\T(z,\bar z ) \frac{\bar z - \bar \zeta}{z-\zeta}\right)dz
   = \bar{\zeta}\zeta\,\hat{\T}(0, 0) +  \bar{ \zeta}\,\partial_{\hat{z}}\hat{\T}(0, 0) +   \zeta\,\partial_{\bar{\hat{z}}}\hat{\T}(0, 0) + \partial_{\hat{z}}\partial_{\bar{\hat{z}}}\hat{\T}(0, 0).
\end{align*}
To see this, first notice that, in the limit that we are considering, of infinite radius for our circular contour, the correction term $\mathcal{O}\big(|z|^{-1}\big)$ in the expression of $\T(z,\bar z)$ decays too fast to contribute to the boundary terms. The other contributions can then be computed by writing $z=r e^{i\theta}$ and performing the integral in $\theta$ at fixed radius $r$.  Therefore
\begin{align}
    &\int \frac{i}{2}dz\wedge d\bar z\;\left( \T(z,\bar z) \,\delta^{(2)}(z-\zeta,\bar z - \bar \zeta) -\frac{1}{\pi}\partial_{\bar z}^2 \T(z,\bar z ) \frac{\bar z - \bar \zeta}{z-\zeta} \right) \nonumber\\
    & \;\;=\;\; \bar{\zeta}\zeta\,\hat{\T}(0, 0) +  \bar{ \zeta}\,\partial_{\hat{z}}\hat{\T}(0, 0) +   \zeta\,\partial_{\bar{\hat{z}}}\hat{\T}(0, 0) + \partial_{\hat{z}}\partial_{\bar{\hat{z}}}\hat{\T}(0, 0) \\
    & \;\;=\;\; \int \frac{i}{2} dz\wedge d\bar z\; \T(z,\bar z) 
    \; q^{\mu}(\zeta,\bar\zeta)\,
    \mathcal{D}_{\mu}\big(\delta^{(2)}(z-\infty)\big).\nonumber
\end{align}

\subsection{Proof of Proposition \ref{Prop: massive hard supermomenta identity}: identity for massive hard supermomenta} \label{proof: identity for massive hard supermomenta}

Let us start from
\begin{equation*}
     \int \frac{i}{2}dz\wedge d\bz\, \T(z,\bar z) \,\frac{m^4}{\pi\big(q(z,\bar z) \cdot p\big)^3}  =  \frac{1}{\pi}\int \frac{i}{2}dz\wedge d\bz \, \T(z,\bar z) \, \partial_{\bar z}^2 \left(\frac{\big(\partial_z q(z,\bar z) \cdot p\big)^2 }{2\, q(z,\bar z) \cdot p}\right)
\end{equation*}
where we used the valid equality \eqref{notquitecorrect2} between functions on the complex plane.
Integrating by part, we obtain
\begin{align}\label{Massive unicity proof: boundary term general expression}
    &\int \frac{i}{2}dz\wedge d\bar z \left( \T(z,\bar z) \,\frac{m^4}{\pi(q \cdot p)^3} -\frac{1}{\pi}\partial_{\bar z}^2 \T(z,\bar z ) \nonumber\frac{(\partial_z q \cdot p)^2 }{2\, q \cdot p}\right) \\   &\; =\; \frac{1}{2\pi i}\oint \left(\T(z,\bar z ) \partial_{\bz}\left(\frac{(\partial_z q \cdot p)^2 }{2\, q \cdot p}\right) - \partial_{\bar z}\T(z,\bar z ) \frac{(\partial_z q \cdot p)^2 }{2\, q \cdot p}\right)dz \,,
\end{align}
where, as in Appendix \ref{proof: identity for massless hard supermomenta}, we implicitly integrate on the left-hand side on an open disk $D\subset\mathbb C$ inside the complex plane while the contour integral on the right-hand side is on the circle $\partial D$.

 Since $\T(z,\bar z ) = |\hat z|^{-2}\hat{\T}(\hat z, \bar{ \hat{z}})$ is a smooth globally-defined weight-one density on the celestial sphere, we have the asymptotic behavior \eqref{Massles unicity proof: asymptotic behavior of T}. On the other hand, making use of
 \begin{equation}\label{proof: identity for massive hard supermomenta, q.p}
     q(z,\bz) \cdot p = (p^3-p^0) + (p^1-i\,p^2) \; z + (p^1+i\,p^2) \;\bar z - (p^0+p^3)\;|z|^2\,,
 \end{equation}
 \begin{equation}
 \partial_zq(z,\bz) \cdot p =  (p^1-i\,p^2) - (p^0+p^3)\;\bar z\,,
 \end{equation}
we obtain
\begin{align}
    \frac{\big(\partial_z q(z,\bar z) \cdot p\big)^2 }{q(z,\bar z) \cdot p} \,=\,\frac{\bz}{z}\,&\Bigg( - (p^0+p^3) + \bar{z}^{-1}(p^1-i\,p^2) - z^{-1}(p^1+i\,p^2) \\
    &+ |z|^{-2}\left( p^0 -p^3\right)  - z^{-2} \frac{(p^1+i\,p^2)^2}{p^0+p^3} + O\left(|z|^{-3}\right) \Bigg).\nonumber
\end{align}
Using the above expansion as well as the expansion \eqref{Massles unicity proof: asymptotic behavior of T} of the supertranslation $\mathcal{T}(z,\bar z)$,
the boundary term on the right-hand side of \eqref{Massive unicity proof: boundary term general expression} can then be explicitly computed in the limit where the boundary becomes an infinitely large circle. First one should check that, in the limit of large radius that we are considering, the subleading terms $O\left(|z|^{-3}\right)$ in these asymptotic expansions do not contribute. It then turns out that several contributions of the leading terms vanish due to the angular integration $\int d\theta e^{i\theta} = \frac{1}{|z|}\oint dz$ in \eqref{Massive unicity proof: boundary term general expression}. Nevertheless, some terms survive the final angular integration and yield a non-zero contribution 
\begin{align}
 &\frac{1}{2\pi i}\oint \left(\T(z,\bar z ) \partial_{\bz}\left(\frac{(\partial_z q \cdot p)^2 }{2\, q \cdot p}\right) - \partial_{\bar z}\T(z,\bar z ) \frac{(\partial_z q \cdot p)^2 }{2\, q \cdot p}\right)dz\\
   &= \frac{p^3-p^0}{2}\,\hat{\T}(0, 0) -  \frac{p^1 -ip^2}{2}\,\partial_{\hat{z}}\hat{\T}(0, 0) - \frac{p^1 +ip^2}{2}  \,\partial_{\bar{\hat{z}}}\hat{\T}(0, 0) -\frac{p^0+p^3}{2}\partial_{\hat{z}}\partial_{\bar{\hat{z}}}\hat{\T}(0, 0). \nonumber
\end{align}
Therfore we conclude that
\begin{align*}
\int\frac{i}{2}dz\wedge d\bar z \left( \T(z,\bar z) \,\frac{-m^4}{\pi(q \cdot p)^3} +\frac{1}{\pi}\partial_{\bar z}^2 \T(z,\bar z ) \frac{(\partial_z q \cdot p)^2 }{2\, q \cdot p}\right) 
     &\;\;=\;\; \int \frac{i}{2}dz\wedge d\bar z\; \T(z,\bar z) \; p^{\mu} \mathcal{D}_{\mu}\big(\delta^{(2)}(z-\infty)\big).\nonumber
\end{align*}

\subsection{Proof of Proposition \ref{Prop: final hard supermomenta identity}: identity for massive and massless momenta} \label{proof: final identity for hard supermomenta}

Since $\partial_z^2q_{\mu}(z,\bz) = 0$ for all $z\in\mathbb{C}$, we have
\begin{equation}
    \partial_{ z}^2 \Big( (q \cdot p)\; \ln|q \cdot p| \Big) = \frac{(\partial_z q \cdot p)^2 }{q \cdot p}
\end{equation}
Thus, making use of Proposition \ref{Prop: massive hard supermomenta identity},
\begin{align*}
   &\int \frac{i}{2} dz \wedge d\bz\, \T(z,\bz) P(z,\bar z)\\ 
   &=-\frac{1}{\pi}\int \frac{i}{2} dz \wedge d\bz\, \partial_{\bar z}^2\T(z,\bz)
 \frac{\big(\partial_zq(z,\bz) \cdot p\big)^2}{2\, q(z,\bz)\cdot p}  +  \Big(p^{\mu}\mathcal{D}_{\mu}\T \Big)(z=\infty)\\
   &=-\frac{1}{2\pi}\int \frac{i}{2} dz \wedge d\bz\, \partial_{\bar z}^2\T(z,\bz)\, \partial_z^2 \Big( \big( q(z,\bz) \cdot p\big)\; \ln\big|q(z,\bz) \cdot p\big| \Big) +  \Big(p^{\mu}\mathcal{D}_{\mu}\T \Big)(z=\infty)
\end{align*}
integrating by parts gives 
\begin{align}\label{proof: uniqueness3, boundary term}
   &\!\!\int \!\frac{i}{2} dz \wedge d\bz \!\left(  \T(z,\bz) P(z,\bar z) + \frac{1}{2\pi}\partial_z^2\partial_{\bar z}^2\T(z,\bz) \, q(z,\bz) \cdot p\, \ln\big|q(z,\bz)\cdot p\big|  \right)  -\Big(p^{\mu}\mathcal{D}_{\mu}\T \Big)(z=\infty) \\
   &=     -\frac{1}{2i}\oint  \Bigg( \partial_{\bz}^2\T(z,\bz)\, \partial_z \Big( \big(q(z,\bz) \cdot p\big)\, \ln\big|q(z,\bz) \cdot p\big| \Big) -  \partial_z\partial_{\bz}^2\T(z,\bz)\, \big(q(z,\bz) \cdot p\big)\; \ln\big|q(z,\bz) \cdot p\big| \Bigg) d\bz.\nonumber
\end{align}
Now, on the one hand \eqref{proof: identity for massive hard supermomenta, q.p} implies that
 \begin{equation}
     \big(q(z,\bz) \cdot p\big)\; \ln\big|q(z,\bz) \cdot p\big|= - \,(p^0+p^3)\;|z|^2 \ln\Big(|p^0+p^3|\;|z|^2 \Big) + \mathcal{O}\Big(|z|\ln(|z|)\Big)
 \end{equation}
and, on the other hand, $\bar{\eth}^2 \T$ is a smooth spin-weighted density of weight $(w=-1, s=2)$ which behaves asymptotically as
\begin{equation}
    \partial_{\bz}^2 \T(z,\bz) =  \frac{z}{\bar{z}^3} \partial_{\bar{\hat{z}}}^2\hat{\T}(0, 0) + \mathcal{O}\Big(|z|^{-3}\Big).
\end{equation}
Here again, in the limit of large radius that we are considering, the subleading terms in these asymptotic expansions do not contribute to the boundary term \eqref{proof: uniqueness3, boundary term}. Finally, a careful calculation shows that the contribution of the leading order terms vanishes due to the angular integration.

\subsection{Proof of Corollary \ref{Corollary: sum of supermomenta}}\label{proofcorollary}

Let $\{ p^{\mu}_i \}$ be a collection of momenta satisfying momentum conservation, $\sum_i p^{\mu}_i =0$. Let us consider the combination $\mathcal{S} = -\frac{1}{2\pi} \sum_i (q\cdot p_i) \ln|q\cdot p_i|$. From \eqref{proof: identity for massive hard supermomenta, q.p}, we can write 
\begin{equation}
  q(z,\bz)\cdot p_i = -(p^0_i + p^3_i)|z^2|\Big(1 + F_i(z,\bz) \Big),
\end{equation} where $F_i(z,\bz)=\mathcal{O}\big(|z|^{-1}\big)$ are functions which are smooth around $z=0$. Therefore,
\begin{align}
    \mathcal{S}(z,\bz) &= -\frac{1}{2\pi} \sum_i q(z,\bz)\cdot p_i\; \ln\Big[( |p^0_i + p^3_i|\,|z^2|\Big( 1 + F_i(z,\bz) \Big)\Big]\\
    &= -\frac{1}{2\pi} \sum_i  q(z,\bz)\cdot p_i\;\ln\Big( |p^0_i + p^3_i|\Big) -\frac{1}{2\pi} \sum_i  q(z,\bz)\cdot p_i\; \ln\Big(1 + F_i(z,\bz)\Big)\label{lastofeqs}
\end{align}
where we used momentum conservation to drop the extra term 
\begin{equation}
\sum_i q(z,\bz)\cdot p_i \ln\big( |z|^2\big)=\ln\left( |z|^2\right) q(z,\bz)\cdot \big(\sum_i p_i\big) =0\,.    
\end{equation}
One sees that both terms in \eqref{lastofeqs} define smooth densities of weight one around $z=\infty$ and, therefore, so does $\mathcal{S}$.

\section{Haar mesure on \texorpdfstring{$SL(2,\mathbb{C})$}{SL(2,C)}}\label{Appendix: Haar mesure} 

We here parametrises elements $g\in SL(2,\mathbb{C})$ as in \eqref{BMS massless rep: SL(2,C) factorisation}.

The Killing metric on $SL(2,\mathbb{C})$ is given by
\begin{align*}
    g_{Killing} = \text{Re} \Big[\text{Tr}\big[ (g^{-1} dg) \odot (g^{-1} dg)\big] \Big] 
\end{align*}
where $\odot$ is the symmetric tensor product (which we most often leave implicit in the expression of the metric, as is common use in physics literature).
We have
\begin{align*}
    \text{Tr}\big[ (g^{-1} dg) \odot (g^{-1} dg)\big]  & = \text{Tr} \big[(N^{-1} dN)^{2}\big]+ \text{Tr} \big[L^{-1} dL)^2\big] +2\, \text{Tr}\big[(N^{-1} dN) \odot (dL \,L^{-1})\big]
\end{align*}
and direct computation gives:
\begin{align}
    N^{-1} dN &= \begin{pmatrix}
        -\frac{1}{2}\omega^{-1}d\omega & \;\omega d\zeta \\0 & \;\frac{1}{2}\omega^{-1}d\omega
    \end{pmatrix}\\
    L^{-1}dL &= \begin{pmatrix}
        \frac{i}2d\alpha & \quad0 \\d \beta - i e^{i\tfrac{\alpha}{2}}\beta\, d \alpha &\quad -\frac{i}2 d\alpha
    \end{pmatrix}\\
     dL \,L^{-1} &= \begin{pmatrix}
        \frac{i}2d\alpha &\; 0 \\ e^{-i \alpha}d \beta & \;-\frac{i}2 d\alpha
    \end{pmatrix}
\end{align}
and thus
\begin{align}
    \text{Tr}\big[(L^{-1} dL)^2\big] &= -\frac12 \,d \alpha ^2\\
    \text{Tr}\big[(N^{-1} dN)^2\big] &= \frac{1}{2} \,\omega^{-2} d \omega ^2\\
    \text{Tr}\big[(N^{-1} dN) \odot (dL L^{-1})\big] &= -\frac{i}2 \omega^{-1}\,d\omega d\alpha + \omega e^{-i\alpha}\,d\zeta d\beta
\end{align}
and therefore
\begin{align*}
    g_{Killing} = -\frac12 \,d \alpha^2 + \frac{1}{2} \,\omega^{-2} d \omega^2 +  \omega \left( e^{-i\alpha}\,d\zeta d\beta + e^{i\alpha}\,d\bar\zeta d\bar\beta \right)
\end{align*}
Therefore
\begin{equation}
    \det(g_{Killing}) = -2 \times \left(\frac12\omega^{-2}\right) \times \omega^4 = \omega^2
\end{equation}
and thus the volume element of the Killing metric reads
\begin{equation}
    \mu_{Haar} = 
    \sqrt{\det\big(g_{Killing}\big)} \,d^6X  = \frac12\,\omega \,d\omega\, d\zeta_1\, d\zeta_2\, d \alpha\, d\beta_1\, d\beta_2\,,
    \end{equation}
where $\zeta=\zeta_1+i\,\zeta_2$ and $\beta=\beta_1+i\,\beta_2$.

\section{Action of supertranslations in generic massless BMS representations}\label{Appendix: transformedsupermomentum}

This appendix provides the proof of Proposition \ref{Action of supertranslations_prop}.

By definition, supertranslations act via a phase $\langle\P,\T\rangle$ where the supermomenta takes the form \eqref{BMS massless rep: generic massless supermomenta}. This produces the action \eqref{BMS massless rep: supertranslation shift}
where the phase $\varphi$ is given by the integral of the supertranslation $\T$ multiplied by the soft part in \eqref{BMS massless rep: generic massless supermomenta}. Integrating by part the Paneitz operator, one obtains
\begin{align}
    \varphi &= \int dz^2 \; \partial_{z}^2 \partial_{\bar z}^2 \T(z,\bar z)\;  \left|\frac{\partial z'}{\partial z}\right|^{-1} \N(z',\bar z'),  
    \nonumber\\
    &= \int dz'^2 \;\partial_{z}^2 \partial_{\bar z}^2 \T(z,\bar z)\; \left|\frac{\partial z}{\partial z'}\right|^{3} \N(z',\bar z'), \nonumber\\
     &=  \omega^{-3}\int dz'^2 \;\partial_{z}^2 \partial_{\bar z}^2 \T(z,\bar z)\; \left|1 + \beta z'\right|^{-6} \N(z',\bar z'), \nonumber
\end{align}
where \eqref{dz/dz'} was used in the last equality.
Moreover, if one does not integrate by part in the phase $\varphi$, one can also write
\begin{align}    \varphi &= \int dz^2 \; \T(z,\bar z)\;  \left|\frac{\partial z'}{\partial z}\right|^{3} \partial_{z'}^2 \partial_{\bar z'}^2 \N(z',\bar z'), & z'&=\frac{\omega(z - \zeta)}{e^{i\alpha} - \beta\, \omega (z-\zeta)} \nonumber\\
    &= \int dz'^2 \; \T(z,\bar z)\;  \left|\frac{\partial z}{\partial z'}\right|^{-1} \partial_{z'}^2 \partial_{\bar z'}^2 \N(z',\bar z'), & z&=\frac{ (e^{i\alpha}+ \omega \zeta \beta) \,z' + \omega\zeta}{\omega( 1 + \beta z')}\nonumber\\
    &= \omega \int dz'^2 \; \T(z,\bar z)\; \left| 1 + \beta z'\right|^{2} \partial_{z'}^2 \partial_{\bar z'}^2 \N(z',\bar z'),\nonumber
\end{align}
where again \eqref{dz/dz'} was used in the last step.
Integrating by part gives
\begin{align}
    \varphi
    = &\, \omega \int dz'^2 \; \partial_{z'}^2\left(\T(z,\bar z)\; \left| 1 + \beta z'\right|^{2}\right)  \partial_{\bar z'}^2 \N(z',\bar z')\nonumber\\
    = &\, \omega \int dz'^2 \; \left( \left| 1 + \beta z'\right|^{2}  \partial_{z'}^2 \T(z,\bar z) + \;2 \beta(1+\bar\beta \bar z') \partial_{z'} \T(z,\bar z) \right)  \partial_{\bar z'}^2 \N(z',\bar z')\nonumber\\
    =& \,\omega \int dz'^2 \; \left| 1 + \beta z'\right|^{2}  \left(  \frac{\partial^2 z}{\partial z'^2}\partial_{z} \T(z,\bar z)+ \left(\frac{\partial z}{\partial z'}\right)^2\partial_{z}^2 \T(z,\bar z)  + \frac{\partial z}{\partial z'}\; \frac{2\beta}{1+\beta z'} \partial_{z} \T(z,z)  \right)  \partial_{\bar z'}^2 \N(z',\bar z') \nonumber\\
    =& \,\omega \int dz'^2 \; \left| 1 + \beta z'\right|^{2}  \left(\frac{\partial z}{\partial z'}\right)^2\; \partial_{z}^2 \T(z,\bar z)  \; \partial_{\bar z'}^2 \N(z',\bar z')\nonumber\\
=&\, \frac{e^{2i\alpha}}{\omega} \int dz'^2 \;\left(\frac{1+ \bar \beta \bar z'}{1+\beta z'}\right)^2\; \partial_{z}^2 \T(z,\bar z)  \; \partial_{\bar z'}^2 \N(z',\bar z') .\nonumber
\end{align}

\section{Continuous-spin representations of the Poincar\'e group}\label{Appendix: CSP} 

We here briefly recap the ``continuous-spin'' UIRs of the Poincar\'e group (for more details, see e.g. the reviews \cite{Bekaert:2006py,Bekaert:2017khg} and refs therein). Before that, it is useful to describe the UIRs of the massless little group, i.e. the UIRs of the Euclidean group (see e.g. \cite[Chapter 9]{Tung} for a basic introduction).

\subsection{Unitary irreducible representations of the Euclidean group}

The Euclidean group $ISO(2)=SO(2)\ltimes\mathbb{R}^2$ is a semidirect product of the rotation group $SO(2)$ acting on the Abelian normal subgroup $\mathbb{R}^2$. Therefore, the corresponding UIRs are again obtained from Wigner's method of induced representations. 

The first step of this procedure is to pick an eigenvalue of the translation generators of the subgroup $\mathbb{R}^2\subset ISO(2)$: this is nothing but a vector $\vec\pi\in\mathbb{R}^2$. The second step amounts to pick a UIR of the stabiliser of this vector. Accordingly, there are two cases to distinguish:
\begin{enumerate}
    \item $\underline{\vec\pi=\vec 0}:$ The degenerate case $\vec\pi=\vec 0$ corresponds to representations which are unfaithful UIRs of $ISO(2)$ but which are faithful UIRs of $SO(2)$ instead. Therefore, the UIRs are labeled by a single number $\lambda \in \mathbb{R}$ and they pick a phase $e^{i\lambda\theta}$ under a rotation of angle $\theta$. This label may be called the ``helicity''. This label $\lambda$ is clearly an integer for single-valued representations and a half-integer for double-valued representations of $ISO(2)$, i.e. $\lambda\in\tfrac{1}{2}\mathbb{Z}$ for UIRs of $\ISO$.
    \item $\underline{\vec\pi\neq\vec 0}:$ The generic case $\vec\pi\neq\vec 0$ corresponds to faithful UIRs of $ISO(2)$. Since the stabiliser in $SO(2)$ of a plane vector is the identity, the second step in the method of induced representations is absent. Therefore, these UIRs of $ISO(2)$ are only labeled by the norm $\mu:=|\vec\pi|\in\mathbb{R}^+$ (since the orbits of a vector in the plane under rotations are labeled by the radius of the vector). 
\end{enumerate}

Several remarks are in order because they will be instrumental for the identification of continuous-spin modes inside wavefunctions of massless BMS particles in Section \ref{Ssection: branching}.

\begin{remark}
    In practice, the UIRs in the second case are carried by square-integrable functions on the circle of radius $\mu$ in the plane of vectors $\vec\pi$. However, as always there are infinitely many equivalent realisations of a given UIR. For instance, the UIR can be represented by wavefunctions $\Psi(\pi_1,\pi_2)$ on the plane with support on the circle $(\pi_1)^2+(\pi_2)^2=\mu^2$. Equivalently, in dual space they are square-integrable solutions $\Psi(\beta_1,\beta_2)$ of the Helmholtz equation 
    \begin{equation}\label{Helmholtz}
        (\partial^2_{\beta_1}+\partial^2_{\beta_2}+\mu^2)\Psi=0\,.
    \end{equation}
\end{remark}

\begin{remark}\label{helicityremark}
Furthermore, one may equivalently consider realisations of the same UIRs of $ISO(2)$ in terms of vector (or tensor) fields on the plane. Therefore, although these fields may have non-vanishing helicities, they would nevertheless carry an equivalent UIR, by assumption. This point will be important later on. For instance, consider for a given $s\in\mathbb N$ the rank-$s$ symmetric tensor fields with indices taking two values,
$\Psi_{i_1\,\cdots\,i_{s}}=\partial_{\beta_{i_1}}\cdots\partial_{\beta_{i_s}}\Psi$, defined as derivatives of the
square-integrable solutions $\Psi(\beta_1,\beta_2)$ of the Helmholtz equation \eqref{Helmholtz}.\footnote{This corresponds to the usual realisation of UIRs of $ISO(2)$ in terms of symmetric tensor fields $\Phi_{i_1\,\cdots\,i_s}(\beta_1,\beta_2)$ which are divergenceless $\partial_{\beta_i}\Phi^{i}{}_{j_1\,\cdots\,j_{s-1}}=0$ and solutions of the Helmholtz equation $(\partial^2_{\beta_1}+\partial^2_{\beta_2}+\mu^2)\Phi_{i_1\,\cdots\,i_s}=0$. Hodge dualising all indices implies that $\Psi_{i_1\,\cdots\,i_s}:=\epsilon_{i_1j_1}\cdots\epsilon_{i_sj_s}\Phi^{j_1\,\cdots\,j_s}$ obeys to
the closure condition $\partial_{[i}\Psi_{j]k_1\,\cdots\,k_{s-1}}=0$. By Poincar\'e lemma, this is equivalent to the exactness condition $\Psi_{i_1\,\cdots\,i_{s}}=\partial_{\beta_{i_1}}\cdots\partial_{\beta_{i_s}}\Psi$.}
The correspondence between the tensor $\Psi_{i_1\,\cdots\,i_{s}}$ and the scalar $\Psi$ is one-to-one because the latter is assumed square-integrable (thus the polynomial solutions of the homogeneous equation $\partial_{\beta_{i_1}}\cdots\partial_{\beta_{i_s}}\Psi=0$ are not admissible). Introducing the complex variable $\beta=\beta_1+i\beta_2$\,, one can equivalently consider the complex field 
$\tilde\Psi_{2s}:=\partial^s_{\bar\beta}\Psi$ defined as the $s$-th complex derivative of the
square-integrable solutions $\Psi(\beta,\bar\beta)$ of the Helmholtz equation $(4\partial_{\beta}\partial_{\bar\beta}+\mu^2)\Psi=0$. Note that this field $\tilde\Psi_{2s}$ transforms as $\tilde\Psi_{2s}\to e^{is\theta}\tilde\Psi_{2s}$ under the rotation $\beta\to e^{i\theta}\beta$. Therefore the field $\Psi_{2s}$ has helicity $s$ but nevertheless carries a representation of $ISO(2)$ equivalent to the representation carried by $\Psi$.
\end{remark}

\begin{remark}\label{Branching: Rm, second realisation of continuous spin}
Under translations $\vec\beta'=\vec\beta+\vec B$, the Fourier transform $\Psi(\vec\pi)$ of the function $\Psi(\vec\beta)$ merely picks a phase 
\begin{equation}\label{translat}
    \Psi'(\vec\pi)=e^{i\vec\pi\cdot\vec B}\Psi(\vec\pi)\,,
\end{equation}while under rotations $\vec\beta'= R_\theta(\vec\beta)$, it transforms as $\Psi'(\vec\pi)=\Psi(\vec\pi')$ where $\vec\pi'=R_\theta(\vec\pi)$. 
For an irreducible representation, the function $\Psi(\vec\pi)$ on the plane actually has support on the circle $S^1_\mu$ of radius $\mu$ in the plane, i.e. the circle $\vec\pi^2=\mu^2$.
It belongs to the Hilbert space $L^2(S^1_\mu)$ of square-integrable functions on the circle $S^1_\mu\subset\mathbb{R}^2$. One can alternatively consider wavefunctions $\Psi(\vec\pi)$ belonging to the Hilbert space $L^2(\mathbb{R}^2)$ of square-integrable functions on the plane and introduce polar coordinates $(\mu,\alpha)$ on the latter. In particular, a wavefunction $\Psi(\vec\pi)$ with support on the circle $S^1_\mu$ is equivalent to a function $\Psi(\alpha;\vec\pi_0)$ of an angle only, while $\vec\pi_0$ is a choice of reference vector on the circle. The angle $\alpha$ is defined by $\vec\pi=R_\alpha(\vec\pi_0)$. Accordingly, rotations act by mere shifts $\alpha'=\alpha+\theta$ of the angle while translations \eqref{translat}
act as 
\begin{equation}
    \Psi'(\alpha;\vec\pi_0)=e^{iR_\alpha(\vec\pi_0)\cdot \vec B}\Psi(\alpha;\vec\pi_0)
\end{equation}
This is again an equivalent realisation of the same UIR of $ISO(2)$.
\end{remark}

\begin{remark}\label{separableISO(2)}
The Hilbert space $L^2(S^1_\mu)$ of square-integrable functions on the circle is of course separable\footnote{\label{separabletheorem}A Hilbert space is separable if and only if it has one countable orthonormal basis (see e.g. \cite[Theorem 5.11]{Brezis} and \cite[Theorem 3.52]{Rynne}). Actually, there is more: a Hilbert space is separable if and only if every orthonormal basis is countable (see e.g. \cite[Problems 7 and 17]{Halmos}).} (cf. the orthonormal basis of Fourier modes), although this is not manifest in terms of the plane-wave basis of solutions $e^{i\vec\pi\cdot\vec\beta}$ of the Helmholtz equation \eqref{Helmholtz} (see e.g. \cite[Sections 9.2-9.3]{Tung} for the explicit relations between these two bases).
The space $L^2(\mathbb{R}^2)$ of square-integrable functions on the plane is a separable Hilbert space which carries a unitary but reducible representation of $ISO(2)$. This representation obviously decomposes into a direct integral of  UIRs of $ISO(2)$ as follows
\begin{equation}\label{decompL2(R2)}
    L^2(\mathbb{R}^2)=\int_0^\infty \hspace{-2.5mm}d\mu\;L^2(S^1_\mu)
\end{equation}
Let us stress that $L^2(\mathbb{R}^2)$ is a separable Hilbert space (cf. the harmonic oscillator basis), although this may look in tension with the above decomposition since there is a continuous sum of UIRs on the right-hand-side.  Nevertheless, there is no contradiction because the decomposition \eqref{decompL2(R2)} is only natural in the plane-wave basis $e^{i\vec\pi\cdot\vec\beta}$ whose elements are not normalisable (so they do no provide an orthonormal basis which would be in contradiction with separability, cf. Footnote \ref{separabletheorem}).
\end{remark}

\subsection{Massless representations of the Poincar\'e group}

The Poincar\'e little group of a null momentum $p_\mu$ is isomorphic to the Euclidean group, $\ell_p\simeq ISO(2)$.
The step 2 in Theorem \ref{Wignertheo} implies that the massless UIRs of $ISO_0(3,1)$ are induced from UIRs of the Euclidean subgroup $ISO(2)\subset SO_0(3,1)$. Accordingly, one may introduce a vector $\vec\pi$ describing the spinning degrees of freedom. As explained above, there are two cases to distinguish:
\begin{enumerate}
    \item $\underline{\vec\pi=\vec 0}:$ These UIRs of the little group $ISO(2)$ are unfaithful. In fact, they are genuine UIRs of $SO(2)$ labeled by a single integer or half-integer $\lambda\in\tfrac12\mathbb Z$ according to whether the representation is single or double valued. They induce UIRs of the Poincar\'e group $ISO_0(3,1)$ which are the usual massless representations, where $\lambda$ is the helicity.
    \item $\underline{\vec\pi\neq\vec 0}:$ These UIRs of $ISO(2)$ are labeled by the norm $\mu=|\vec\pi|\in\mathbb{R}^+$. The corresponding UIRs of the Poincar\'e group $ISO_0(3,1)$ are the continuous-spin representations. Their name originates from the fact that they are labeled by a continuous parameter: $\mu$. The only extra information is whether the UIR of $ISO_0(3,1)$ is single or double valued.  
\end{enumerate}

\newpage

\bibliographystyle{utphys}
\bibliography{BMSUIR}

\end{document}